\documentclass{aa}
\usepackage{lineno}
\usepackage[breaklinks=true]{hyperref}
\hypersetup{colorlinks,linkcolor=blue,citecolor=blue,urlcolor=blue}
\usepackage{natbib}
\bibpunct{(}{)}{;}{a}{}{,}
\usepackage{graphicx}
\usepackage{txfonts}
\usepackage{lscape}
\usepackage{longtable}
\usepackage{multirow}
\usepackage{multicol}
\usepackage{amssymb}
\usepackage{pifont}
\usepackage{geometry}
\usepackage{pdflscape}
\usepackage[fleqn]{amsmath}
\usepackage{color,soul}
\usepackage{comment}
\DeclareUnicodeCharacter{2212}{-}
\begin{document}

\title{OB runaway stars originating in the Vel OB1 association}

 \author{N. Azatyan{\inst1} \and L. Kaper{\inst2}  \and A. Samsonyan{\inst1}  \and M. Stoop{\inst2}  \and D. Andreasyan{\inst1} \and J. van den Eijnden{\inst{2}} \and E. Nikoghosyan{\inst1}}

\institute{{\inst1}Byurakan Astrophysical Observatory, 0213, Aragatsotn prov., Armenia\\ \email{nayazatyan@bao.sci.am}\\
{\inst2}Anton Pannekoek Institute for Astronomy, University of Amsterdam, Science Park 904, 1098 XH Amsterdam, the Netherlands\\}

   \date{Received ; accepted }

 \abstract
{OB~runaway stars are massive stars moving through interstellar space with a high velocity (up to 200\,km~s$^{-1}$). They are produced by dynamical ejections in young massive clusters or by supernova explosions in massive binaries. OB~runaways can travel several hundred parsec before exploding as supernovae, affecting the dynamical and chemical evolution of the Galaxy.}
{The Vel\,OB1~association is one of the largest OB~associations, hosting about 20\,O-type and more than 50\,B-type~stars. Our aim is to find OB~runaway stars in this region. By quantifying their number and identifying their parent clusters, we seek to better understand their production channels and their impact on the surrounding medium.}
{We used \textit{Gaia}\,DR3 coordinates, parallaxes, and proper motions of massive stars in the field centred on Vel\,OB1 to identify OB~runaways by measuring their peculiar velocity. Under suitable physical conditions, OB~runaways create observable bow shocks in the interstellar medium~(ISM). We inspected infrared WISE~images to identify wind bow shocks and their associated OB~runaways. By reconstructing their path, we tried to locate their parent cluster and estimate their travel times.}
{We identified six~young stellar clusters hosting most of the massive-star population in Vel~OB1 (distance $1.6-2.1$~kpc; age $1-10$~Myr). From the tangential velocity distribution of the members, we derived a threshold velocity of 15~km~s$^{-1}$ to classify a star as a runaway. We identified 25\,OB~runaways (including the high-mass X-ray binary Vela~X-1) and one F-type~runaway. We detected 16\,arc-like~features, four for the first time, and six of the features are associated with OB~runaways selected by peculiar velocity. Ten bow shocks are aligned with the proper motion of the runaways. Parent clusters are identified for seven runaways. Most likely, the majority of these runaways are produced by dynamical ejection.} 
{The runaway fraction of the young stellar population in Vel\,OB1 is about 30~\%. Many OB~runaways, even some far above the Galactic plane, produce wind bow shocks, which consequently reveal information on local ISM conditions.}
  

\maketitle
\nolinenumbers

\section{Introduction}
\label{In}
Massive stars in the Milky Way are predominantly found in organised stellar groupings rather than being uniformly distributed across the Galactic disc. Observational surveys have indicated that roughly 80\% of O-type stars reside in clusters or associations associated with spiral arms \citep{brown99}. The remaining population displays kinematic properties that are inconsistent with in situ formation, implying that these stars originated in clustered environments but were subsequently ejected; such objects are commonly referred to as runaway stars \citep{blaauw93,Wit05,Gvaramadze12}. Two main physical mechanisms have been proposed to account for the origin of OB runaways and their large peculiar velocities. In one channel, close gravitational encounters in young, dense clusters dynamically expel massive stars at high velocities \citep{poveda67,gvaramadze09,fujii11,Stoop24b}. In the other, a supernova explosion in a massive binary system disrupts the orbit and accelerates the surviving companion \citep{blaauw61,vanrensbergen96,kaper97}. In this supernova-driven scenario, which assumes symmetric mass loss during the explosion, the ejection velocity of the secondary is set by its pre-explosion orbital motion. The binary remains gravitationally bound only if the fraction of mass removed is less than half of the total system mass \citep{blaauw61,boersma61}.

A growing body of observational and theoretical work suggests that both mechanisms operate in nature \citep{gies87,hoogerwerf00,kaper00,gualandris04,boubert17}, although their relative importance remains uncertain. In addition, combined pathways have been proposed in which a massive binary is first ejected from its natal cluster through dynamical interactions and subsequently experiences an additional velocity boost following a supernova explosion of one component \citep{pflamm-altenburg10,gvaramadze18,Dorigo20}. In turn, the study of OB runaways provides insight into the physical conditions of young massive clusters.

The velocity threshold historically adopted to define O- and B-type runaway stars is $\sim$~30\,km\,s$^{-1}$, as the typical (radial) velocity of massive stars is of the order of 10~km~s$^{^-1}$ with respect to the local standard of rest (LSR), as was the accuracy of these measurements back then \citep[e.g. ][]{blaauw61,hoogerwerf00,Wit05,Eldridge11,Boubert18}.More recent binary evolution models predict that a significant fraction of ejected systems acquire substantially lower velocities, leading to the introduction of the term `walkaway stars' for these objects \citep{Renzo19}. But there is no physical reason why a threshold velocity of $30$~km~s$^{-1}$ has been introduced, and the escape velocity of the parent cluster (a few km~s$^{-1}$) may be more appropriate.

\begin{figure*}
\centering
\includegraphics[width=0.9\linewidth]{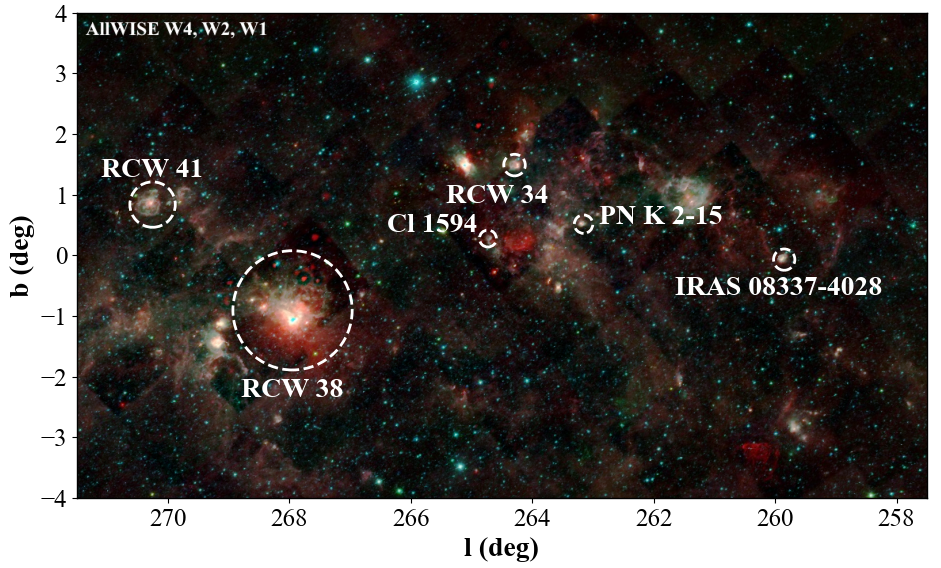}
\caption{Vel~OB1 association: the extension on the sky according to the limits proposed by \citet{reed00}. Indicated are several young stellar clusters embedded in this region, within the distance interval of about 1.6 to 2.5~kpc. Some star-forming regions and clusters (not labelled) are in the foreground to Vel~OB1, e.g. Vel~OB2 (just below the plotted region at 410~pc) and RCW~36 ($l = 265.1, b = 1.4$, at a distance of 700~pc). The background image is a colour composite of the AllWISE survey in bands W1, W2, and W4. Milky Way Star Cluster N\textsuperscript{\underline{o}}\,1594 ([KPS2012]\,MWSC\,1594, \citet{kharchenko13}) is indicated as Cl\,1594.}
\label{fig:VelOB1}
\end{figure*}

In addition to their anomalously large space velocities, massive runaway stars are often characterised by the presence of wind-driven bow shocks \citep[e.g. ][]{buren88,gvaramadze08,peri12,Mackey2023}. These bow shocks typically extend over parsec scales and emit strongly at infrared wavelengths, which makes them observable over large distances despite the effects of interstellar extinction \citep[e.g. ][]{gvaramadze10,gvaramadze11}. While early studies focused primarily on their mid-infrared signatures, subsequent observations have revealed that bow shocks associated with runaway stars can also be detected across a broader range of wavelengths, including optical and radio bands \citep{kaper97,gvaramadze08,gvaramadze14,benaglia10,eijnden22,Moutzouri25}.

OB runaway stars provide a valuable means of tracing the properties of the surrounding interstellar medium (ISM), as their strong stellar winds interact directly with the ambient gas during their motion. When a runaway star travels at supersonic velocity relative to the local medium, the momentum flux of the stellar wind can balance the external ram pressure, leading to the formation of a wind-driven bow shock \citep{baranov71,comeron98}. Observational searches have revealed such structures around a fraction of OB runaways, with \citet{buren95} reporting a detection rate of roughly 30\%. The absence of detectable bow shocks in many cases can be attributed to variations in ISM conditions. For instance, a high sound speed in the surrounding gas may prevent the development of a supersonic flow, or a low ambient density may reduce the observability of the interaction \citep{huthoff02}. Moreover, the alignment of the bow shock with the (peculiar) velocity of the runaway depends on the relative velocity of the system with respect to the ISM \citep{Bodensteiner18}.

OB runaways provide important information on the dynamical evolution of stellar clusters (dynamical ejection scenario), and on binary evolution (supernova scenario). A key step in such analyses is the identification of the parent cluster or association, as this enables an estimate of the kinematic age of the runaway. When this timescale is comparable to the age of the host cluster, a dynamical ejection origin is favoured. Conversely, a substantially shorter kinematic age points towards a supernova-related ejection from a massive binary. In this case, the cluster age at the moment of ejection constrains the initial mass of the progenitor that underwent core collapse, while the present-day velocity of the runaway provides information on the amount of mass ejected during the explosion \citep{ankay01,vandermeij21}. As a natural consequence of the supernova channel, high-mass X-ray binaries (HMXBs)—systems composed of a massive donor star and a compact remnant—are expected to exhibit runaway characteristics \citep{vandenheuvel00}. \citet{Carretero23} confirmed seven HMXBs as runaways out of a sample of 152 systems \citep{Fortin23}. This sample includes both OB-supergiant systems (52) and Be/X-ray binaries (74). When interpreting these results, one needs to take into account that \citet{vandenheuvel00} propose that OB-supergiant systems get a substantially higher space velocity (about 50~km~s$^{-1}$) than the lower-mass Be/X-ray binaries (about 15~km~s$^{-1}$). Furthermore, \citet{Renzo19} predict systematically lower space velocities for these systems. Thus, the value of the threshold velocity above which systems are identified as runaway candidates is also important. In systems hosting a black hole, such as Cyg~X-1, the mass lost during the supernova event is expected to be small, potentially resulting in negligible recoil velocities \citep{mirabel03,rao20}.

On larger scales, the ejection of massive stars from their natal environments has important implications for star cluster evolution and Galactic ecology. Numerical studies have indicated that up to 20–30\% of the most massive stars formed in clusters may escape within a few million years \citep{Stoop24b}, an effect that must be accounted for when constraining the high-mass end of the initial mass function.  Also, massive stars have a dominant impact on their surrounding medium (UV radiation, stellar wind, supernova explosion). As OB runaway stars may travel several tens to hundreds of parsec away from their birth place, they reach locations where the ISM is much more tenuous than inside the dense molecular clouds confined to the Galactic plane. Therefore, the study of OB runaways is of key importance to our understanding of the dynamical and chemical evolution of the Galaxy \citep{Andersson20}.

The astrometric and photometric data provided by the {\itshape Gaia} mission make it possible to search for runaways in order to determine membership and the age of their potential parent clusters, and to reconstruct the kinematic and evolutionary history of these systems. The target of our study is the Vel~OB1 association. Vel~OB1, at a distance of 1.6-2.5\,kpc, is one of the largest OB associations known  \citep{humphreys78,bassino82,sahu92}. \citet{humphreys78} derived an angular extent of Vel\,OB1 ($l=262^{\circ}$ to 268$^{\circ}$, $b=-2.7^{\circ}$ to +1.4$^{\circ}$) and identified 17 probable members. Later, \citet{reed00} expanded the size of the association ($l=255^{\circ}$ to 275$^{\circ}$, $b=-5^{\circ}$ to +5$^{\circ}$)\footnote{This roughly corresponds to a range in RA from 08h12 to 09h24 and in DEC from $-36^{\circ}$ to $-53^{\circ}$.} and extended the list of candidate members of Vel\,OB1 to 70 stars, including HD77581, the B supergiant companion to the X-ray pulsar Vela\,X-1. Fig.~\ref{fig:VelOB1} shows the region on the sky including Vel~OB1. In the foreground several other star-forming regions and stellar clusters are present, such as Vel~OB2 (not included in Fig.~\ref{fig:VelOB1}) at a distance of 410~pc \citep{armstrong22} and RCW36 at a distance of 700~pc \citep{ellerbroek13}. Tab.~\ref{tab:par1} lists the proposed members of Vel~OB1 collected from the literature. The table lists the location, spectral type, and radial velocity ($V_{r}$) of these candidate members. We include the {\it Gaia} DR3 parallax, proper motion, and G-band magnitude for these stars. The mean radial velocity of Vel\,OB1 is 23\,$\pm$\,4\,km\,s$^{-1}$, and the median proper motion along the Galactic longitude and latitude are $-6.78 \pm 1.07$~mas\,yr$^{-1}$ and  $-1.69 \pm 0.25$~mas\,yr$^{-1}$, respectively \citep[see also][]{melnik17}. 

Star formation in Vel\,OB1 has been continuous between 4 and 20\,Myrs ago \citep{rensbergen96}. The region contains several young stellar clusters, such as RCW34, RCW38, and RCW41 (Fig.~\ref{fig:VelOB1}, \ref{fig:parallax}, Tab.~\ref{tab:parori}). The top-left panel in Fig.~\ref{fig:disOB1_run} displays the {\it Gaia} DR3 distance distribution of the candidate members of Vel~OB1 according to \citet{humphreys78} (orange) and \citet{reed00} (blue). This provides a rough estimate of the distance range of Vel~OB1. In the lower panel of Fig.~\ref{fig:disOB1_run} we plot the proper motion of the candidate members \citep[Table \ref{tab:par1}, ][]{humphreys78, reed00} of the Vel\,OB1 association (circles). Thus, the main fraction of the candidate members is concentrated in the range $\mu_{l^{*}}$= (-8.6;-4.9)\,mas\,yr$^{-1}$ and $\mu_{b}$\,=\,(-3.1;0.5)\,mas\,yr$^{-1}$. This information serves as an additional aid in reducing the \textit{Gaia}\,DR3 sample (Section \ref{DR3}) and the identification of other members of the Vel\,OB1 association. 

The aim of this paper is to search for OB runaways in the region contained by the OB association Vel OB1. We have organised the paper as follows: Section~\ref{data} presents the membership analysis of six young clusters in Vel~OB1 based on {\it Gaia} DR3 data and OB stars listed as members of Vel~OB1 in \cite{reed00}. We used these young clusters as a representation of the young stellar population of Vel~OB1 and searched for OB runaways potentially originating in and moving with high velocity with respect to these clusters. In Section~\ref{OBrunaways} we describe our search for OB runaways in Vel~OB1, either based on their peculiar velocity and/or the presence of a wind bow shock. We also summarise their properties. In Section~\ref{origin} we identify the parent cluster of a small subset of the OB runaways. In Section 5 we analyse the detected arc-like features and investigate whether the stellar wind and ambient medium parameters are consistent with the interpretation that these features are due to a wind bow shock. In Section~\ref{discussion} we compare our results with earlier work, discuss the possible scenarios producing these OB runaways, and determine the conditions of the ISM in case a wind bow shock has been detected. In section~\ref{conclusions} we summarise our conclusions.

\section{Young stellar clusters in Vel~OB1}
\label{data}

In this section, we describe the procedure for selecting stars from the Gaia DR3 database in order to identify members of the Vel OB1 association and the stellar clusters located within this region. Our analysis reveals seventeen stellar overdensities, six of which are young stellar clusters that constitute the main components of Vel OB1. In the following, we focus on the search for OB runaway stars originating from Vel OB1, and in particular from these young clusters.

\subsection{Gaia DR3 analysis}
\label{DR3}

Based on the historical definition of Vel~OB1, we queried the {\it Gaia} database \citep[DR3;][]{Gaia22} in the range of Galactic longitude $l=255^{\circ}$ to 275$^{\circ}$, and Galactic latitude $b=-5^{\circ}$ to +5$^{\circ}$. The initial dataset includes $\sim$15\,million sources. We corrected the parallax taking into account the zero-point offset estimated from quasars \citep{Lindegren21}. 

The initial sample was refined using a series of astrometric quality criteria. First, we excluded sources with a Renormalised Unit Weight Error (\texttt{ruwe}) exceeding 1.4, which typically signals unreliable astrometric solutions and is often associated with unresolved multiplicity. Second, we removed objects with fewer than ten visibility periods contributing to the astrometric fit (\texttt{visibility\_periods\_used} < 10), as such cases may suffer from systematic astrometric or photometric effects \citep{Gaia21A}. Third, sources with an image parameter determination goodness-of-fit harmonic amplitude (\texttt{ipd\_gof\_harmonic\_amplitude}) larger than 0.15 were discarded, since elevated values are indicative of crowding or binarity \citep{Gaia21A}. In addition, we excluded stars for which multiple peaks were detected in the image parameter determination (\texttt{ipd\_frac\_multi\_peak}), as well as sources flagged as duplicated (\texttt{duplicated\_source}), both of which point to potentially problematic astrometric solutions. Finally, we required the parallax to be measured with sufficient precision by retaining only sources with a fractional parallax uncertainty more than 5 (\texttt{parallax/parallax\_error} > 5).

The remaining sample consists of approximately two million sources. The next step is to search for stellar over-densities in Galactic longitude, Galactic latitude, parallax and proper motion using the Tool for OPerations on Catalogues And Tables \citep[TOPCAT; ][]{Possel20}, until we obtain samples that correspond to stellar clusters included in the Vel\,OB1 association, and field stars. The result of that selection is shown in Fig.~\ref{fig:parallax} where we present the parallax distribution as a function of Galactic longitude and the distribution of stellar over-densities in Galactic coordinates, respectively. The parallax distribution indicates an excess of stars in the range from about 0.40\,mas to 0.65\,mas, which coincides well with the distance of the candidate members of the Vel\,OB1 association. The position and putative size of the young stellar clusters are shown (see Sect.~\ref{members} below for a membership analysis).

\subsection{Membership analysis young stellar clusters}
\label{members}

We carefully inspected the stellar over-densities in the region contained by Vel~OB1 (in the parallax range 0.40--0.65~mas). Apart from six young stellar clusters (Tab.~\ref{tab:parori}), we identified 11 stellar clusters corresponding to open clusters listed in the {\it Simbad} astronomical database.

Cluster membership was determined using the \textsc{pyupmask} package, a \textsc{python}-based implementation of the unsupervised clustering algorithm \textsc{upmask} \citep{Krone14,Pera21}. The analysis was performed in a five-dimensional astrometric parameter space defined by sky position, parallax, and proper motion. Radial velocities were intentionally excluded, as they are available for only a subset of sources in \textit{Gaia}\,DR3 \citep{Katz23}; including them would introduce systematic differences between stars with and without measured radial velocities. The algorithm proceeds through a two-level iterative scheme. In the internal clustering step, groups of approximately 25 stars are identified using a Gaussian mixture model, which has been shown to provide optimal performance for this method \citep{Pera21}. In the subsequent resampling stage, observational uncertainties are incorporated by repeatedly drawing astrometric parameters from their Gaussian error distributions, while correlations between parameters and between stars are neglected. We performed 10,000 resampling iterations and employed a Gaussian–uniform mixture model to reduce contamination by spurious members. The \textsc{pyupmask} analysis was restricted to regions containing more than 25 sources (i.e. the RCW clusters), whereas membership in sparser regions was assessed manually.

For each star, a membership probability \textit{p} was assigned based on the fraction of iterations in which it was classified as a cluster member. The threshold value of \textit{p} used to distinguish members from field stars was not fixed globally, but instead selected individually for each region after inspecting the noise characteristics of the resulting probability distribution.

Table~\ref{tab:parori} lists the parameters of the young stellar clusters in Vel~OB1 based on the membership analysis. The zoomed area in the top-right panel of Fig.~\ref{fig:disOB1_run} shows the proper motion distribution of the members of these clusters. As can be seen, their proper motions are in the same range as the main fraction of the candidate members of Vel\,OB1 (Tab.~\ref{tab:par1}). 

\begin{table*}
\centering
\caption[]{Young stellar clusters in Vel~OB1.}
\resizebox{1\textwidth}{!}{
\label{tab:parori}
\renewcommand{\arraystretch}{1.2} 
\begin{tabular}{l *{11}{c}}
\hline\hline\noalign{\smallskip}
Name	&	l  & b & 	N & Parallax	&	$\mu_{l^{*}}$	&	$\mu_{b}$ &	V$_{LSR}$ & D	& Age & A$_V$ & \textit{p}\\
 $-$&	(deg) & (deg)	& $-$& (mas)		&	(mas\,yr$^{-1}$)	&	(mas\,yr$^{-1}$) & (km\,s$^{-1}$)  &		(pc)	& (Myr) & (mag) & $-$\\
\hline\noalign{\smallskip}
(1)     &       (2)     &       (3)     &       (4) 	&	 (5) 	& (6)	&	(7) & (8) & (9) & (10) & (11) & (12)\\
\hline \noalign{\smallskip}
Cl\,1594	& 264.73 & 0.27 & 21 & 0.480$\pm$0.010		& -6.498$\pm$0.452 & -1.203$\pm$0.589   & 6.8$\pm$3.9 & 2084$^{+43.0}_{-41.3}$ & 2.6$-$6.6
  & 5 & $-$\\
IRAS\,08337-4028	&	 	259.95& 0.02 & 14 & 0.624$\pm$0.012		& -5.979$\pm$0.833 &  -2.108$\pm$0.982  & 10.8$\pm$5.2 & 1603$^{+31.7}_{-30.5}$	& 5.0$-$8.9 & 5 & $-$\\
PN\,K\,2-15	&  	263.24 & 0.52 & 12& 0.560$\pm$0.007		& -7.125$\pm$0.402 &  -1.074$\pm$0.533  & 8.3$\pm$2.9
 & 1787$^{+23.0}_{-22.4}$	& 0.5$-$3.0 & 5 & $-$\\
RCW\,34	&	 	264.35 & 1.44 & 48 & 0.509$\pm$0.005	& -6.773$\pm$0.487 & -0.826$\pm$0.763 & 8.2$\pm$4.6
 & 1967$^{+18.4}_{-18.0}$ & 4.7$-$9.9 & 4 & 0.85 \\
RCW\,38	&	267.94 & -1.06  & 378 & 0.560$\pm$0.001	& -6.682$\pm$0.359 &  -1.052$\pm$0.346 & 4.3$\pm$3.6 & 1785$^{+4.2}_{-4.1}$	& 3.2$-$7.1 & 3 & 0.9\\
RCW\,41&  	270.24& 0.83  & 35 & 0.645$\pm$0.006		& -6.682$\pm$0.367 &  -0.913$\pm$0.370  & 4.5$\pm$2.3 & 1550$^{+14.6}_{-14.3}$	& 0.7$-$1.0  & 7 & 0.7\\
\hline
\end{tabular}
}
\tablefoot{
(1) Name of the region, (2), (3) Galactic coordinates, (4) number of members based on the membership probability \textit{p}, (5) \textit{Gaia}\,DR3  parallaxes with 1$\sigma$ uncertainties, (6-7) mean proper motion with their standard deviation, (8) tangential velocity of the cluster with respect to the LSR with 1$\sigma$ uncertainty, (9) distance from \textit{Gaia}\,DR3 parallax with 1$\sigma$ uncertainties, (10) and (11) the age range based on 100 iterations and a fixed extinction, respectively, and (12) the membership probability. The [KPS2012]\,MWSC\,1594 open cluster is indicated as Cl\,1594.}
\end{table*}

\subsubsection{Membership and distance of the young clusters}
\label{members1}

We analysed the parallax distribution of the members of each cluster against their G-magnitude (the individual panels are included in Appendix~\ref{appA}, Fig.~\ref{fig:plxG_sep}). We determined the distance to each cluster through a log-likelihood function as in \citet{Stoop24}. For an assumed distance, we computed the likelihood of the observed parallaxes using Gaussian uncertainties without explicitly normalising the probability distribution. The most probable distance corresponds to the maximum of the resulting log-likelihood function, while the associated 1$\sigma$ confidence interval was derived from the 16th and 84th percentiles of the distance distribution implied by the likelihood. The determined best-fit parallax, distance and their 1$\sigma$ uncertainties are in the range 1.6 to 2.1~kpc (Tab.~\ref{tab:parori}), with RCW~38 the richest young cluster at 1.8~kpc. The sky distribution of the cluster members can be found in Appendix~\ref{appA} (Fig.~\ref{fig:lb_sep}). 

If we take the six young clusters to define the current distance and motion of Vel~OB1, it is located at a distance of $1.85$~kpc with a mean proper motion $\mu_l^* = -6.76\pm0.49
$~mas~yr$^{-1}$ and $\mu_b = -1.03\pm0.53$~mas~yr$^{-1}$, which is in good agreement with the proper motion $\mu_l^* = -6.78\pm 1.07
$~mas~yr$^{-1}$ and $\mu_b = -1.09\pm0.25$~mas~yr$^{-1}$ reported in \citet{melnik17} (Fig.~\ref{fig:disOB1_run}). The mean radial velocity of Vel~OB1 is $25 \pm 15$~km~s$^{-1}$ (based on the members of Vel~OB1 with a measured radial velocity), which is also in good correspondence with the mean radial velocity of $23 \pm 4$~km~s$^{-1}$ reported in \citet{melnik17}. The peculiar velocity of Vel~OB1 with respect to its LSR is $9.6 \pm 3.9$~km~s$^{-1}$; for more details we refer to Sect.~\ref{nBS}. 

In column (13) of Tab.~\ref{tab:par1} we reassess the membership of Vel~OB1 based on the {\it Gaia} DR3 membership analysis. If the candidate (according to \citet{humphreys78} and \citet{reed00}) is member of one of the young clusters, we list the name of the respective cluster. If the candidate's parallax is in the range between 0.4 and 0.65~marcsec and the proper motion $\mu_l^*$ between -8.6 and
-4.9~mas~yr$^{-1}$ and $\mu_b$ between -3.1 and 0.5~mas~yr$^{-1}$, we assign it as a member of Vel~OB1 (Y). These limits are based on the observed range in parallax and proper motion of the young clusters (Tab.~\ref{tab:parori}).

\subsubsection{Age of the young clusters}

For each young cluster, the colour-absolute magnitude diagram is used to constrain the age (Appendix~\ref{appA}, Fig.~\ref{fig:isoch}). To estimate the age of the individual clusters, we generated \texttt{PARSEC (v2.0) + COLIBRI (S\_37 + S\_35 + PR16)} (Padova and Trieste Stellar Evolution Code) isochrones\footnote{\url{http://stev.oapd.inaf.it/cgi-bin/cmd}} considering an age range from 0.2\,Myr to 10\,Myr with steps of 0.1\,Myr \citep{Bressan12,Marigo13,Rosenfield16,Marigo17,Pastorelli19,Pastorelli20,Nguyen22}. To identify the best-fit isochrone for our clusters, we used the statistical estimation and likelihood maximisation standard methods described in \citet{vandermeij21}. To estimate the age dispersion for each cluster, we use a bootstrap method. We repeat the age determination 100 times on randomly selected samples, which contain 90\% of all cluster members. Columns (10) and (11) of Tab.~\ref{tab:parori} list the thus obtained age range based on 100 iterations and a fixed
extinction; to estimate the (average) extinction A$_{\rm V}$ we used the near-infrared counterparts of the members, and their colours (adopting R$_{\rm V} = 3.1$). The derived ages are in accordance with previous estimates and range from 0.5 to 10~Myr. We defer a more detailed discussion on the young massive clusters included in Vel~OB1 to a future paper.

\begin{figure*} 
\centering
\includegraphics[width=0.4\linewidth]{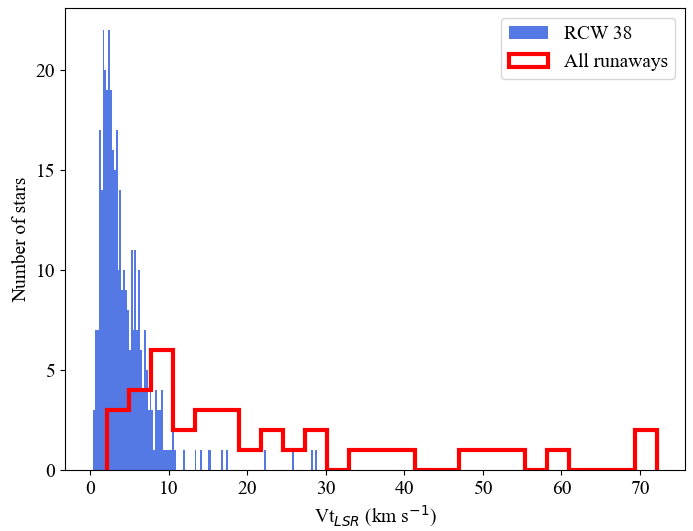}
\includegraphics[width=0.4\linewidth]{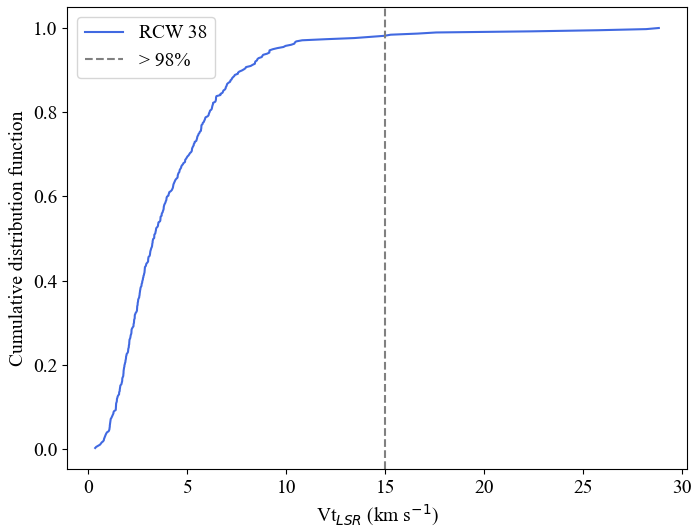}
\caption{OB-runaway peculiar tangential velocity threshold. \textit{Left panel:} Tangential velocity of the members (blue) of the young cluster RCW38 with respect to the LSR velocity: the peculiar tangential velocity $v^{t}_{\rm LSR}$. The red histogram shows $v^{t}_{\rm LSR}$ for all OB runaways identified in this region. \textit{Right panel:} Cumulative distribution of $v^{t}_{\rm LSR}$ for the members of RCW38. We note that 98~\% of the members have a peculiar tangential velocity less than 15~km~s$^{-1}$. We take this value as the threshold for the tangential peculiar velocity to identify the star as a runaway.}
\label{fig:threshold}
\end{figure*}

\section{OB runaways in Vel~OB1}
\label{OBrunaways}

The main aim of this paper is to search for OB runaways in the region contained by the OB association Vel~OB1. One way to identify OB runaways is by estimating their peculiar velocity, i.e. by comparing their proper motion (and/or radial velocity) to the LSR velocity, or to the mean velocity of the young clusters contained in Vel~OB1 (Sect.~\ref{nBS}). The former is obtained from a model describing the differential Galactic rotation and by taking into account the peculiar solar motion and the distance (the latter in Sect.~\ref{members}). An alternative approach is to look for the presence of a wind bow shock produced by the supersonic motion of the OB runaway, and the interaction of its stellar wind with the ambient medium (Sect.~\ref{BS}). 

The peculiar velocity threshold of $30$~km~s$^{-1}$ is rather arbitrary; in order to escape from a stellar cluster the OB runaway velocity should exceed the escape velocity of the cluster which is of the order of $5 - 10$~km~s$^{-1}$ for a young massive cluster; the tangential velocity dispersion in these clusters is in the range $1 - 2$~km~s$^{-1}$ (Stoop et al., in prep). For the formation of a stellar wind bow shock, however, it is sufficient that the stellar velocity exceeds the local sound speed in the ISM, which can be as low as the escape velocity of young clusters. \citet{Renzo19} introduced the term `walkaway' to emphasise the point that OB runaways (in this particular case produced by the binary supernova scenario) are also predicted to travel with a space velocity lower than $30$~km~s$^{-1}$. In fact, some of the OB runaways for which we observe a wind bow shock have a space velocity consistent with those of massive stars in clusters (see Fig.~\ref{fig:threshold}). 

\begin{figure}
\centering
\includegraphics[width=0.9\linewidth]{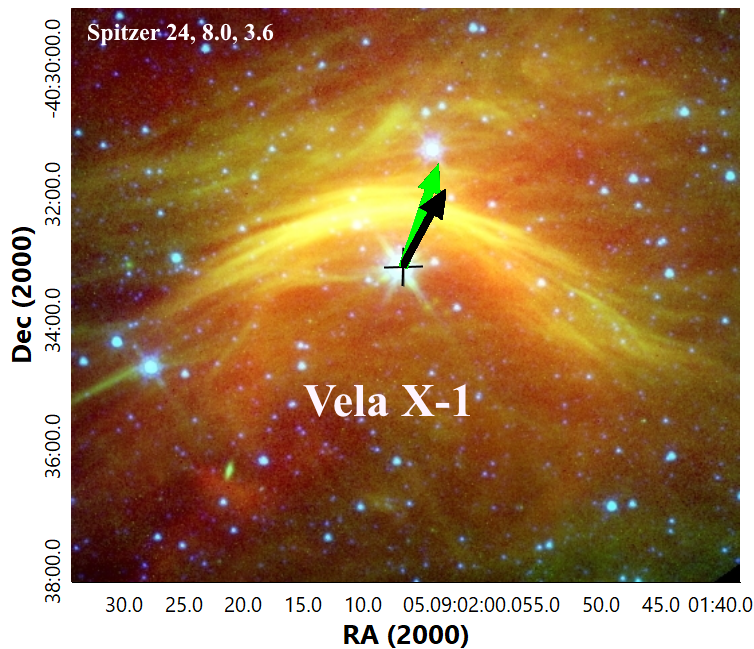}
\caption{\textit{Spitzer} colour-composite image of the Vela\,X-1 system. The runaway star is marked with a cross. The black arrow indicates the direction of the current proper motion of the system. The green arrow exhibits the orientation of the wind bow shock, i.e. the direction from the star to the peak of the 22\,$\mu$m emission.}
\label{fig:runawayVelax-1}
\end{figure}

\begin{figure*}
\centering
\includegraphics[width=0.9\linewidth]{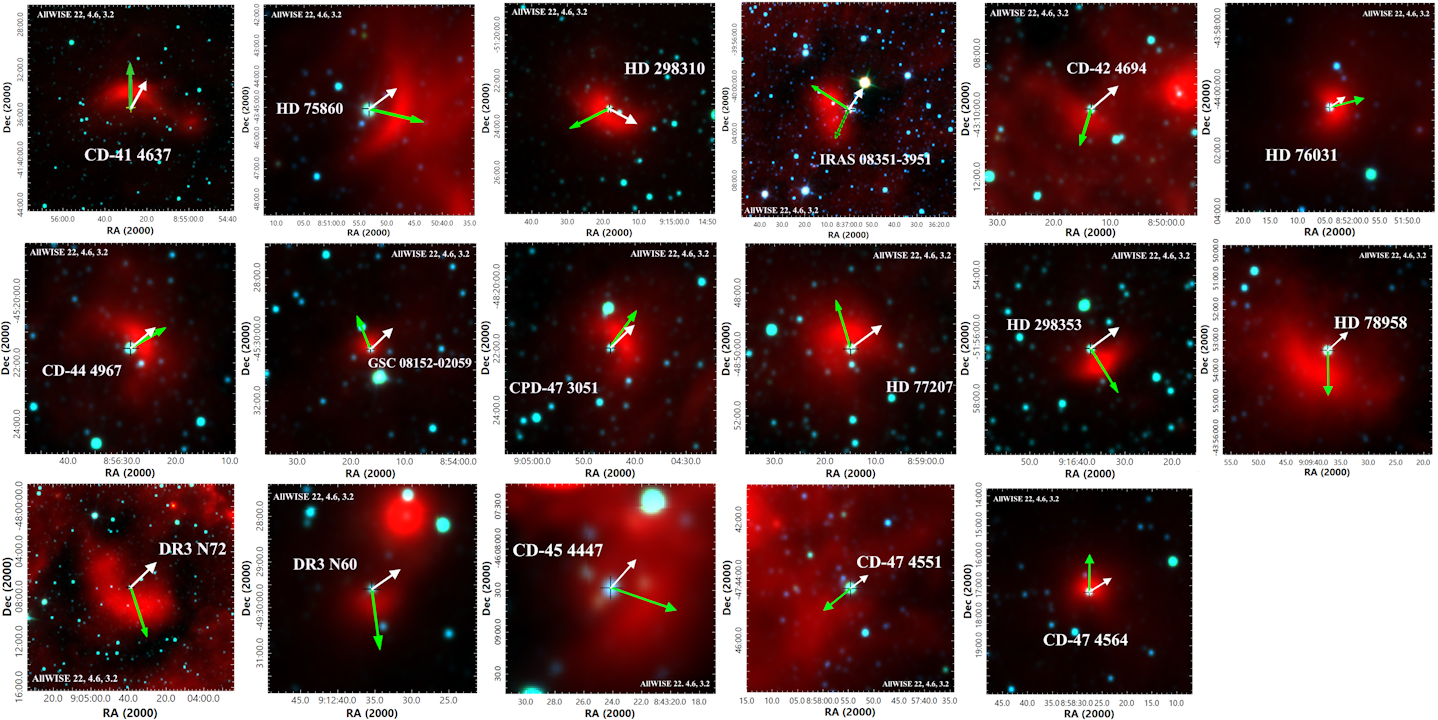}
\caption{AllWISE colour-composite images of arc-like structures associated with (OB-type) stars in the Vel\,OB1 region. The star is identified with the white label (and a cross), and the infrared flux is represented by the heat colour scale. The white arrow shows the direction of the current proper motion of each star. The green arrow indicates the orientation of the wind bow shock, i.e. the direction from the star to the peak of the 22\,$\mu$m emission. In ten cases (including HD\,77581 / Vela~X-1, Tab.~\ref{tab:parrun}), the arc-like structure is well aligned with the direction of the proper motion, demonstrating that these structures relate to a wind bow shock (and confirming the runaway nature of the star). 
}
\label{fig:runaways}
\end{figure*}

\subsection{Runaways identified by their peculiar velocity}
\label{nBS}

To investigate the runaway nature of the OB stars in Vel\,OB1 (Tab. \ref{tab:par1}), we calculate the peculiar velocity $V_{\rm pec}$ of each star relative to the LSR. To achieve this, we first used the Galactocentric transformations of \citet{johnson87} to determine the mean Galactocentric velocity $(U, V, W)$ components of nearby stars within 100\,pc of each OB star, using the same approach as for the OB stars themselves. Specifically, we subtracted the predicted Galactic rotation velocity at the location of the star \citep{Binney98,Binney08} from its observed proper motion and radial velocity (if available):
\begin{equation}
\label{equ:1}
V_{\rm pec} = \sqrt{(U_*-U_{\rm sys})^2+ (V_*-V_{\rm sys})^2+(W_*-W_{\rm sys})^2}   
\end{equation}
where $U$ is the Galactic radial-velocity component positive towards the Galactic centre, $V$ is the Galactic rotational velocity component positive towards $l=90^{\circ}$, and $W$ is positive towards the Galactic north pole. For circular rotation around the Galactic centre, the $W$ component of the velocity is expected to be zero. We adopted $\alpha_{GC} = \,$ 17:45:37.2, $\delta_{GC} = \,$−28:56:10.23 for the Galactic centre and $\alpha_{GNP} = \,$12:51:26.3, $\delta_{GNP} = \,$27:07:42.01 for the Galactic north pole \citep{reid04}. We used the solar peculiar motion relative to the LSR of ($U_{\sun}=10.6$, $V_{\sun}=10.7$, $W_{\sun}=7.6$)\,km\,s$^{-1}$, the velocity of the LSR relative to the Galactic centre of $236$~km\,s$^{-1}$, and a solar Galactocentric distance of 8.15\,kpc based on the A5 fit provided by \citet{Reid19}. For the computation of the peculiar velocity, for each star the position, distance, proper motion and radial velocity (if available) were used. If the radial velocity was not available, we adopted zero for the peculiar radial-velocity component; this may lead to an underestimation of $V_{\rm pec}$.

We also calculated the motion of the young clusters with respect to the LSR; one would expect that they are at rest with respect to the Galactic rotation model, but e.g. \citet{comerontorra98} showed that the young stellar population in the solar environment can deviate from this model by up to 20~km~s$^{-1}$. We applied the same approach to the young clusters as we did for the OB stars, calculating the mean relative velocity of their members. Since the radial velocities of the clusters are unavailable, we computed their mean tangential velocities relative to the LSR, along with 1$\sigma$ uncertainties (see Column 8 of Tab.~\ref{tab:parori}). The results indicate that the clusters' motion relative to the LSR is similar to that of Vel\,OB1 ($\sim 10$~km~s$^{-1}$, Section \ref{members1}), which is another indication that the young clusters are part of Vel\,OB1.

Of the 77 objects listed in Tab.~\ref{tab:par1} we determine a peculiar velocity $v_{\rm pec} \succsim 30$~km~s$^{-1}$ for 12 of them. The relatively flat distribution of tangential velocities among the runaway stars (Fig.~\ref{fig:threshold} right panel) provides the basis for proposing a revised threshold for the peculiar velocity. Adopting $v_{\rm pec} \succsim 15$~km~s$^{-1}$, the number of candidate runaways increases to 22. For some objects $v_{\rm pec}$ differs depending on whether this velocity is measured with respect to the LSR velocity or to Vel~OB1. If the object is located relatively far away from Vel~OB1, one should take the value of the peculiar velocity with respect to the LSR there. We confirm the runaway nature of the 7 objects in common with \cite{Carretero23} (bold face in Tab.~\ref{tab:par1}; we note that HD298429 has $v_{\rm pec} \simeq 20$~km~s$^{-1}$).

\subsection{Runaways associated with a wind bow shock}
\label{BS}

We have searched for arc-like features in the region of Vel~OB1 that may be due to a wind bow shock produced by an OB runaway star.
These bow shocks are relatively large features (arcminute size, parsec scale) and an indication for the (supersonic) motion of the OB star with respect to the ISM \citep{buren88}. Wind bow shocks associated with massive stars radiate primarily at mid-infrared wavelengths, where thermal emission from dust grains heated by the intense ultraviolet (UV) radiation of the OB star dominates \citep{comeron98,Meyer14,Henney19}. A well-studied illustration of this phenomenon is the bow shock surrounding the high-mass X-ray binary HD\,77581 (Vela~X-1), which has been detected and characterised across multiple wavelength regimes \citep{kaper97,eijnden22}. Fig.~\ref{fig:runawayVelax-1} presents a \textit{Spitzer} colour-composite image of this runaway system \citep{Iping07}. The wind bow shock around Vela\,X-1 is aligned with the direction of proper motion. Apart from {\itshape Spitzer} Space Telescope data using the {\itshape Spitzer} Infrared Array Camera \citep[IRAC,][]{fazio04} centred at 3.6, 4.5, 5.8, and 8.0\,$\mu$m and the Multiband Infrared Photometer of {\itshape Spitzer} (MIPS) at 24\,$\mu$m \citep{carey09}, we also used Wide-field Infrared Survey Explorer \citep[WISE,][]{wright10} data in the 3.4, 4.6, 12, and 22\,$\mu$m band pass. 

We explored the AllWISE survey in the field contained by Vel~OB1 for the presence of arc-like structures that could be due to a wind bow shock. Fig.~\ref{fig:runaways} presents the infrared colour-composite AllWISE images of 17 arc-like features (excluding Vela~X-1) that often turn out to be associated with an OB star in Vel~OB1 (though not all included in the compilation of \citet{reed00}, cf.\ Tab.~\ref{tab:par1}). Four of these arc-like features — HD~76031, GSC~08152-02059, DR3~N72, and DR3~N60 — are reported here for the first time (see Appendix~\ref{appD} for notes on individual objects. The white arrow indicates the direction of the observed proper motion of the stars. The green arrow highlights the orientation of the bow shock, i.e. the direction from the star to the peak of the 22\,$\mu$m emission. Often these two arrows are not precisely aligned, i.e. the apparent motion of the star is not aligned with the opening angle of the arc-like structure. Obviously projection effects could play a role, as well as the bulk velocity of the ISM. Based on {\it Hipparcos} observations \citet{kobulnicky16} measured a relatively wide distribution of misalignment (up to 180~degrees) for a sample of 286 infrared bowshocks. Similarly \citet{Bodensteiner18} found that 64\% of 94 bow shocks are misaligned (>\,20\,$^{\circ}$). We will return to this point in Sect.~\ref{DisBS}.

For each bow shock we looked for a stellar counterpart (Tab.~\ref{tab:parrun}); the large majority are early-type stars in the distance range occupied by the Vel~OB1 association. Some objects may be too distant to be related to Vel~OB1. The peculiar or transverse velocity of the stars associated with a bow shock (column 10 in Tab.~\ref{tab:parrun}) indicates that some of them are "classical" OB runaway stars \citep[e.g. ][]{blaauw61, Boubert18} with a large, and very likely supersonic space velocity such that a wind bow shock is produced. Others may be termed walkaways \citep[<\,30\,km\,s$^{-1}$,][]{Renzo19}; the inclusion of the radial velocity may change this result. Some of the OB runaway candidates associated with a bow shock have a modest space velocity ($\simeq 10$~km~s$^{-1}$), but that may still be supersonic (Sect.~\ref{DisBS}).

\begin{figure}
\centering
\includegraphics[width=0.98\linewidth]{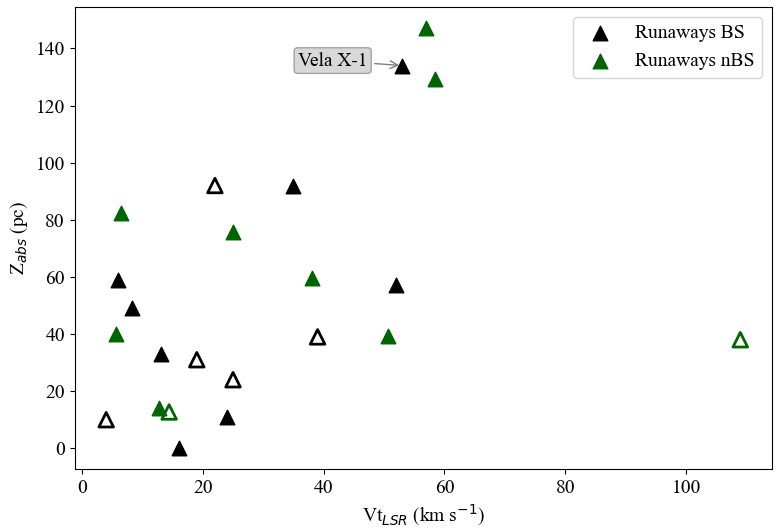}
\caption{Galactic height Z$_{abs}$ versus the peculiar transverse velocity of the OB runaways in Vel~OB1. Open triangles refer to candidate runaways (see Tab. \ref{tab:parrun}). It shows the expected trend that the fastest runaways reach the largest distance from the Galactic plane. Ten of the OB runaways with an observed bow shock are within 60~pc from the Galactic plane where most of the gas and dust is present. Vela~X-1 is at a height of 134~pc and still produces a prominent wind bow shock.}
\label{fig:transvel}
\end{figure}

\begin{figure}
\centering
\includegraphics[width=0.9\linewidth]{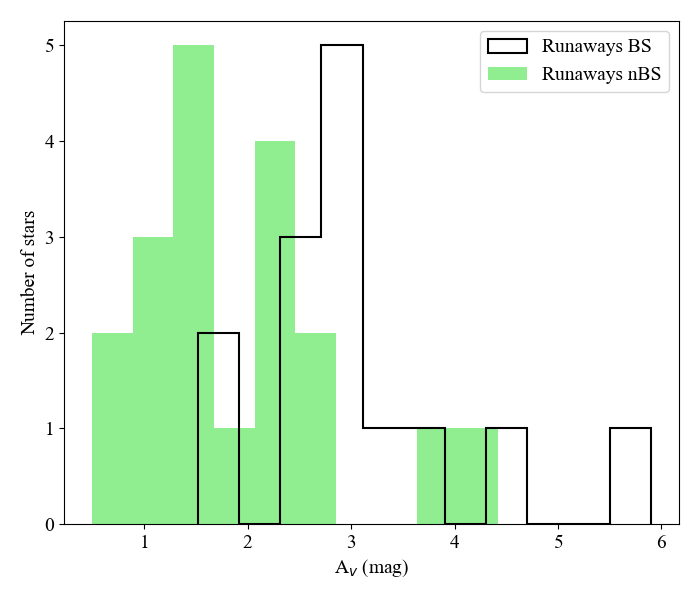}
\caption{Extinction, $A_{V}$, measured in the sight lines towards the OB runaways with (black histogram) and without an associated wind bow shock (green histogram). The OB runaways with a bow shock experience a higher extinction than those without a bow shock.}
\label{fig:histAv}
\end{figure}

Based on the images in Figs.~\ref{fig:runawayVelax-1} and \ref{fig:runaways} and the information in Tab.~\ref{tab:parrun}, we conclude that for 10 objects the arc-like feature most likely refers to the presence of a wind bow shock. Among the confirmed OB runaways in Vel~OB1 (excluding candidate runaways in the projected field), the aligned cases are predominantly associated with higher peculiar velocities, while misaligned cases tend to have lower peculiar velocities (Fig.~\ref{fig:BS}). This trend is a direct physical prediction of bow shock theory: at higher Mach numbers, the ram pressure of stellar motion dominates over ISM thermal and turbulent pressure, producing a well-defined shock tightly aligned with the space velocity. Conversely, at lower peculiar velocities — where the star is only marginally supersonic — ISM density gradients, bulk flows, and turbulence can distort the shock orientation, naturally producing a wide range of misalignment angles. Crucially, this V$_{pec}$–alignment correlation would not be expected if the arc-like features were randomly oriented ISM structures coincidentally aligned with stellar motion in the aligned cases. A Fisher's exact test comparing the fraction of aligned bow shocks among high-velocity (V$_{pec}$ $\geq$ 15~km~s$^{-1}$) versus low-velocity systems gives p $\approx$ 0.17, which does not reach formal significance given the small sample of 16 objects. Nevertheless, the direction of the trend — four out of five high-velocity systems aligned versus six out of 11 low-velocity systems — is consistent with the physical prediction of bow shock theory, and the limited statistical power reflects the sample size rather than the absence of a real physical effect.

Additional physical support for the bow shock interpretation comes from the consistency between the ISM number density derived independently (see Section~\ref{DisBS}) from the standoff distance (using the \citet{baranov71} pressure balance, Eq.~\ref{equ:2}) and from the Str\"omgren sphere radius (Eq.~\ref{equ:3}). For most sources in Tab.~\ref{tab:estpar1}, these two independent estimates agree to within a factor of a few, indicating that the same star is responsible for both the ionisation of the surrounding medium and the wind-ISM interaction producing the arc-like feature. This consistency provides strong evidence that the observed structures are genuine wind bow shocks rather than unrelated ISM features.
The remaining 8 arc-like features that are misaligned with the stellar proper motion (angle > 70$^{\circ}$) may be attributed to: (1) in situ bow shocks shaped by bulk ISM flows rather than stellar motion, as proposed by \citet{Povich08, Bodensteiner18} and observed in e.g. \citet{Guarcello25}; (2) wind-ISM interactions distorted by density gradients perpendicular to the stellar velocity, which cause the bow shock to become asymmetric \citep{Wilkin00}; or (3) unrelated ISM structures, as appears to be the case for CD-47~4551 and IRAS~08351-3951 (labelled ISM in Tab.~\ref{tab:parrun}), which are included in the count of 18 arc-like features but excluded from the bow shock alignment statistics. Two additional objects — HD~71649 and HD~78927 — show arc-like emission in the AllWISE images but no reliable bow shock angle could be measured and they are not included in the bow shock sample. Excluding CD-47~4551 and IRAS~08351-3951 reduces the bow shock sample to 16, with ten aligned and six genuinely misaligned cases. We note that several confirmed runaways — including HD~76031, HD~77207, CPD-47~3051, and CD-47~4564 — have peculiar tangential velocities below the 15~km~s$^{-1}$ threshold and are identified as runaways primarily on the basis of their associated bow shock. Indeed, some of these stars have peculiar velocities below the local ISM sound speed (10–15~km~s$^{-1}$), meaning that their stellar motion alone is insufficient to drive a supersonic bow shock. In these cases, large-scale flows in the ISM — potentially driven by cluster winds — must provide the additional relative velocity between the star and its environment \citep{Povich08,Bodensteiner18}, and the misalignment of the arc-like feature with the stellar motion is then a natural consequence. Nevertheless, for three of these stars, the consistency between the standoff distance and Strömgren sphere density estimates in Tab.~\ref{tab:estpar1} provides independent physical confirmation that these stars are indeed moving supersonically through the local ISM, likely due to such large-scale ISM flows contributing to the relative velocity.  In the Appendix (Fig.~\ref{fig:noBow}) we show AllWISE images centred on the OB runaways that we identified based on their peculiar velocity (and are not displayed in Fig.~\ref{fig:runaways}). Thus, a significant number of OB runaways apparently do not produce a wind bow shock; see also \citet{huthoff02,carretero25}.

\begin{figure*}
\centering
\includegraphics[width=0.4\linewidth]{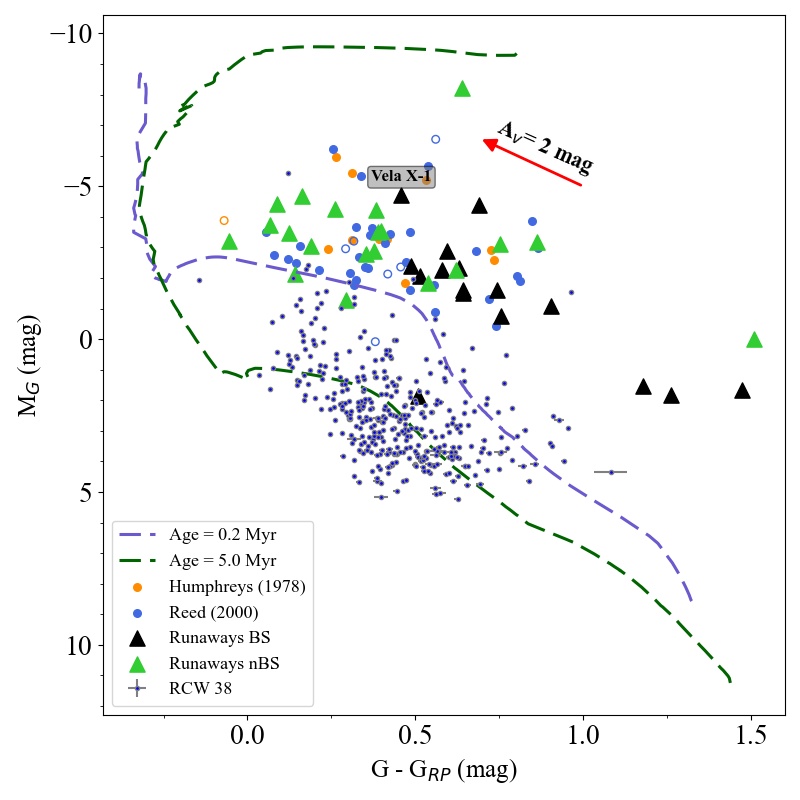}
\includegraphics[width=0.4\linewidth]{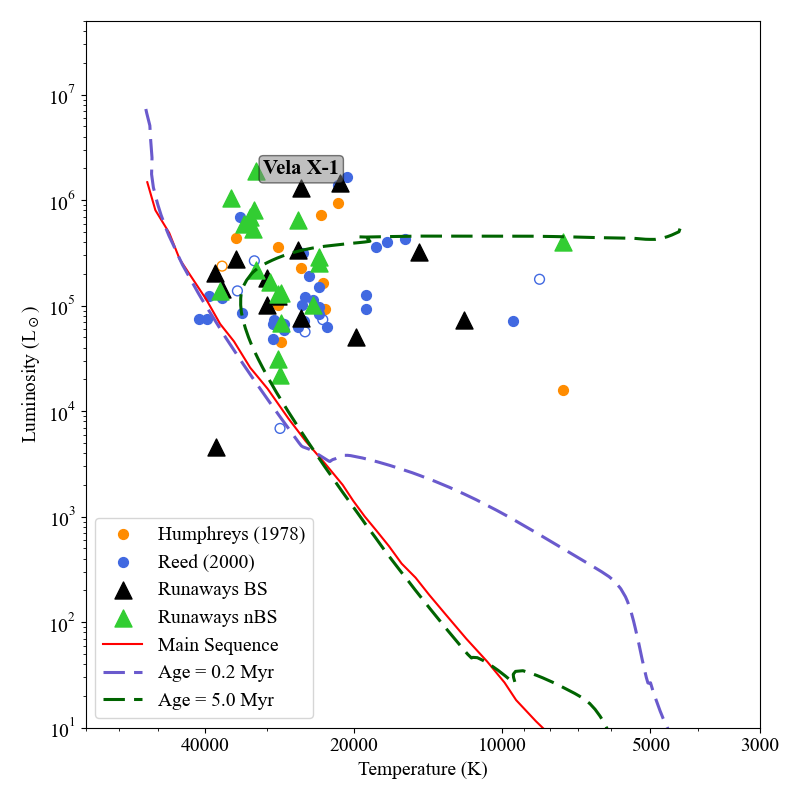}
\caption{\textit{Left panel:} OB runaways originating in Vel~OB1 and the candidate members of Vel~OB1 (neither has been corrected for interstellar extinction) plotted in the colour-absolute magnitude diagram of RCW~38. The OB runaways associated with a wind bow shock (black triangles) are more reddened than those without a bow shock (green triangles). The red arrow represents the dereddening vector. The PARSEC isochrones (dashed lines) are shown for an adopted reddening of $A_V = 3$~mag). \textit{Right panel:} OB runaways and the candidate members in the HRD (with the effective temperature based on spectral type, luminosity based on distance, bolometric correction included, and corrected for extinction). The red line represents the zero-age main sequence from the solar metallicity grid in \citet{Maiz13}; the dashed lines are PARSEC isochrones corresponding to the age indicated in the legend. The location of the OB runaways suggests that most of them are still in the hydrogen-burning phase; some others are on their way to become a red supergiant.}
\label{fig:HRD}
\end{figure*}

\subsection{Properties of the OB runaways}
\label{OBrunprop}

Table~\ref{tab:parrun} lists the astrometric and kinematic properties of the OB runaway stars. Their peculiar velocity ranges between a few to over a hundred km~s$^{-1}$ (Column 10 in Tab. \ref{tab:parrun}). The tangential velocity distribution of the OB runaways is relatively flat (Fig.~\ref{fig:threshold}) and similar to that of the runaways presented in \cite{Carretero23} and Stoop et al. (in prep). We conclude that 8 out of 77 OB stars in Vel\,OB1 are runaways with a peculiar tangential velocity exceeding 30\,km\,s$^{-1}$; 15 runaways move faster than 15\,km\,s$^{-1}$, i.e. with respect to the LSR (and thus also relative to the young clusters in Vel~OB1). If we also take into account the radial velocity (if available) we find that 14 or 21 (+1 F-type) OB runaways have a peculiar velocity greater than 30 or 15~km~s$^{-1}$, respectively, yielding a runaway fraction of 18 or 29~\%, respectively. Seven of the 26 identified OB runaways (including HD77581, the companion to the X-ray pulsar Vela\,X-1) are in common with the list of Galactic OB runaways in \citet{Carretero23}, i.e. all seven OB runaways that they report in the field of Vel~OB1 that we studied.

Most OB runaways identified in the Vel~OB1 region are of spectral type O (8/26) or B (17/26); only HD 74920 is a luminous F2 supergiant located more than 100~pc below the Galactic plane. The runaway fraction as a function of spectral type then becomes 38~\% for the O stars (8/21) and 32~\% (17/52) for the B stars. If we count the B supergiants only, we arrive at a runaway fraction of 24~\% (6/25). Four out of eleven OB supergiants are associated with a wind bow shock.

Two OB runaways are member of a HMXB: HD77581 (BP~Cru) with its X-ray pulsar companion Vela~X-1 \citep{kaper97} and HD74194 (LM~Vel) that is associated with the supergiant fast X-ray transient IGR~J08408-4503 \citep{Gamen15}. HMXBs are expected to be runaway systems produced by the binary supernova scenario \citep{vandenheuvel00}; both systems are included in the studies of \citet{Carretero23} and \citet{fortin22} and are confirmed as runaways.

The Galactic height $Z$ of the OB runaways ranges from 0 to almost 150~pc above the plane. The OB runaways at a larger distance from the plane typically have a larger space velocity (Fig.~\ref{fig:transvel}). Also, most (10) OB runaways with an associated bow shock are found within 60~pc from the Galactic plane where the interstellar gas and dust are predominantly concentrated. An exception is Vela~X-1 which is located at 134~pc above the plane and still manages to produce a prominent wind bow shock. It should be noted, however, that large-scale infrared and radio images suggest that Vela~X-1 passed through an over-dense ridge in the ISM $\sim$10$^5$ years ago, where it may have swept up a significant amount of gas \citep{gvaramadze18,eijnden22}.

Despite their high space velocity, we did not observe a bow shock for the majority of the OB runaway stars identified by their high peculiar velocity (Tab. \ref{tab:parrun} and Fig.~\ref{fig:noBow}). This could be related to the conditions of the ambient medium through which they travel (cf. Sect.~\ref{DisBS}).

For the runaways with an accurate spectral type we derive the intrinsic colour $(B-V)_0$ and, with this information, estimate the interstellar extinction. Fig.~\ref{fig:histAv} presents a histogram of the visual extinction $A_{V}$ (adopting $R_{V}$ = 3.1) demonstrating that $A_{V}$ in the line of sight of these runaways is in the range between $0.5$ to $6$~mag. The OB runaways for which a wind bow shock has been detected have an on average higher extinction than those without a bow shock. This may indicate that a wind bow shock is formed if the OB runaway moves through denser regions of the ISM, and is also consistent with the observation that OB runaways with a bow shock are more confined towards the Galactic plane (Sect.~\ref{DisBS}).

Fig.~\ref{fig:HRD} displays the location of the OB runaways (triangles) in the colour-magnitude diagram with respect to the young cluster RCW~38 (blue) as a reference. Clearly, the OB runaways are more luminous than most of the cluster members; correction for the interstellar extinction would move them to the top-left side of the diagram. The right panel in Fig.~\ref{fig:HRD} shows the position of the OB runaways in the Hertzsprung-Russell diagram (HRD). The effective temperature is derived from the spectral type, and the luminosity has been determined using the distance ({\it Gaia} parallax), the bolometric correction and the extinction. Most OB runaways are likely still in the hydrogen-burning phase; some others may already be on their way to become a red supergiant. 

As soon as a parent cluster has been identified, one would know the age of the runaway (assuming that it was born in the cluster) if the runaway is the result of dynamical ejection.  If the binary supernova scenario produced the runaway, the evolutionary state of the OB runaway depends on whether mass transfer occurred prior to the supernova explosion. In cases where the primary transferred a significant amount of mass to the secondary before undergoing core collapse, the secondary gains mass and luminosity, making it appear younger than the cluster population — a so-called blue straggler \citep[rejuvenation; e.g. ][]{vandenheuvel00,vandermeij21}. In systems where no mass transfer took place, the companion is simply accelerated by the supernova kick and remains a normal star consistent with the cluster age. For Vela X-1, \citet{rensbergen96} proposed a phase of mass transfer prior to the supernova, so one would expect that its B-supergiant companion HD77581 is located at a more luminous and bluer location than the isochrone corresponding to the age of the parent cluster.

\begin{figure*}
\centering
\includegraphics[width=1\linewidth]{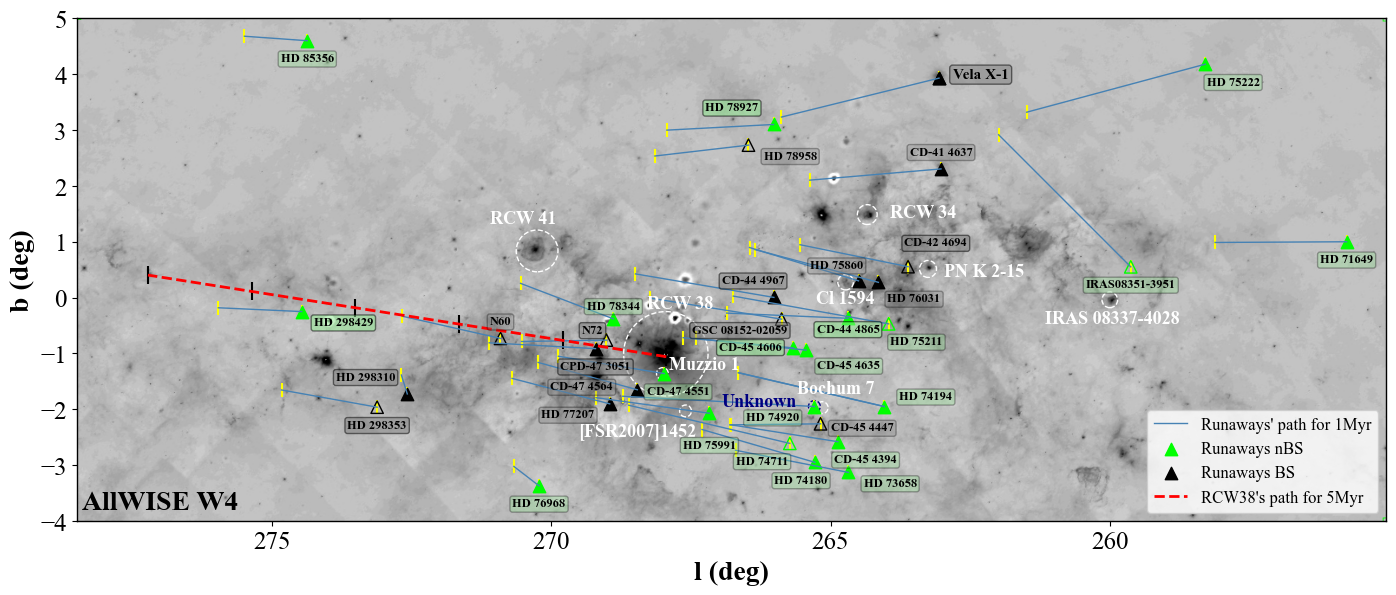}
\caption{AllWISE\,W4 image of the Vel\,OB1 association showing the positions of the OB~runaways with (black triangles) and without an associated bow shock (green triangles). The OB~runaways and the potential parent clusters are labelled (white). The dashed red line shows the path of RCW\,38 traced back over 5\,Myr. This motion can assumed to be representative for the motion of the other young clusters in this association. The blue lines display the path of each OB runaway up to 1\,Myr ago. The [KPS2012]\,MWSC\,1594 open cluster is refered to as Cl\,1594. For a handful of OB runaways an origin in one the young clusters is likely; most OB runaways seem to originate from the Galactic plane.}
\label{fig:VelaOB1}
\end{figure*}

\section{Identifying the parent cluster of the OB runaway stars}
\label{origin}

In order to identify the potential parent cluster of an OB runaway we use its current motion to find out whether its path originates in one of the (young) clusters embedded in Vel~OB1. One also needs to take into account the motion of the cluster.
For each OB runaway Fig.~\ref{fig:VelaOB1} shows the path that was followed during the past Myr; here we assume that the Galactic potential has not changed the direction of motion significantly. The path of RCW~38 is indicated by the dashed red line. The direction and pace along this path is representative for all young clusters listed in Tab.~\ref{tab:parori}.

The general conclusion based on Fig.~\ref{fig:VelaOB1} would be that the large majority of the OB runaways originate in the Galactic plane. In a few cases, we think that we can identify the parent cluster of the OB runaway (Tab.~\ref{tab:parrunori}). Recent observational studies suggest that a large fraction of OB runaways originate in young stellar clusters located in the Galactic plane and are ejected during, or shortly after, the cluster formation phase \citep[e.g. ][]{gvaramadze11,Drew21,maiz22,Stoop23,Stoop24}. For these objects, the time elapsed since ejection—referred to as the kinematical age—is therefore expected to be comparable to the age of the parent cluster. In contrast, if a runaway star is produced through the binary supernova channel, the kinematical age is typically much shorter than the cluster age, since the system must first evolve until the supernova occurs; an illustrative example is discussed by \citet{vandermeij21}.

For five OB runaways, we identified the potential parent cluster (top five systems in Tab.~\ref{tab:parrunori}): HD75222 (Bochum~7), CD-41 4637 (RCW34), GCS 08152-02059 (RCW38), HD76968 (Bochum~7), and HD298310 (RCW38). For more information on these systems, see App.~\ref{appD}.

\citet{fortin22} searched for the birth place of a sample of HMXBs, including HD74194 / IGR~J08408-4503 and HD77581 / Vela~X-1. For both systems they cannot identify a parent cluster. Tracing back the path of these systems shows that they would have intersected with the Local arm several tens of Myr ago, too long ago compared to the expected age of these systems. \citet{fortin22} conclude that these systems may have formed in isolation. Alternatively, the path of HD74194 (LM~Vel) may originate in the open cluster [FSR2007]\,1452, which would result in a kinematical age of about 4~Myr; however, \citet{Tarricq21} report a cluster age of $\sim$260\,Myr which would be too old to be the parent cluster of this HMXB. We return to the case of Vela~X-1 in Sect.~\ref{velaX}.

Possibly the pulsar PSR\,J0858-4419 has been the companion to one of the identifed OB runaways in Vel~OB1 before the supernova explosion. The galactic coordinates of PSR\,J0858-4419 are 265.45724;+00.82166; the proper motion of the pulsar has not been measured, but the distance estimate of 1.9~kpc \citep{romani11,krishnakumar15} is in accordance with Vel~OB1 (though the alternative distance of 6.5\,kpc reported by \citet{damico98} would not fit to this scenario).

As already mentioned, most OB runaways in Vel~OB1 move away from the Galactic plane. In Tab.~\ref{tab:parrunori} we list the time it took to cross Galactic latitude $b$ = 0, 1, and 2~degrees: the conclusion is that these OB runaways, if born in the Galactic plane, escaped typically a few Myr ago to reach their current position.

\section{Wind bow shocks and ISM conditions}
\label{DisBS}

For eight out of the 25 OB~runaways (32~\%) in Vel~OB1 an arc-like feature has been detected most likely due to the presence of a wind bow shock. This percentage is in line with the findings of \citet{carretero25} that $\sim24$~\% of their O-type runaways produce a bow shock. Another observation is that the OB~runaways producing a bow shock experience a higher visual extinction (Fig.~\ref{fig:histAv}) indicating that their ambient medium has a higher density (of dust). This is not unexpected as, besides the higher than average dust density near the Galactic plane, the formation of a wind bow shock depends on the ISM density. 

\begin{figure}
\centering
\includegraphics[width=0.9\linewidth]{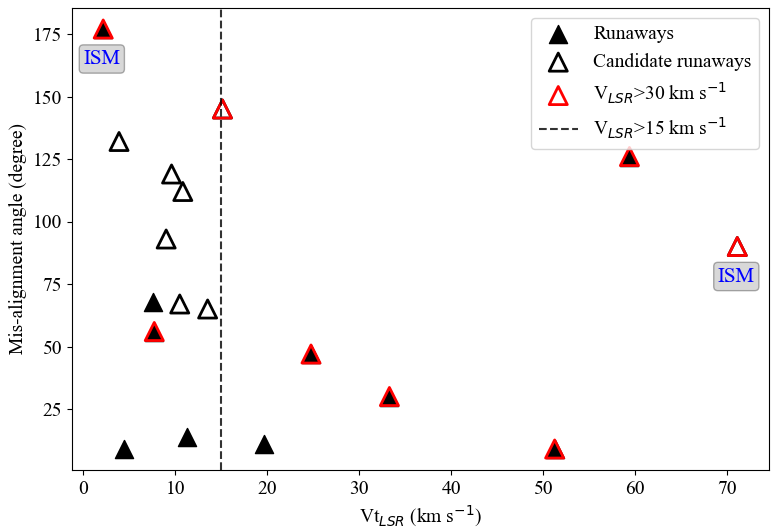}
\caption{Alignment (or misalignment) angle between the proper motion vector of the system and the symmetry axis of the bow shock as a function of the peculiar tangential motion. We have classified the bow shock as `aligned', `aligned?', and `misaligned' following the criteria listed in \citet{kobulnicky16} (see text);  CD-47~4551 and IRAS~08351-3951 are identified as unrelated ISM structures and excluded from the bow shock sample (see text). The misalignment angle shows a wide range for those systems with a low peculiar velocity.}
\label{fig:BS}
\end{figure}

We also measured the (mis)alignment angle between the (proper) motion of the system and the symmetry axis of the bow shock (Fig.~\ref{fig:BS}). The uncertainty of the misalignment angle is difficult to judge: it largely depends on the observed symmetry of the arc-like feature and the assumption that the maximum 22~$\mu$m emission refers to the top of the bow shock. Assuming a 5$^{\circ}$ uncertainty for the bow-shock position angles from \citet{kobulnicky16} and propagating Gaia proper-motion uncertainties including covariance terms, we obtain a typical misalignment-angle uncertainty of $\sim$5$^{\circ}$. In contrast, \citet{kobulnicky16} applied a conservative cut on the uncertainty of the proper-motion position angle (<45$^{\circ}$) rather than estimating the uncertainty in the misalignment angle itself. It turns out that OB runaways with a relatively low peculiar velocity show a wide range in alignment (up to total misalignment), while the bow shocks associated with systems with a high peculiar velocity show better alignment with the proper motion vector. We divided the observed bow shocks in two morphology groups: aligned and misaligned, based on the measured misalignment angle. If the misalignment angle is $<\,45^{\circ}$, the bow shock is considered aligned \citep[e.g.][]{kobulnicky16,Bodensteiner18}; if the misalignment angle is $>\,45^{\circ}$ and $<\,70^{\circ}$, the bow shock is considered aligned with a question mark (considering the contribution of the proper motion errors). In other cases ($>\,70^{\circ}$), it is mis-aligned. This means that the apparent motion of the star is not aligned with the opening angle of the nebula. Of the 18 arc-like features, two objects — CD-47 4551 (misalignment angle 177$^{\circ}$) and IRAS 08351-3951 (90$^{\circ}$) — are identified as unrelated ISM structures (labelled ISM in Tab.~\ref{tab:parrun}) and are excluded from the bow shock alignment statistics. Thus, among the remaining 16 bow shock candidates, we have ten aligned (including those with a question mark) and six misaligned bow shocks, respectively.

The observed stand-off distance of a wind bow shock is determined by the equilibrium between the momentum flux of the stellar wind and the ram pressure of the surrounding ISM \citep{baranov71}. We can confirm whether the observed arc-like feature is due to a wind bow shock based on the stellar parameters and the local ISM conditions. This is relevant, as several of our OB~runaways are identified on the basis of the association with a bow shock, while their peculiar velocity is lower than our tangential velocity threshold of 15~km~s$^{-1}$. In these cases, large-scale flows in the ISM—potentially driven by cluster winds—may produce a substantial relative velocity between the star and its environment \citep{Povich08,Bodensteiner18}.

\begin{figure}
\centering
\includegraphics[width=0.9\linewidth]{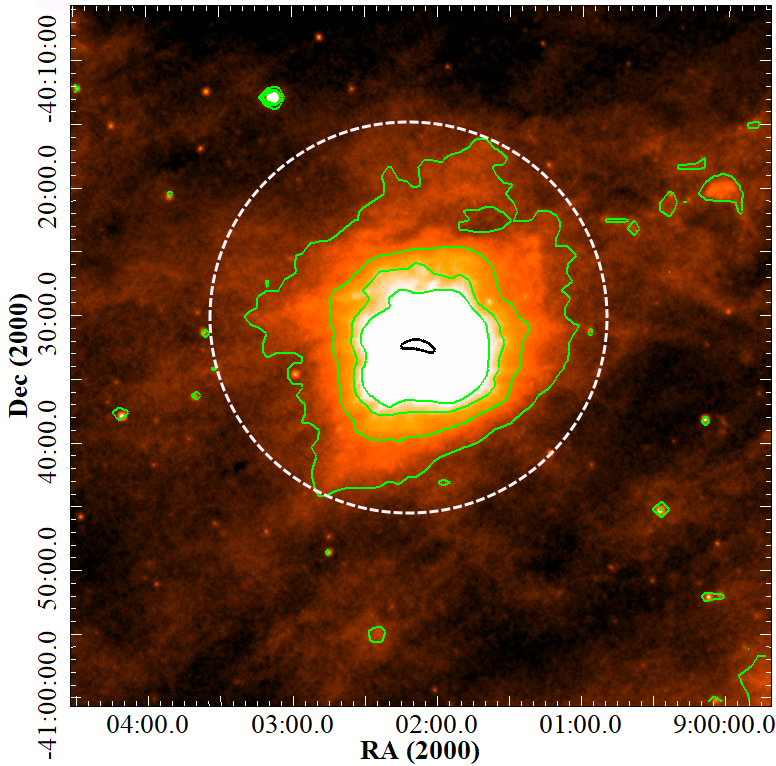}
\caption{\textit{WISE}\,22\,$\mu$m image of the field containing Vela\,X-1. The position of the bow shock generated by Vela X-1 is indicated by a black contour. The area surrounded by a dashed white circle of radius of 0.27$^{\circ}$ corresponds to the putative Str\"omgren zone produced by Vela X-1. The external green contour corresponds to 3$\sigma$ of background surface brightness.}
\label{fig:Strom}
\end{figure}

Using the spectral type of the OB star, we estimate the wind mass-loss rate and terminal velocity from empirical prescriptions \citep{howarth89,Kudritzki00}. Given the stellar wind properties, the measured stand-off distance, and the peculiar velocity, the ambient ISM density remains the only unconstrained parameter \citep{baranov71}:
\begin{equation}
\label{equ:2}
\rho_{\rm ISM} = \frac{\dot{M}V_{\rm w}}{4\pi V_{\rm pec}^2 R^2} \, ,\\   
\end{equation}
where $\rho_{\rm ISM}$ is the density of the ISM, equal to 1.4~$m_{\rm H} n_{\rm ISM}$, $m_{\rm H}$ is the mass of the hydrogen atom, $\dot{M}$ is the stellar wind mass-loss rate, \textit{V$_{\rm w}$} is the terminal wind velocity, \textit{V$_{\rm pec}$} is the (peculiar) velocity of the star with respect to ISM and $R$ is the stand-off distance, i.e. the projected closest distance between the star and the bow shock. For five runaway stars \textit{V$_{\rm w}$} and $\dot{M}$ have been reported in literature (see Tab. \ref{tab:estpar1}). For the remaining stars\textit{V$_{\rm w}$} and $\dot{M}$ are estimated by taking the average value for stars of the same spectral type \citep{kobulnicky19, mokiem07}. As a result, we obtain a value of \textit{n$_{\rm ISM}$} (Tab. \ref{tab:estpar1}).   

Another way to estimate \textit{n$_{\rm ISM}$} is through the size of the extended H~{\sc ii} region powered by the UV emission of the star; this provides a consistency check on the derived value of \textit{n$_{\rm ISM}$} based on the stand-off distance of the bow shock. The number density of the ambient medium can be obtained using the relationship \citep{lequeux05}
\begin{equation}
\label{equ:3}
R_{\rm Str} = \left[ \frac{3S(0)}{4\pi n_{\rm ISM}^2a} \right]^{1/3} \, ,\\   
\end{equation}
where \textit{R$_{\rm Str}$} is the radius of Str\"omgren sphere \citep{Stromgren39}, $S(0)$ the ionising flux emitted by the star \citep{panagia73} and $a$ the recombination coefficient ($a=3.4 \times 10^{-13}$~cm$^{3}$~s$^{−1}$ for T$_{e} = 7000$~K typical for H~{\sc ii} regions, see \citet{lequeux05}). 

As an example we take the case of HD77581 / Vela~X-1: Figure \ref{fig:Strom} shows the \textit{WISE}\,22\,$\mu$m image of the region including the B supergiant HD77581 with the position of the putative Str\"omgren sphere indicated by a dashed white circle \citep[see also][]{gvaramadze18}. It corresponds to the angular radius of the green contour where the 22\,$\mu$m flux exceeds the background surface brightness by 3$\sigma$ (a bit larger area than noted by \citet{kaper97}). The position of the bow shock generated by HD77581 is indicated by a black contour. If this extended infrared source is indeed an ionisation-bounded H~{\sc ii} region created by the UV emission of the OB runaway star, we obtain $R_{\rm Str} = 0.27^{\circ}$ and $n_{\rm ISM} = 3.7$~cm$^{-3}$ (Eq.~\ref{equ:3}). When we adopt the values for $\dot{M}$ and V$_w$ reported for HD77581 \citep{gimenez16} and the observed stand-off distance of the wind bow shock, we arrive at $n_{\rm ISM} = 1.0$~cm$^{-3}$, in reasonable agreement with the value based on the estimate of $R_{\rm Str}$ (Tab. \ref{tab:parrun}) and the ambient interstellar number density adopted by \citet{kaper97}.

\begin{table*}
\centering
\caption[]{Adopted and estimated parameters of the observed bow shock around the runaway stars.}
\resizebox{0.9\textwidth}{!}{
\label{tab:estpar1}
\begin{tabular}{l *{19}{c}}
\hline\hline\noalign{\smallskip}
\centering
Runaway	&	l & b & V$_w$	& $\dot{M}$ 
&	n$_{\rm ISM}^{\rm BS}$ & n$_{\rm ISM}^{\rm St}$ &	R$_{\rm St}$	& R\\
 $-$ &	(deg) & (deg)	 &(km\,s$^{-1}$)	
 & (M$_\sun$\,yr$^{-1}$) & (cm$^{-3}$)	&  (cm$^{-3}$) & ($^{\circ}$)  &	(arcsec)\\
\hline\noalign{\smallskip}
(1)     &       (2)     &       (3)     &       (4) 	&	 (5) 	& (6)	&	(7)		& (8)	&	(9)  \\

\hline \noalign{\smallskip}

CD-41 4637	&	263.02 & 2.30 	 & 2250\tablefootmark{(e)} & 6.9\,$\times$\,10$^{-6}$\tablefootmark{(e)}
& 30 & 28 & 0.26 & 90  \\
HD\,77581/Vela\,X-1 &	263.06 & 3.93	 &700\tablefootmark{(a)}& 6.3\,$\times$\,10$^{-7}$\tablefootmark{(a)} & 1.0 & 3.7	& 0.27 & 53\\
HD\,75860	& 264.14 & 0.27 &  (700$-$800)\tablefootmark{(d)} & (1.6\,$\times$\,10$^{-7}$$-$7.6\,$\times$\,10$^{-6}$)\tablefootmark{(d)}  
&  (0.6$-$34)  &  4.3 &  \textit{0.08} & 59\\
HD \,76031 & 264.50 & 0.30 &  (300$-$1200)\tablefootmark{(f)} & (10$^{-9}$$-$10$^{-8}$)\tablefootmark{(f)} &(0.7$-$26)& 17 & \textit{0.05}& 11\\ 
CD-44\,4967 &	266.01 & 0.00  & $-$ & $-$ &  $-$ & $-$ & 0.18	 &	 29	\\
CD-47\,4564 & 268.46 & -1.65 &(900$-$1100)\tablefootmark{(d)}& 6.4\,$\times$\,10$^{-8}$\tablefootmark{(d)}& (31$-$38)& 13 & 0.07& 8\tablefootmark{**}\\
HD\,77207 &	268.96 & -1.90   & 500\tablefootmark{(b)} & 1.5\,$\times$\,10$^{-7}$\tablefootmark{(b)} 		 & 35  & $-$ & 	0.10 &	32 	\\
CPD-47\,3051 &	269.21 & -0.91   & 1200\tablefootmark{(b)} & 7.7\,$\times$\,10$^{-9}$\tablefootmark{(b)}	
&  13.5 & 14 &  0.10	 &	 28\\
HD\,298310	&	272.58 & -1.72  & 1200\tablefootmark{(b)} & 3.4\,$\times$\,10$^{-7}$\tablefootmark{(b)}
&  4.0 & 8.2 & 0.16	& 21\\
\hline
CD-42\,4694 &	263.61 & 0.55    & (1140$-$1300)\tablefootmark{(e)}	& (10$^{-7}$\,$-$\,10$^{-9}$)\tablefootmark{(e)} 
&  (0.3$-$25) & \textit{70} & \textit{0.05}	 & 18\\
CD-45\,4447 & 265.18 & -2.26 & (1745$-$2390)\tablefootmark{(c)}	 & (3.6\,$\times$\,10$^{-8}$$-$1.1\,$\times$\,10$^{-7}$)\tablefootmark{(d)}& (130$-$560) & \textit{91} & \textit{0.02} & 22\tablefootmark{**}\\ 
GSC\,08152-02059 &	265.88 & -0.39   &  $-$	& $-$ &  $-$ & $-$ &  \textit{0.06}	 &	15\tablefootmark{**}	\\
HD\,78958 & 266.47 & 2.73 &(500$-$2000)\tablefootmark{(c)}	 &3.0\,$\times$\,10$^{-6}$\tablefootmark{(e)}&  (809$-$3236)? & 14 & 0.10& 40\tablefootmark{**} \\
DR3\,N72 &	269.01 & -0.76  & $-$  	& $-$ & $-$  & $-$ & 0.13	 &	34\tablefootmark{**}	\\
DR3\,N60 &	270.91 & -0.73  &  $-$  & $-$	&  $-$ & $-$ & \textit{0.10}	 &	19\tablefootmark{**}	\\
HD\,298353 &	273.12 & -1.96    &2500\tablefootmark{(b)} & 6.0\,$\times$\,10$^{-8}$\tablefootmark{(b)}	
&  25 &  41 & 	0.10 &	32	\\
\hline
\end{tabular}
}
\tablefoot{
(1) name of star, (2), (3) Galactic coordinates of star, (4) adopted terminal wind velocity, (5)  adopted mass-loss rate, (6) ambient interstellar number density determined by the balance between the ram pressures of the wind and the ambient medium \citep{baranov71} computed in this work, (7) ambient interstellar number density determined by the relationship of \citet{lequeux05} computed in this work, (8) the putative Str\"omgren radius obtained from \textit{WISE}\,22\,$\mu$m image of the region, (9) standoff distance from \citet{kobulnicky16}. The parameters are taken from \citet{gimenez16} (a); \citet{kobulnicky19} (b); \citet{Prinja90} (c); the parameters are based on the spectral type  \citet{kobulnicky19} (d); \citet{mokiem07} (e); typical value for early-B subgiants \citet{Krtivcka14} (f); (**) we measured the standoff distance in the WISE\,22\,$\mu$m images. Two objects are indicated as DR3\,N72 and N60 based on their last numbers of \texttt{source\_id} from \textit{Gaia} DR3. Suspicious results are in \textit{italic}.  Candidate runaway systems are listed in the bottom part of the table.}
\end{table*}

We used \textit{WISE}\,22\,$\mu$m images for the identification of the putative Str\"omgren sphere for the other stars. In some cases, the identification of the putative Str\"omgren sphere is not clear because the object is embedded in a complex region. Tab. \ref{tab:estpar1} lists the results; for some runaways, we lack information. In most cases we find reasonable agreement between the two estimates of the ISM number density, suggesting that the Str\"omgren sphere and the bow shock are produced by the same star. In other cases the discrepancy could be due to the underestimated relative velocity as we assumed that the ISM is at rest with respect to the Galactic rotation. The derived values of the ambient-medium density around HD77581, CD-41\,4637, HD\,77207, and HD\,298353 stars are fairly high for what one would expect at these Galactic heights (see Columns 8-9 in Tab. \ref{tab:estpar1}). The observations at longer wavelengths (>\,22\,$\mu$m), however, indicate the presence of a relatively dense medium surrounding these stars. In general we find good agreement between the two determinations of the ISM number density and the measurements of \citet{kobulnicky19} for individual stars. We conclude that the measured ISM number density is in the range of 0.1\,$-$\,40\,cm$^{-3}$. We derived the mass-loss rate of the OB runaways using \textit{n$_{ISM}^{St}$} in Eq.~\ref{equ:2} (see Column 7 in Tab. \ref{tab:estpar1}). The derived mass-loss rate is compatible with the adopted mass-loss rate. This is in accordance with the findings of \citet{Patten25} regarding the parameters of the central stars of 103 infrared bow shock nebulae, but see Sect.~\ref{BS}.

The agreement between n$_{\rm ISM}^{\rm BS}$ and n$_{\rm ISM}^{\rm St}$ for most sources (Tab.~\ref{tab:estpar1}) provides an independent consistency check on the bow shock interpretation. If the arc-like feature were an unrelated ISM structure, there would be no reason to expect its standoff distance to be consistent with the pressure balance between the known stellar wind parameters and the local ISM density inferred from the Str\"omgren sphere. Eight of the 11 sources with both estimates available agree within a factor of three, strongly supporting the identification of these features as genuine stellar wind bow shocks.

\section{Discussion}
\label{discussion}

We have searched for OB runaways in the region contained by the Vel~OB1 association, host of close to a hundred OB-type stars, at a distance of about 1.8~kpc. We characterised Vel~OB1 by performing a membership analysis of a few young clusters included in the association. We identified OB runaways based on their peculiar motion (with respect to the LSR and to the mean motion of the young clusters, in some cases including the radial-velocity component) and/or the association with a wind bow shock. The definition of an OB~runaway depends on the value of the peculiar-velocity threshold. We argue that a threshold of 15~km~s$^{-1}$, which is still higher than the escape velocity of a stellar cluster, is sufficient to identify an OB~runaway in this region.

\subsection{Comparison to earlier work}

\subsubsection{Kinematical properties of the OB runaway stars}

The OB stars in the sample studied by \citet{Carretero23} (Galactic O-Star Catalogue and the Be Star Spectra Database) show a Gaussian peculiar tangential velocity distribution with a dispersion of 5--10~km~s$^{-1}$. Their OB~runaways have a (tangential) peculiar velocity in the range from 5 to about 150~km~s$^{-1}$. \citet{StoopPhD} demonstrate that their sample of over 400 OB~runaways show a transverse velocity distribution (with tangential velocities in the range from 5 to 300~km~s$^{-1}$ with respect to their parent cluster) that can be modelled with a power law. The majority of the OB~runaways has a tangential peculiar velocity in the range 5 to 20~km~s$^{-1}$; note that these OB~runaways are most likely produced by dynamical ejection.

Our Vel~OB1 runaway sample shows a similar peculiar velocity distribution, rather flat (Fig.~\ref{fig:transvel}) and in the range between 5 and 70~km~s$^{-1}$. The OB~runaways identified by the association with a bow shock do not need to meet the tangential peculiar velocity threshold of 15~km~s$^{-1}$, as long as they move at supersonic velocity. As argued in Sect.~\ref{DisBS}, these arc-like features are often misaligned with the direction of motion and may be the result of the interaction between a bulk flow of the ISM with the OB-star wind.

The distinction between runaway and walkaway stars \citep{Renzo19} is not sharply defined. While escape from the natal environment is ultimately governed by the cluster escape velocity, the traditionally adopted velocity threshold of 30\,km\,s$^{-1}$ does not have a strong physical basis. Binary evolution models show that the supernova ejection mechanism can produce a substantial number of systems with velocities well below this classical limit \citep{Renzo19}, blurring the boundary between the two populations. Thus, all OB~runaways observed in Vel~OB1 can be produced by both mechanisms.

\subsubsection{The OB runaway fraction}

In Sect.~\ref{OBrunprop} we report on 21 OB~runaways on a total of 70 OB~stars in Vel~OB1 resulting in a runaway fraction of 29~\% (and 18~\% if we take 30~km~s$^{-1}$ as the threshold). The OB runaway fraction is higher for the O-type stars (38~\%) than for the B-type stars (32~\%). \citet{Carretero23} report a percentage of 25~\% for O-type stars and 5.2~\% for Be stars. For about 24~\% of the O~runaways they detect a wind bow shock and for just about 3~\% of the Be~runaways.

In a recent study of 15 young massive clusters \citet{StoopPhD} find an OB-runaway fraction of 19~\% for the combined population of WN(h) and O-type stars, but this could be as high as $\sim$~33~\% as a significant fraction of the OB~runaways is lacking a spectral classification. As these OB~runaways have a kinematical age less than 3~Myr and have an identified parent cluster, they are produced by dynamical ejection.

\citet{Pantaleoni25} present the third release of the Alma catalogue of OB stars \citep{reed03}, cross-matched with \textit{Gaia}\,DR3 and an extension based on new spectral classifications. Their list contains 583 stars in the direction of the Vel~OB1 region; of them, 314 stars are at the same distance as Vel\,OB1. We compared Tab.~\ref{tab:par1} with their list and all our 77 stars are in common. We also compared the OB runaway stars in Tab.~\ref{tab:parrun} (not included in Tab. \ref{tab:par1}) with the list of \citet{Pantaleoni25} and 5 out of 8 stars are in common. Three stars, namely IRAS\,08351-3951, DR3\,N72, and DR3\,N60 are not included in their list. It should be noted that IRAS\,08351-3951 has spectral type M8 and that the spectral type of DR3\,N72 and DR3\,N60 are not known. Excluding the stars in common with the Alma catalogue, we searched for runaways with category "M"  (likely massive stars) by measuring their tangential peculiar velocity. It turns out that 30 stars out of 211 have a tangential peculiar velocity higher than 15\,km\,s$^{-1}$; these runaways are of spectral type O (2), B (13), A (2), G (1) and unknown (12). Adding them to our list of OB~runaways (also from Tab.~\ref{tab:parrun}), the runaway fraction as a function of spectral type then becomes 29~\% for the O stars (12/(21+20)) and 21~\% for the B stars (27/50+80). 

We inspected the regions surrounding the new OB~runaways based on the Alma catalogue and only ALS\,1327 has a small arc-like feature associated with it. This star has a tangential peculiar velocity 63\,km\,s$^{-1}$.

\subsection{Dynamical ejection or binary supernova scenario}

Two of the OB~runaways are the optical counterpart of a HMXB: HD77581 (Vela~X-1) and HD74194 (IGR~J08408-4503). In some HMXBs, the system remains gravitationally bound following the supernova event that produces the compact object, while acquiring a systemic recoil velocity as a consequence of the mass lost during the explosion \citep{vandenheuvel00}. For both systems no convincing parent cluster has been identified prohibiting a reconstruction of the evolutionary history of the system such as for HD153919 (4U1700-37) \citep{vandermeij21}. In exceptional cases an OB~runaway can be traced back to a radio pulsar, indicating that the OB~runaway was produced by the binary supernova scenario leading to the disruption of the binary progenitor system. Notable examples include the association of the OB~runaway $\zeta$~Oph with PSR B1706-16 \citep[e.g. ][]{hoogerwerf00,neuhauser20}, as well as the massive runaway star HD\,254577 with the neutron star CXOU J61705.3+222127 \citep{dincel26}.

For the OB~runaways in Vel~OB1 that could be linked to a parent cluster, the kinematical age is consistent with the estimated age of the parent cluster. For these runaways the dynamical ejection scenario is the most likely production mechanism. Thus, for the majority of the OB runaways in Vel~OB1 the production mechanism cannot be determined. Both the dynamical ejection and the binary supernova scenario are at work, but in some cases also a hybrid scenario (a massive binary, dynamically ejected from a cluster, undergoes a supernova explosion at a later stage) could provide the explanation for the observed motion of the runaway.

\subsection{A hybrid scenario for Vela X-1?}
\label{velaX}

The formation history of the HMXB Vela~X-1 remains uncertain. \citet{rensbergen96} proposed that the initially more massive component, with an estimated mass of 25\,M$_\odot$, underwent a supernova after at least 9\,Myr, leaving behind a neutron star and the present-day 24\,M$_\odot$ B-type supergiant companion. In this scenario, the system was ejected from the Vel~OB1 association less than 3\,Myr ago, implying a total system age of approximately 12\,Myr. Using {\it Hipparcos} astrometry, \citet{kaper97} reached a similar conclusion regarding an origin in Vel~OB1. Alternatively, \citet{gvaramadze18} suggested a two-step ejection process, in which the binary was first dynamically removed from its natal cluster through few-body interactions and later received an additional velocity kick from the supernova explosion after it had already travelled far from its birthplace.

Based on the Galactic latitude of Vela\,X-1 (b=+3.9$^{\circ}$) and distance, its current location is 134\,pc above the Galactic plane. We traced back the path of Vela\,X-1 HMXB and considered various scenarios to explain its current position and velocity. One of reconstructed paths of Vela\,X-1 reveals that its position on the sky was close the Vela\,R2 association about 2\,Myr ago. However, the distance to Vela\,R2 is 850\,pc \citep{eggen82}, inconsistent with the current distance of Vela~X-1 ($\sim$1.9~kpc). 
Another option is that Vela~X-1 left the RCW\,38 region about 4.5\,Myr ago; but then the total age of the system would be 
about 13\,Myr \citep{rensbergen96}. This is inconsistent with the age range of RCW\,38 (3.2-7.1~Myr), unless this young cluster also hosts an older population \citep[like M16,][]{Stoop23}. Therefore, we consider a hybrid scenario as proposed by \citet{gvaramadze18} as the most likely scenario for Vela~X-1. This is also concluded in the recent paper by \citet{Patten25} where they looked at the fundamental parameters of the central (OB) stars of 103 infrared bow shock nebulae (Vela~X-1 corresponds to BS386).

\subsection{Impact on the surrounding medium}
\label{Impact}

Massive OB stars strongly influence their surroundings through intense ultraviolet radiation, powerful stellar winds, and eventual supernova explosions, regardless of whether they remain in their natal clusters or migrate into the field. It is important to quantify the number and production rate of OB runaways. In this paper we attempt to obtain a census of the OB runaways in the massive star-forming region Vel~OB1 using {\it Gaia} astrometric data and the association with a (wind) bow shock. While we identify approximately a dozen arc-like infrared features, not all of these structures necessarily arise from the supersonic motion of a runaway star. In some cases, the interaction between a fast bulk flow in the ISM and the wind of a relatively stationary OB star can produce similar morphologies \citep{Bodensteiner18,kobulnicky16}. A comparable situation was reported by \citet{Guarcello25}, who identified a red supergiant in Westerlund~1 with a bow shock oriented away from the cluster centre, most likely shaped by large-scale ISM dynamics. Indeed we find indications that several arc-like features do not align with the measured space motion of a potential OB runaway, especially in case of runaways with a low peculiar velocity. Therefore, we need to be careful in identifying OB runaways based on their association with an arc-like ISM feature. 

As runaway OB stars can traverse distances of hundreds of parsecs before exploding as supernovae, their feedback may be deposited far from their birth sites. In such low-density environments, supernova remnants can expand more freely, potentially allowing nucleosynthetically enriched material to escape the Galactic disc. The contribution of runaway stars to the heating, structuring, and chemical enrichment of the ISM--and possibly to Galactic outflows—may therefore be more significant than previously assumed \citep{Andersson20}.

\section{Summary and conclusions}
\label{conclusions}

The aim of this work was to search for OB runaway stars (and their associated wind bow shocks) in the Vel\,OB1 association, one of the largest OB associations known. We analysed six young stellar clusters contained in the Vel\,OB1 region, which make the majority of the massive-star population in this part of the Galaxy. We were able to identify 26 runaways by measuring their peculiar velocity with respect to the association. Moreover, we detected 16 arc-like features (bow shocks) in this region, four of them for the first time. We traced back the path of these objects to understand both the most likely cause for their acceleration (supernova of the companion star or the dynamical ejection from the parent cluster) and from which part of the Vel\,OB1 association the stars were kicked out. We identified the parent cluster for three runaways with observed bow shocks and four runaways without observed bow shocks. Ten of the 16 bow shocks are well aligned with the proper motion of the runaway. The aligned and misaligned bow shocks found in the mid-infrared images around the runaway or walkaway stars suggest either supersonic velocities of stars with strong stellar winds relative to the local ISM or the presence of large-scale motions within the ISM. 

\begin{acknowledgements}
\small
We thank the anonymous referee for their helpful report that improved clarity of this work. This work was partially supported by a research grant number N\textsuperscript{\underline{o}}\,21AG-1C044 from Higher Education and Science Committee of Ministry of Education, Science, Culture and Sport RA. This work presents results from the European Space Agency (ESA) space mission Gaia. Gaia data are being processed by the Gaia Data Processing and Analysis Consortium (DPAC). Funding for the DPAC is provided by national institutions, in particular the institutions participating in the Gaia MultiLateral Agreement (MLA). This publication also makes use of data products from the Wide-field Infrared Survey Explorer, which is a joint project of the University of California, Los Angeles, and the Jet Propulsion Laboratory/California Institute of Technology, funded by the National Aeronautics and Space Administration.
\end{acknowledgements}

\bibliographystyle{aa}
\bibliography{iras05}

\appendix

\section{Members of Vel OB1 collected from literature, runaways, and parent clusters}
\label{appA}
This section presents supporting material related to the identification of candidate members of the Vel~OB1 association and the characterisation of associated OB runaway stars. Table~\ref{tab:par1} lists the candidate Vel~OB1 members compiled from the literature together with their \textit{Gaia}~DR3 astrometric parameters.

\begin{figure*}
\centering
\includegraphics[width=0.8\linewidth]{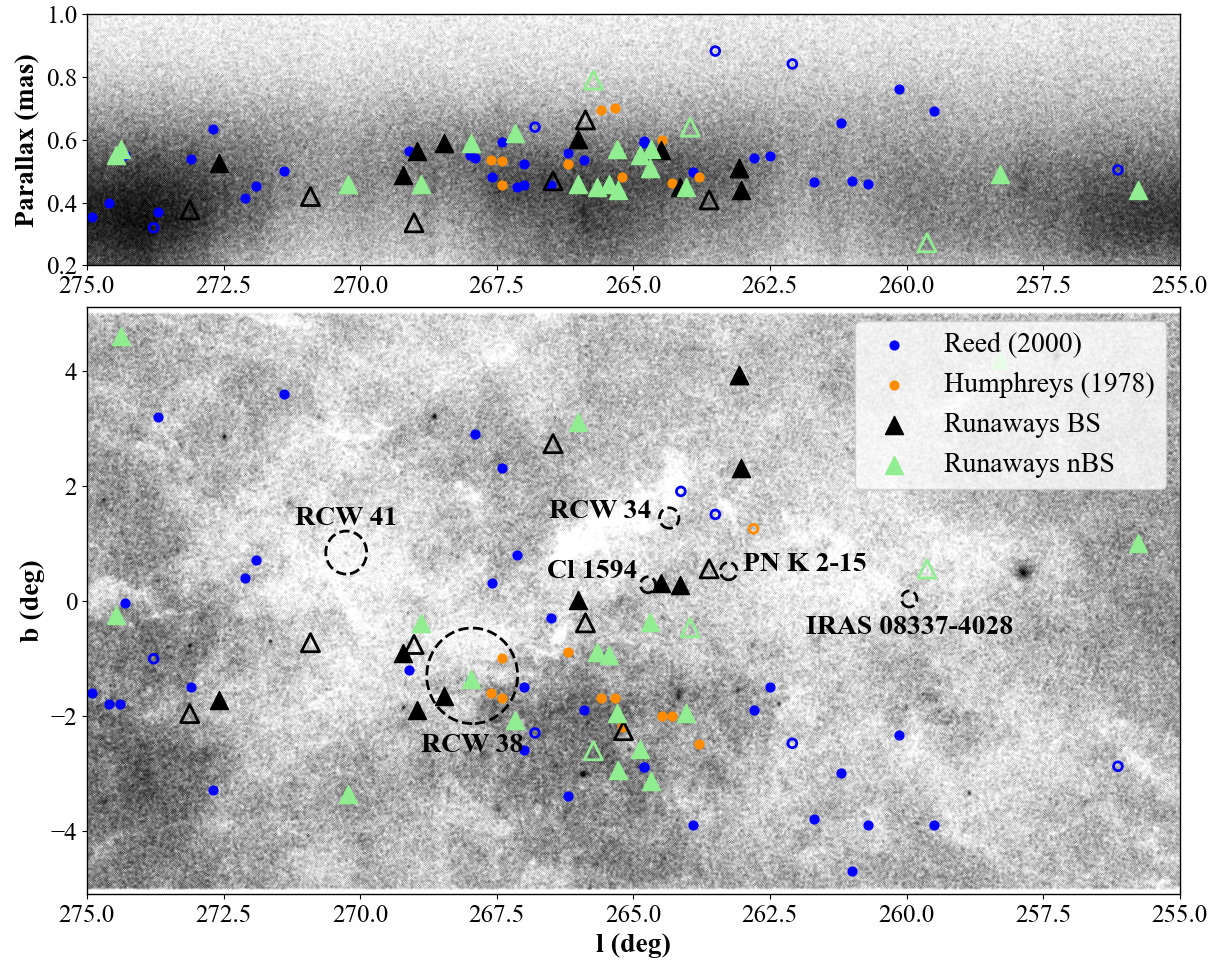}
\caption{The about 2 million {\it Gaia} DR3 sources in the region contained by Vel~OB1. The members proposed by \citet{humphreys78} (orange) and \citet{reed00} (blue) are indicated. Open circles refer to non-members (see Tab. \ref{tab:par1}). The black and green triangles represent the runaways with (Runaways BS) and without (Runaways nBS) an observed bow shock. Open triangles refer to candidate runaways (see Tab. \ref{tab:parrun}). \textit{Upper panel:} The parallax distribution as a function of Galactic longitude. \textit{Lower panel:} The position of known young clusters in the distance range of Vel~OB1 in Galactic coordinates. The approximate size of the young clusters in this region is marked by a dashed circle and labelled. The foreground cluster RCW~36 (0.7~kpc) is located at $l = 265.1$ and $b = 1.4$ and has a reddening $A_{V} \geq 8$~mag \citep{ellerbroek13}; the latter explains the low stellar density in this area. The [KPS2012]\,MWSC\,1594 open cluster is referred to as Cl\,1594.}
\label{fig:parallax}
\end{figure*}

\begin{figure*}
\centering
\includegraphics[width=0.8\linewidth]{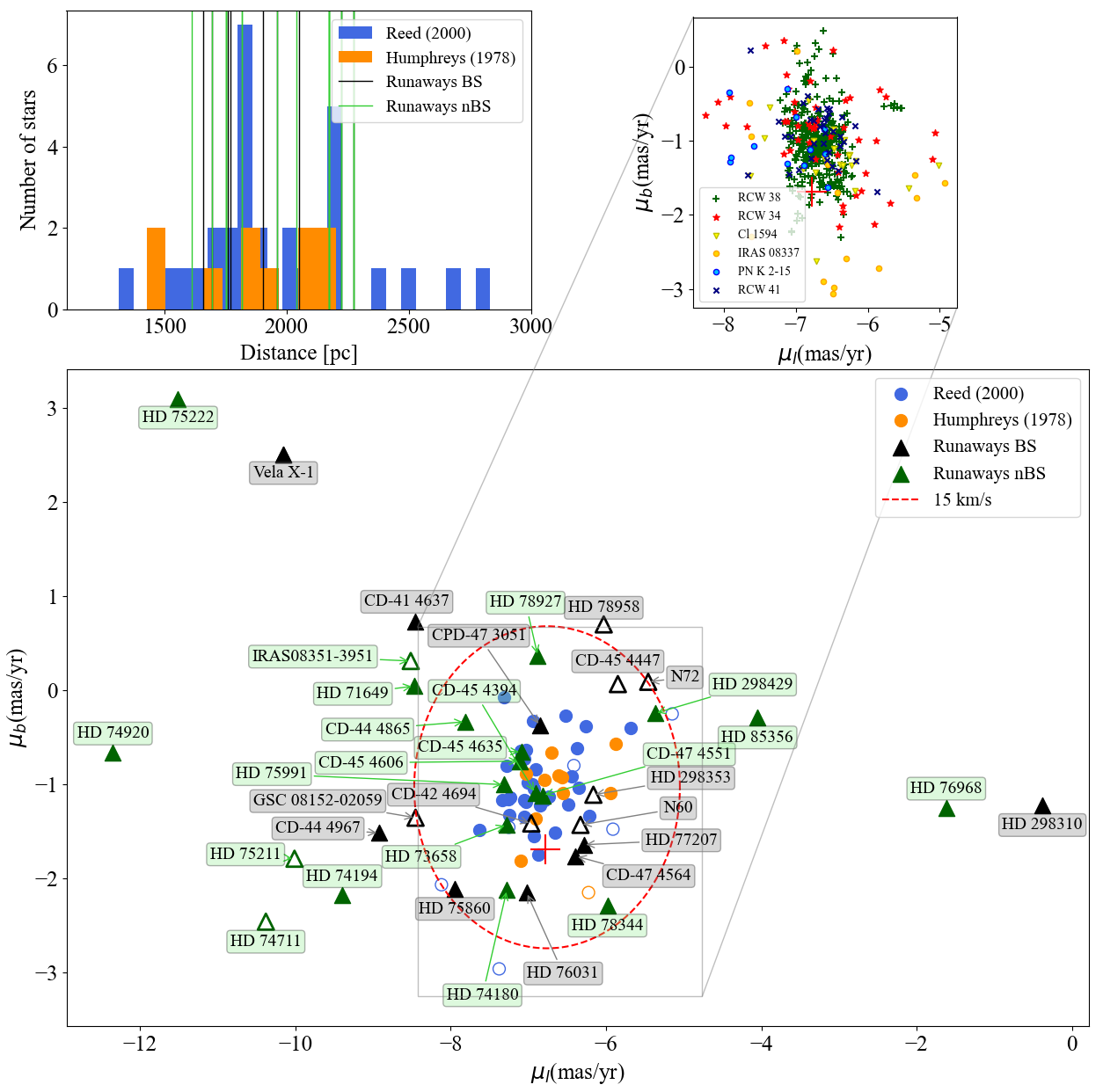}
\caption{\textit{Top-left panel:} The {\it Gaia} DR3 distance distribution of the candidate members of Vel\,OB1 (cf.\ Tab.~\ref{tab:par1}); candidate members proposed by, respectively, \citet{humphreys78} (orange) and \citet{reed00} (blue) are indicated. A black vertical line represents the distance to a runaway with an observed bow shock (Runaways BS), and the green vertical lines to those without an observed bow shock (Runaways nBS). \textit{Lower panel:} The \textit{Gaia} DR3 proper motion of the candidate members of Vel\,OB1 (coloured circles) and runaways (triangles). Runaway stars with and without an associated bow shock are labelled black and green, respectively. Open circles are non-members (see Tab. \ref{tab:par1}). Open triangles refer to candidate runaways (see Tab. \ref{tab:parrun}). The red cross indicates the proper motion of Vel\,OB1 as reported by \citet{melnik17}. The red dashed ellipse indicates a tangential peculiar velocity of 15~km~s$^{-1}$ with respect to the mean proper motion of the young massive clusters in Vel~OB1. \textit{Top-right panel:} The proper motion distribution of the members of young clusters in Vel~OB1. The [KPS2012]\,MWSC\,1594 open cluster and IRAS\,08337-4028 are indicated as Cl\,1594 and  IRAS\,08337, respectively.}
\label{fig:disOB1_run}
\end{figure*}

Figure~\ref{fig:parallax} provides an overview of the parallax distribution of \textit{Gaia}~DR3 sources in the Vel~OB1 region and the spatial distribution of young stellar over-densities in Galactic coordinates, placing the candidate members, clusters, and runaway stars in their distance and spatial context. Figure~\ref{fig:disOB1_run} further illustrates the distance and proper-motion distributions of the candidate members and runaways, and compares them with the mean kinematics of Vel~OB1 and the young clusters.

The astrometric and kinematic properties of the OB runaway stars are summarised in Table~\ref{tab:parrun}, while Table~\ref{tab:parrunori} presents estimates of their kinematical ages and, where possible, their likely parent clusters, distinguishing confirmed from candidate runaways. 

\begin{table*}[th!]
\centering
\caption[]{Candidate members of Vel~OB1 according to literature.}
\resizebox{0.75\textwidth}{!}{
\label{tab:par1}
\begin{tabular}{l r r l *{8}{c} ll}
\hline\hline\noalign{\smallskip}
\centering
Star	&	l	&	b &	Sp. Type		&	V$_r$ &	Parallax	&	$\mu_{l^{*}}$ &	$\mu_{b}$	&	G	&	D & \multicolumn{2}{c}{V$_{\rm pec}$}	& Member	& Ref (SpT,V$_r$)	\\
    &   &   &  &  &  &  &  & & & LSR & Vel OB1 & & \\
$-$ &	(deg)	&	(deg)	& $-$  &	(km\,s$^{-1}$)			&		(mas)		&	(mas\,yr$^{-1}$)	&	(mas\,yr$^{-1}$)	&	(mag)	&	(pc)	& (km\,s$^{-1}$) & (km\,s$^{-1}$)	& $-$&	$-$ \\
\hline\noalign{\smallskip}
(1)     &       (2)     &       (3)     &       (4) 	&	 (5) 	& (6)	&	(7)		& (8)	&	(9)			& (10)	& (11) &  (12) & (13) & (14) \\
\hline \noalign{\smallskip}
HD 71649\tablefootmark{**}	&	255.76	&	1.00	&			B1 IIIn &		$-$	&	0.44	$\pm$	0.02	&	-8.47	$\pm$	0.02	&	0.05	$\pm$	0.02	&	8.7	&	2256	&  39.6  & 36.8 &	N,R	&	9			\\	
CD-38 4168\tablefootmark{**}	&	256.13	&	-2.88	&			O8.5 III	&	$-$	&	0.50	$\pm$	0.02	&	-5.15 $\pm$	0.02	&	-0.25	$\pm$	0.02	&	9.1	&	1986	& 3.7  & 15.9 &	N?	&	9			\\	
\textbf{HD 75222}\tablefootmark{**} &		258.29	&	4.18	&			B0 Ia &		64	&	0.49	$\pm$	0.02	&	-11.51 $\pm$	0.02	&	3.10	$\pm$	0.02	&	7.3	&	2017	& \textit{74.6} & \textit{78.8}  &	N,R &	9	/	18	\\	
HD 69882\tablefootmark{**}	&	259.50	&	-3.91	&			B0.7 Ib &			25.9	&	0.70	$\pm$	0.02	&	-6.93
 $\pm$	0.02	&	-1.04 $\pm$	0.02	&	7.1	&	1427	& \textit{9.3} & \textit{10.9} &	Y	& 9	/	19	\\
HD 71528\tablefootmark{**}	&	260.13	&	-2.33	&			B1.5 III &		$-$	&	0.76 $\pm$ 0.04	&	-6.85
 $\pm$ 0.04	&	-1.22 $\pm$ 0.05	&	7.8	&	1331	& 5.9 & 17.1 &	Y? & 9			\\	
HD 70583\tablefootmark{**}	&	260.78	&	-3.96	&			B9 Ia &		$-$		&	0.47	$\pm$	0.02	&	-6.45
	$\pm$	0.02	&	-0.92 $\pm$	0.02 	&	7.8	&	2149	& 9.0 & 8.5 &	Y& 9			\\
HD 70122\tablefootmark{**}	&	261.07	&	-4.71	&			B1 II &		$-$		&	0.48	$\pm$	0.01	&	-6.22
	$\pm$	0.01	&	-1.34	$\pm$	0.01	&	9.1	&	2094	& 9.4 & 5.9 &	Y	& 9			\\
HD 71609\tablefootmark{**}	&	261.20	&	-3.01	&			B3 II &			25	&	0.66	$\pm$	0.02	&	-7.26
 $\pm$	0.02	&	-1.15 $\pm$	0.02	&	7.8	&	1501	& \textit{10.4} & \textit{6.8} &	Y? & 9	/	2	\\
HD 71304\tablefootmark{**}	&	261.76	&	-3.77	&			O9 II &			25	&	0.48	$\pm$	0.02	&	-6.48
	$\pm$	0.02	&	-1.22	$\pm$	0.02	&	8.1	&	2096	& \textit{14.2} & \textit{8.1} &	Y	& 3	/	16,2	\\
HD 72576\tablefootmark{**}	&	262.09	&	-2.48	&			B2 II/III &		$-$		&	0.84	$\pm$	0.02	&	-7.25
	$\pm$	0.02	&	-1.46	$\pm$	0.02	&	8.2	&	1190	& 5.3 & 18.3 &	N & 9			\\
CD-43 4488\tablefootmark{**}	&	262.53	&	-1.46	&			O9/B0	&	$-$	&	0.56	$\pm$	0.01	&	-6.93
	$\pm$	0.01	&	-1.55 $\pm$	0.02	&	10.4	&	1792	&  9.1 & 3.0 &	Y	& 5			\\	
HD 75759\tablefootmark{*}	&	262.80	&	1.25	&	O9 Vn	&		23.4		&	1.08 $\pm$ 0.04	&	-6.23
 $\pm$ 0.04	&	-2.15 $\pm$ 0.05	&	6.0	&	934	& \textit{5.0} & \textit{30.2} & N		& $-$			\\
HD 73420\tablefootmark{**}	&	262.81	&	-1.95	&			B0 III &		$-$		&	0.55	$\pm$	0.02	&	-7.24
	$\pm$	0.02	&	-1.14 $\pm$	0.02	&	8.8	&	1819	&  9.9 & 4.9 &	Y	& 9			\\
\textbf{HD 77581}\tablefootmark{**} &	263.06 & 3.93 &			B0.5 Ia	& -3.2 & 0.51	$\pm$	0.02 & -10.15
	$\pm$	0.02 	&	2.51 $\pm$	0.02 &	6.7 & 1960 &	\textit{64.7} &	\textit{60}	& N,R,B &   22/19 \\
HD 76341\tablefootmark{**}	&	263.53	&	1.52	&			O9.2 IV &			26	&	0.90 $\pm$ 0.05	&	-5.91	
 $\pm$ 0.05	&	-1.48 $\pm$ 0.05	&	7.0	&	1137	& \textit{5.3} & \textit{27.3} &	N & 3	/	14	\\
CD-42 4694\tablefootmark{**} &	263.61 & 0.55 &	O9 V	& $-$ & 0.41 $\pm$ 0.03  & -6.97 $\pm$ 0.03	&	-1.41 $\pm$ 0.03	& 10.4 & 2466 &	   13.4 & 22.4 &  Y?,R?,B? &	10\\
HD 73568\tablefootmark{*}	&	263.80	&	-2.50	&	B0.5 III	&		$-$		&	0.50	$\pm$	0.03	&	-5.87	$\pm$	0.03	&	-0.57	$\pm$	0.04	&	8.3	&	2023	& 2.9 &   7.2	& Y & 1,4			\\
HD 72554\tablefootmark{**}	&	263.94	&	-3.86	&			B1.5 Ib & $-$ &	0.51 $\pm$	0.02	& -7.05	$\pm$	0.02 &	-1.34 	$\pm$	0.02	&	8.1	&	1984	& 10.9 
&  7.2 &	Y& 9			\\				
\textbf{HD 75211}\tablefootmark{*}  &		263.96	&	-0.47	&	O9 lb	&		20		&	0.64	$\pm$	0.02	&	-10.01
$\pm$	0.02	&	-1.79	$\pm$	0.02	&	7.4	&	1550	& \textit{30.4} & \textit{17.1}  &		N,R?	& $-$			\\
\textbf{HD 74194}\tablefootmark{*} &		264.04	&	-1.95	&	O8.5 lb &					-2.7	&	0.45	$\pm$	0.02	&	-9.40	$\pm$	0.02	&	-2.18		$\pm$	0.02	&	7.5	&	2220	& \textit{54.4} & \textit{51.3} &	N,R 	& 1			\\
CD-42 4819\tablefootmark{**}	& 264.13 & 1.90 & B2 Vp & $-$ &  1.24 $\pm$	0.03 & -7.38 $\pm$	0.02 &	-2.96 $\pm$	0.02 &  9.6 & 808 & 3.7  &  29.9   & N & 23\\
HD 75860\tablefootmark{*}& 264.14 & 0.27&	B1.5 Iabp	& 16.8 & 0.45 $\pm$ 0.04	 & -7.94 $\pm$ 0.04	&	-2.11 $\pm$ 0.05	& 7.3 & 2276 	&	\textit{31.1} &	\textit{28.7} 	 & Y?,R,B & 9\\
CD-44 4691\tablefootmark{*}	&	264.26	&	-2.02	&	B2.5 II		&		39		&	0.47	$\pm$	0.02	&	-6.55	$\pm$	0.02	&	-1.09 $\pm$	0.02	&	8.4	&	2121	 & \textit{8.5}  & \textit{15.1} &	Y &  1	/	1	\\
HD 74371\tablefootmark{*}	&	264.44	&	-2.01	&	B5 lab/b &				24.6		&	0.61	$\pm$	0.07	&	-7.10	$\pm$	0.07	&	-1.81	 $\pm$	0.07	&	5.1	&	1787	& \textit{11.6} & \textit{4.5} &	Y & 	/	1	\\
HD 76031\tablefootmark{**}	&	264.50	&	0.30	&			B0.5 IV &	$-$	&	0.57	$\pm$	0.02	&	-7.02	$\pm$	0.03	&	-2.15	$\pm$	0.03	&	8.8	&	1752	&	11.3 & 6.1 & Y,R,B & 9			\\
HD 73658\tablefootmark{**}	&	264.68	&	-3.13	&			B1 II &		52	&	0.57	$\pm$	0.02	&	-7.28	$\pm$	0.03	&	-1.43	$\pm$	0.02	&	6.8	&	1783	& \textit{22.8} & \textit{26.3} &	Y,R	& 9	/	14	\\	
CD-44 4865\tablefootmark{*}	&	264.69	&	-0.37	&	O9.5 lb	&		39		&	0.51	$\pm$	0.03	&	-7.81	$\pm$	0.03	&	-0.33	$\pm$	0.03	&	9.2	&	1937	& \textit{14.6} & \textit{21.6} &	R	& 1			\\	
HD 73919\tablefootmark{**}	&	264.78	&	-2.90	&			B1 V &			$-$	&	0.61	$\pm$	0.01	&	-7.24	$\pm$	0.02	&	-1.33	$\pm$	0.02	&	8.8	&	1656	& 7.6 & 2.9 &	Y	& 9			\\
CD-45 4355\tablefootmark{**}	&	264.80	&	-2.91	&			B1 Vne		&	$-$	&	0.60	$\pm$	0.01	&	-7.63	$\pm$	0.01	&	-1.48	$\pm$	0.02	&	9.3	&	1671	& 11.0 & 1.7 &	Y	& 12			\\
CD-45 4394\tablefootmark{**}	&	264.87	&	-2.58	&			B2 Vne &			78	&	0.55	$\pm$	0.01	&	-6.91	$\pm$	0.01	&	-1.09		$\pm$	0.01	&	10.0	&	1826	& \textit{46.4} & \textit{52.5} &	N?,R & 11	/	21	\\
HD 74677\tablefootmark{*} & 265.17 & -2.20 & B2 II  & 23 & 0.49 $\pm$	0.01 	&	-5.94 $\pm$	0.02 	& -1.09 $\pm$ 	0.02 & 8.6 & 2063 & \textit{12.3} & \textit{2.2} & Y & 4/17\\
CD-45 4447\tablefootmark{**}	&	265.18	&	-2.26	&			O7.5 V	&	$-$	&	1.53 $\pm$ 0.39	&	-5.85 $\pm$ 0.45	&	0.07 $\pm$ 0.45	&	10.9	&	888	& 10.5 & 42.3 &	N,R?,B? & 8			\\	
HD 74180\tablefootmark{*} & 265.28 & -2.95 & F2 Ia & 31 & 0.43 $\pm$	0.10  & -7.28	$\pm$	0.11 	& -2.12  $\pm$	0.12 &  3.6 & 2532 & \textit{22.3} & \textit{25.5} & Y?,R & 4/24\\
\textbf{HD 74920}\tablefootmark{**}  &		265.29	&	-1.95	&			O7.5 IVn((f)) 	&		$-$	&	0.57	$\pm$	0.03	&	-12.35	$\pm$	0.03	&	-0.66	$\pm$	0.03	&	7.5	&	1753	& 49.6 & 44.5  &	N,R	& 3			\\
HD 75149\tablefootmark{*}	&	265.33	&	-1.69	&	B3 la	&	25.2		&	0.70	$\pm$	0.05	&	-6.57	$\pm$	0.06	&	-0.93		$\pm$	0.06	&	5.3	&	1465	& \textit{3.3} & \textit{15.2} &	Y	& 	/	1	\\		
CD-45 4606\tablefootmark{*}	&	265.44	&	-0.94	&	B0.5 V	&			6	&	0.46	$\pm$	0.01	&	-7.11 $\pm$	0.01	&	-0.75	$\pm$	0.01	&	8.9	&	2149	& \textit{29.7} & \textit{23.9} &	Y,R	& 1			\\	
HD 75276\tablefootmark{*}	&	265.61	&	-1.73	&	F2 lab	&		32		&	0.70	$\pm$	0.04	&	-6.61	$\pm$	0.04	&	-0.91		$\pm$	0.04	&	5.6	&	1443	& \textit{4.3} & \textit{16.8} &			Y? & $-$		\\
CD-45 4635\tablefootmark{*}	&	265.67	&	-0.90	&	B0.5 III		&	18		&	0.45	$\pm$	0.01	&	-7.09	$\pm$	0.02	&	-0.66	$\pm$	0.02	&	8.9	&	2235	& \textit{19.2} & \textit{17.6} &	Y,R & 1	/	1	\\	
HD 74711\tablefootmark{**}	&	265.73	&	-2.61	&			B1 III &		$-$	&	0.79	$\pm$	0.02	&	-10.38	$\pm$	0.02	&	-2.46		$\pm$	0.03	&	7.1	&	1272	& 23.1 & 4.7 &	N,R? & 9			\\	
HD 75309\tablefootmark{**}	&	265.90	&	-1.90	&			B1 IIp &		$-$	&	0.55	$\pm$	0.04	&	-6.96	$\pm$	0.05	&	-0.99		$\pm$	0.04	&	7.8	&	1931	& 5.0  & 3.5  &	Y& 9			\\
HD 78927\tablefootmark{**}	&	266.02	&	3.10	&			B1 II/III &			27	&	0.46 $\pm$	0.02	&	-6.88	$\pm$	0.02	&	0.36	 $\pm$	0.02	&	8.2	&	2169	& \textit{16.5}  & \textit{21.1} &	Y,R	& 4	/	2	\\	
CD-45 4676\tablefootmark{*}	&	266.18	&	-0.85	&	B0.5 III		&	28		&	0.54	$\pm$	0.01	&	-7.03 $\pm$	0.02	&	-0.89	$\pm$	0.02	&	8.8	&	1875	& \textit{6.3} & \textit{6.3} &		Y& $-$		\\	
HD 74401\tablefootmark{**}	&	266.23	&	-3.38	&			B1.5 IIIne	&		$-$	&	0.57	$\pm$	0.01	&	-7.34	
$\pm$	0.02	&	-1.16	$\pm$	0.02	&	8.9	&	1762	&  8.0 & 3.3 &	Y& 9			\\
CD-45 4719\tablefootmark{**}	&	266.46	&	-0.31	&			B0 &		$-$		&	0.47	$\pm$	0.01	&	-7.10 $\pm$	0.01	&	-0.64		$\pm$	0.02	&	9.6	&	2134	& 7.6 &  14.1 &	Y	& 5			\\
HD 78958\tablefootmark{**}	&	266.47	&	2.73	&			B0/1 Iab &			35	&	0.47	$\pm$	0.01	&	-6.03	$\pm$	0.01	&	-0.70	$\pm$	0.01	&	8.8	&	2133	& \textit{4.7}  & \textit{13.8} &	Y,R?,B? & 4	/	14	\\
HD 75658\tablefootmark{**}	&	266.87	&	-2.29	&			B1 IVe 	&		$-$	&	0.65	$\pm$	0.02	&	-8.12	$\pm$	0.02	&	-2.07		$\pm$	0.02	&	8.0	&	1536	&  13.4 & 3.6 &	N? & 9			\\
HD 75534\tablefootmark{**}	&	267.01	&	-2.56	&			B1 Ib &		$-$	&	0.53	$\pm$	0.02	&	-6.66	$\pm$	0.02	&	-1.51	$\pm$	0.02	&	7.7	&	1876	&  7.5 & 2.6 &	Y	& 9			\\	
CD-45 4826\tablefootmark{**} & 267.13 & 0.80 & O9 & $-$ & 0.47 $\pm$	0.01 &	-7.05 $\pm$	0.01 &	-0.72 $\pm$	0.01 &  9.9 & 2143 & 8.0 & 16.4 & Y & 2\\
HD 75991\tablefootmark{**}	&	267.17	&	-2.07	&			B0.5 III 	&		9	&	0.62	$\pm$	0.01	&	-7.31	$\pm$	0.02	&	-1.00 $\pm$	0.02	&	8.9	&	1618	& \textit{19.9} & \textit{14.9} &	Y,R	& 9	/	2	\\
CD-46 4786\tablefootmark{*}	&	267.35	&	-1.03	&	B0.5 lb	&		23		&	0.47	$\pm$	0.01	&	-6.79	$\pm$	0.02	&	-0.96	$\pm$	0.01	&	8.7	&	2132	& \textit{9.3} & \textit{9.1} &	RCW38 & 1,6			\\	
HD 79186\tablefootmark{**}	&	267.36	&	2.25	&			B5 Ia &			28	&	0.61	$\pm$	0.07	&	-7.05	$\pm$	0.07	&	-1.17	$\pm$	0.08	&	4.9	&	1811	& \textit{3.0} & \textit{8.7} &	Y	& 4	/	14	\\
CD-47 4490\tablefootmark{*}	&	267.40	&	-1.67	&	Bl III	&			$-$	&	0.54	$\pm$	0.01	&	-6.71 $\pm$	0.01	&	-0.66	$\pm$	0.02	&	9.5	&	1844	& 1.7 & 6.5 &	Y & 1,7			\\	
HD 76535\tablefootmark{**}	&	267.42	&	-1.52	&			O9 & 		25	&	0.47 $\pm$	0.02	&	-7.01	$\pm$	0.02	&	-0.98	$\pm$	0.02	&	8.5	&	2174	& \textit{9.1} & \textit{11.4} &	RCW38	& 4	/	20	\\		
HD 76556\tablefootmark{*}	&	267.58	&	-1.63	&			O6 IV(n)((f))p 	&		28.2	&	0.55	$\pm$	0.03	&	-6.91	$\pm$	0.03	&	-1.36	$\pm$	0.04	&	8.0	&	1820	& \textit{7.0} & \textit{5.7} &	Y	& 3	/	15	\\
CD-45 4820\tablefootmark{**}&  	267.59 & 0.31 &  	OB & $-$ & 0.50 $\pm$	0.01 	& 	-6.94 $\pm$	0.02 &	-0.33 $\pm$	0.02 &  11.1 & 2013 & 6.9 & 13.4 & Y & 2 \\
CD-47 4551\tablefootmark{**} & 267.98 & -1.36 &	O5 Ifc & 136 & 0.59 $\pm$	0.03 	&	-6.82 $\pm$	0.03 &	-1.12 $\pm$	0.03 	& 8.0 & 1697 & \textit{107.0} & \textit{114.1} & R & 2/25\\
HD 80057\tablefootmark{**}	&	267.99	&	2.87	&			A0 Iab &			25.7	&	0.56	$\pm$	0.02	&	-7.31 $\pm$	0.02	&	-0.07		$\pm$	0.03	&	5.9	&	1798	& \textit{12.6} & \textit{13.5}  &	Y	& 4	/	14	\\
CD-47 4550\tablefootmark{**}	&	268.00	&	-1.38	&			O7 V((f))z &		$-$	&	0.56	$\pm$	0.01	&	-6.74	$\pm$	0.01	&	-1.13		$\pm$	0.01	&	9.9	&	1789	& 2.6 & 4.0 &	RCW38	& 3			\\	
CD-47 4564\tablefootmark{**}	&	268.46	&	-1.65	&			B0.5 III &			-2.00	&	0.59	$\pm$	0.02	&	-6.39	$\pm$	0.02	&	-1.76	$\pm$	0.02	&	9.5	&	1718	& \textit{31.4} & \textit{25.3} &	R,B 	& 7	/	17	\\
HD 78344\tablefootmark{**}	&	268.89	&	-0.38	&			O9.5/B0 (Ib) 	&		-5.70	&	0.46	$\pm$	0.02	&	-5.98	$\pm$	0.02	&	-2.29	$\pm$	0.02	&	8.5	&	2169	& \textit{42.0} & \textit{29.1} &	R	& 4	/	2	\\
HD\,77207\tablefootmark{**}  &	268.96 & -1.90   & B7 Iab+	& $-$ & 0.56 $\pm$ 0.02  & -6.29 $\pm$ 0.02	&	-1.64 $\pm$ 0.02 & 9.2 & 1784 & 7.8 &  6.8 &  Y,R,B  & 4 \\
HD 77852\tablefootmark{**}	&	269.15	&	-1.20	&			B8/9 Iab/b 	&		$-$	&	0.57	$\pm$	0.02	&	-7.03	$\pm$	0.02	&	-0.64	$\pm$	0.02	&	8.3	&	1747	& 3.1 & 7.8 &	Y	& 4			\\
CPD-47\,3051\tablefootmark{**} &	 269.21 & -0.91  & B0 V	& 	$-$		& 0.49 $\pm$ 0.02 & -6.86 $\pm$ 0.02	&	-0.37 $\pm$ 0.02	& 10.8 & 2069 & 4.1	& 12.2 & Y,R,B 	&  26 \\
\textbf{HD 76968}\tablefootmark{**} &	270.22	&	-3.37	&			O9.2 Ib 	&		-22	&	0.46	$\pm$	0.02	&	-1.62	$\pm$	0.02	&	-1.25	$\pm$	0.02	&	7.0	&	2199	& \textit{74.3} & \textit{60.9} &	N,R 	& 3	/	16	\\
HD 82830\tablefootmark{**}	&	271.43	&	3.62	&			B0/2 Ie &		$-$		&	0.51	$\pm$	0.01	&	-6.87	$\pm$	0.02	&	-1.75		$\pm$	0.01	&	9.1	&	1968	& 6.1 & 10.5 &	Y	& 4			\\
CD-48 4654\tablefootmark{**}	&	271.90	&	0.70	&			B2 Ib &		$-$	&	0.47	$\pm$	0.01	&	-7.27	$\pm$	0.02	&	-0.80 $\pm$	0.01	&	9.7	&	2325	& 4.6	& 17.2 & Y	&  7			\\
HD 297433\tablefootmark{**}	&	272.07	&	0.44	&			O8/9 &		$-$	&	0.43	$\pm$	0.01	&	-6.52 $\pm$	0.01	&	-0.27	$\pm$	0.01	&	8.9	&	2323	&   5.9 & 17.9 &	Y	& 5			\\	
HD 298298\tablefootmark{**}	&	272.73	&	-3.31	&			B0	 &  $-$ &				0.64	$\pm$	0.01	&	-6.35	$\pm$	0.01	&	-1.04	$\pm$	0.02	&	9.0	&	1559	&  6.9 & 14.0 &	Y& 13			\\
HD 80558\tablefootmark{**}	&	273.07	&	-1.47	&			B6 Ia 	&		21.7	&	0.54	$\pm$	0.03	&	-7.03	$\pm$	0.03	&	-1.19	$\pm$	0.03	&	5.7	&	1862	& \textit{4.7} & \textit{5.2} &	Y	& 4	/	18	\\
HD 84136\tablefootmark{**}	&	273.72	&	3.19	&			B1/2 II &	$-$ &			0.38	$\pm$	0.01	&	-5.68	$\pm$	0.01	&	-0.40	$\pm$	0.01	&	8.7	&	2634	&  12.6 & 17.3 &	Y	& 4			\\	
HD 81471\tablefootmark{**}	&	273.78	&	-1.01	&			A5 Ia/ab &		28	&	0.32	$\pm$	0.02	&	-6.41	$\pm$	0.03	&	-0.80	$\pm$	0.03	&	5.9	&	3060	& \textit{6.5} & \textit{35.4} &	N	& 4	/	17	\\	
HD 298425\tablefootmark{**}	&	274.34	&	0.04	&			O9 V &			10	&	0.56	$\pm$	0.01	&	-6.91	$\pm$	0.01	&	-0.84	$\pm$	0.01	&	9.6	&	1782	& \textit{13.4} & \textit{7.9} &	Y & 11	/ 17	\\	
HD 85356\tablefootmark{**}	& 274.37 & 4.60 &  B2Ib & $-$ & 0.57 $\pm$	0.03 	&	-4.05 $\pm$	0.04 &	-0.29 $\pm$	0.03 & 8.0 & 1754 & 27.4 & 27.7 & N?,R & 2 \\
HD 81370\tablefootmark{**}	&	274.41	&	-1.81	&			B1 (III)n &		13	&	0.55	$\pm$	0.02	&	-6.80	$\pm$	0.02	&	-1.17	$\pm$	0.02	&	8.7	&	1808	& \textit{11.0} & \textit{4.3} &	Y& 4	/	17	\\	
\textbf{HD 298429\tablefootmark{**}}	&	274.47	&	-0.25	&			O8.5 V &			25	&	0.55	$\pm$	0.01	&	-5.37	$\pm$	0.01	&	0.24	$\pm$	0.01	&	9.5	&	1831	& \textit{20.1} & \textit{23.4} &	Y,R& 3	/	16	\\
HD 298420\tablefootmark{**}	&	274.58	&	-1.81	&			B1/2 &		$-$	&	0.41	$\pm$	0.02	&	-6.37	$\pm$	0.02	&	-0.62	$\pm$	0.02	&	9.8	&	2439	&  4.9 & 19.1 &	Y	& 5			\\	
HD 298448\tablefootmark{**}	&	274.90	&	-1.65	&			B1/3 &		$-$	&	0.36	$\pm$	0.01	&	-6.27	$\pm$	0.01	&	-0.38	$\pm$	0.01	&	9.5	&	2733	& 4.5 & 27.6 &	Y?	& 5			\\	
\hline
\end{tabular}
}
\tablebib{
1. \citet{humphreys78}; 2. \citet{reed00}; 3. \citet{sota14}; 4. \citet{houk78}; 5. \citet{reed03}; 6. \citet{macconnell76}; 7. \citet{bassino82}; 8. \citet{corti07}; 9. \citet{garrison77}; 10. \citet{vijapurkar93}; 11. \citet{feast61}; 12. \citet{fitzgerald79}; 13. \citet{nesterov95}; 14. \citet{gontcharov06}; 15. \citet{williams11}; 16. \citet{holgado18}; 17. \citet{evans67}; 18. \citet{kharchenko07}; 19. \citet{pourbaix04}; 20. \citet{crampton72}; 21. \citet{wilson53}; 22.\citet{apellaiz18}; 23. \citet{parthasarathy12}; 24. \citet{borisov23}; 25. \citet{Tarricq21}; 26. \citet{kobulnicky19}.
}
\tablefoot{
(1) Stellar ID, (2),(3) Galactic coordinates, (4) Spectral type, (5) Radial velocity, (6)-(9) \textit{Gaia}\,DR3 corrected parallax, proper motion in Galactic coordinates and G magnitude, respectively, (10) Distance \citep{bailer21}, (11) Peculiar space (\textit{italic}) or transverse velocity relative to LSR (see Section \ref{nBS}), (12) Peculiar (\textit{italic}) or transverse velocity relative to Vel\,OB1 (see Section \ref{nBS}), (13)  Confirmed member of Vel\,OB1 (or of young cluster), or runaway R, with bow shock B, (14) Reference for spectral type/radial velocity. \tablefoottext{*}{Object from \citet{humphreys78}}, \tablefoottext{**}{Object from \citet{reed00}}. Objects in common with the list of OB runaways in \citet{Carretero23} are in bold.}
\end{table*}

\begin{table*}
\caption[]{Astrometric and kinematic properties of the OB runaway stars in the area of Vel~OB1.}
\resizebox{1\textwidth}{!}{
\label{tab:parrun}
\begin{tabular}{l *{15}{c}}
\hline\hline\noalign{\smallskip}
\centering
Name	&	l & b &	Sp. Type 	&	Parallax	&	D & $\mu_{l^{*}}$ &	$\mu_{b}$	 & V$_r$ & \multicolumn{3}{c}{V$_{\rm pec}$}	 & Z & Bow shock& Aligned & Angle \\
&&&&&&&&& 2D (LSR) & 3D (LSR) &  (Vel OB1) &&&\\
	$-$ &	(deg) & (deg)	&	$-$ & (mas) & (pc) &(mas\,yr$^{-1}$)&		(mas\,yr$^{-1}$)	&	(km\,s$^{-1}$) &	(km\,s$^{-1}$) & (km\,s$^{-1}$)		& (km\,s$^{-1}$)		 & (pc) & $-$ & $-$ &(deg)\\
\hline\noalign{\smallskip}
(1)     &       (2)     &       (3)     &       (4) 	&	 (5) 	& (6)	&	(7)		& (8)	&	(9) & (10) &(11) & (12) & (13) & (14) & (15)& (16) \\
\hline \noalign{\smallskip}
\hline \noalign{\smallskip}
\multicolumn{15}{c}{\textbf{OB runaways with a high peculiar velocity ($>15$~km~s$^{-1}$) and/or a bow shock}}\\
\hline \noalign{\smallskip}
HD\,71649 &     255.76 & 1.00 & B1 IIIn       	&	0.442$\pm$0.016
& 	2256$^{+80}_{-73}$ &	-8.473$\pm$0.015& 	0.045$\pm$0.015& $-$ & 40 & $-$ & 37 & 39 & ISM?  & $-$&$-$\\
\textbf{HD\,75222} &     258.29 & 4.18 & B0 Ia     	&	0.494$\pm$0.017& 	2017$^{+84}_{-76}$ &	-11.511$\pm$0.015 & 	3.096$\pm$0.016 & 64 & 72 & 75 & \textit{79} & 147 & N & $-$ &$-$\\
CD-41\,4637$^{(P)}$	&	263.02 & 2.30	&			O6Ib(f)(n)	& 0.439$\pm$0.013	 & 2280$^{+76}_{-62}$  & 	-8.456$\pm$0.013 	&	0.725$\pm$0.013	&	$-$&  33 & $-$& 38 & 92 & Y	& Y & 30\\
\textbf{HD\,77581 (Vela\,X-1)}$^{(P)}$	&	263.06 & 3.93 &			B0.5Ia	& 0.510$\pm$0.015	 & 1960$^{+57}_{-53}$ & -10.152$\pm$0.015 	&	2.509$\pm$0.016 &	-3.2	&	51 & 65		& \textit{60}		& 134 & Y & Y & 9\\
\textbf{HD\,74194} &        264.04 & -1.95 & O8.5 lb     	&	0.455$\pm$0.017	& 	2220$^{+125}_{-73}$ &	-9.395$\pm$0.020 & -2.177$\pm$0.019	& -2.7 & 38 & 54 & \textit{51}& -76  & N  & $-$&$-$\\
HD\,75860$^{(P)}$	& 264.14 & 0.27&	B1.5Iabp	& 0.450$\pm$0.042	 & 2276$^{+227}_{-201}$ & -7.944$\pm$0.043	&	-2.111$\pm$0.048	&	16.8	&	25 & 31 &	\textit{29} 	 & 11 & Y & Y? & 47\\
HD 76031$^{(P)}$ &	264.50	&	0.30	&			B0.5 IV &	0.568$\pm$0.025	&	1752$^{+64}_{-63}$	&	-7.024$\pm$0.028	&	-2.147$\pm$0.027	&	$-$	&	11 & $-$ & 6 & 8 & Y& Y		& 14\\
HD 73658 &	264.68	&	-3.13	&			B1 II &		0.567$\pm$0.022	&	1783$^{+86}_{-66}$	& -7.276$\pm$0.023	&	-1.427$\pm$0.023	&	52	&	10  & 23 & \textit{26} &	-96 & N	& $-$	&$-$\\	
CD-44 4865	&	264.69	&	-0.37	&	O9.5 lb	&	0.514$\pm$0.026	&	1937$^{+82}_{-79}$	& -7.812$\pm$0.027	&	-0.334$\pm$0.028	&	39		& 14  & 15 & \textit{22} &	-13 & N	& $-$		&$-$	\\	
CD-45\,4394 &         264.87 & -2.58 & B2 Vne    	&		0.547$\pm$0.011& 	1826$^{+37}_{-34}$ &	-6.910$\pm$0.012 & 	-1.093$\pm$0.013 & 78 & 6 & 46 & \textit{53} & -82 &N & $-$  &$-$\\
HD 74180 & 265.28 & -2.95 & F2 Ia & 0.440$\pm$0.101  & 2532$^{+600}_{-397}$	 & -7.277$\pm$0.107 	& -2.118$\pm$0.122 & 31 & 23  & 22 & \textit{26} & -117 & N & $-$&$-$\\
\textbf{HD\,74920} &        265.29 & -1.95 & O7.5IVn((f))    	&		0.574$\pm$0.026& 	1753$^{+63}_{-68}$ &	-12.351$\pm$0.029 & -0.661$\pm$0.032	&$-$ & 50& $-$ & 45 & -60 & N & $-$ &$-$\\
CD-45 4606	&	265.44	&	-0.94	&	B0.5 V	&	0.463$\pm$0.012	&	2149$^{+57}_{-60}$	& -7.109$\pm$0.012	&	-0.752$\pm$0.014	&	6	& 9 & 30 & \textit{24} &	-35 & N	& $-$			&$-$\\
CD-45 4635	&	265.67	&	-0.90	&	B0.5 III		&	0.449$\pm$0.014	&	2235$^{+65}_{-67}$	& -7.090$\pm$0.015	&	-0.657$\pm$0.016	&	18	& 9 & 19 & \textit{18} &	-35 & N& $-$	&$-$\\	
CD-44\,4967 &	266.01 & 0.00   & OB+	& 0.603$\pm$0.010 & 1663$^{+30}_{-25}$ & -8.819$\pm$0.012	&	-2.031$\pm$0.013	& $-$	& 20 &	$-$&	12	 & 0.1 & Y & Y & 11\\
HD 78927	&	266.02	&	3.10	&			B1 II/III &	0.464$\pm$0.020 &	2169$^{+76}_{-87}$	& -6.879$\pm$0.020	&	0.365$\pm$0.022	&	27	&	16 & 17  & \textit{21} & 117 &	ISM? & $-$	&$-$\\	
HD 75991 &	267.17	&	-2.07	&			B0.5 III 	&	0.619$\pm$0.015	&	1618$^{+39}_{-39}$	& -7.313$\pm$0.016	&	-1.002$\pm$0.016	&	9	& 5 & 20 & \textit{15} &	-58 & N & $-$&$-$\\
CD-47 4551 & 267.98 & -1.36 &	O5 Ifc &  0.588$\pm$0.028 	& 1697$^{+95}_{-75}$ &	-6.816$\pm$0.032 &		-1.115$\pm$0.033 	& 136 & 2 & 107 &  \textit{114} & -40 & ISM & N & 177\\
CD-47\,4564	&	268.46	&	-1.65	&			B0.5 III &			0.574$\pm$0.017	&	1718$^{+44}_{-52}$	 & -6.393$\pm$0.019	&	-1.761$\pm$0.020	&	-2	&	8 & 31 &	\textit{25} &	-49 & Y & Y? & 56 \\
HD 78344&	268.89	&	-0.38	&			O9.5/B0 (Ib) 	&			0.452$\pm$0.016	&	2169$^{+82}_{-83}$	 & -5.982$\pm$0.017 &	-2.286$\pm$0.016	&	-5.7	&		17 & 42 &		\textit{29} &	-14	& N & $-$ &$-$\\
HD\,77207  &	268.96 & -1.90   & B7Iab+	& 0.564$\pm$0.017  & 1784$^{+40}_{-50}$ &	 -6.292$\pm$0.018	&	-1.638$\pm$0.017 &		$-$		& 8 & $-$ & 7 & -59 & Y & Y? & 68\\
CPD-47\,3051 &	 269.21 & -0.91  & B0V	& 0.487$\pm$0.019 & 2069$^{+70}_{-77}$ &	-6.856$\pm$0.021	&	-0.372$\pm$0.021	 &	$-$			& 4 & $-$ & 12	& -33 & Y & Y & 9\\
\textbf{HD\,76968} &        270.22  & -3.37 & O9.2Ib   	&		0.455$\pm$0.019& 	2199$^{+97}_{-83}$ &	-1.619$\pm$0.022& -1.247$\pm$0.021	& -22 & 53 & 74 & \textit{61} & -129 & N & $-$ &$-$\\
HD\,298310	&	272.58 & -1.72 &	B0V	& 0.525$\pm$0.011 & 1900$^{+37}_{-35}$ &	-0.468$\pm$0.012	&	-1.088$\pm$0.013	&	-27		&		59 & 80&		\textit{72} 	 & -57 & Y	& N & 126\\
HD 85356	& 274.37 & 4.60 &  B2Ib &  0.571$\pm$0.034 	&	1754$^{+103}_{-96}$ & -4.049$\pm$0.036 &	-0.288$\pm$0.033 & $-$ &  27 & $-$ & 28 & 141 & N & $-$&$-$ \\
\textbf{HD 298429}	&	274.47	&	-0.25	&	O8.5 V &	0.548$\pm$0.011	&	1831$^{+46}_{-44}$	& -5.369$\pm$0.013	&	0.245$\pm$0.013	&	25	& 18 &20 & \textit{23} &	-8 & N & $-$	&$-$\\
\hline \noalign{\smallskip}
\multicolumn{15}{c}{\textbf{Candidate OB runaways in the field centered on Vel~OB1}}\\
\hline \noalign{\smallskip}
IRAS\,08351-3951 &	259.62 & 0.55  &	M8	& 0.272$\pm$0.050	 & 3975$^{+916}_{-599}$  &	3.736$\pm$0.050	&	7.659$\pm$0.055	& 	35		& 71 &	74 &	\textit{94}	 & 38  & ISM & N & 90\\
CD-42\,4694 &	263.61 & 0.55 &	O9V	& 0.409$\pm$0.027 & 2466$^{+148}_{-137}$ &	-6.966$\pm$0.027	&	-1.413$\pm$0.028	 &		$-$		&	13& $-$&	22 & 24 & Y? &	N & 145\\
\textbf{HD\,75211} &  263.96       & -0.47 &  O9 Ib  & 0.633$\pm$0.015 	&	1550$^{+31}_{-36}$& 	-10.013$\pm$0.016 & -1.789$\pm$0.015	& 20	& 27 & 30 & \textit{17}& -13 & N & $-$ &$-$\\
CD-45 4447 &	265.18	&	-2.26	&			O7.5 V	&	1.528$\pm$0.394	&	888$^{+337}_{-221}$	& -5.848$\pm$0.449	&	0.066$\pm$0.451	&	$-$	& 11 & $-$ & 42 &	-26 & Y? & Y?  & 67\\
HD 74711	&	265.73	&	-2.61	&			B1 III &	0.785$\pm$0.022 &	1272$^{+37}_{-35}$	&  -10.381$\pm$0.023	&	-2.458$\pm$0.025	&	$-$	& 23 & $-$& 5 &	-58 & N & $-$			&$-$\\	
GSC\,08152-02059 &	265.88 & -0.39   & OB	& 0.668$\pm$0.018 & 1507$^{+43}_{-40}$  & -8.453$\pm$0.019	&	-1.354$\pm$0.022	& $-$	 & 14 &	$-$ &	3 & -10 & Y? & Y? & 65\\
HD 78958	&	266.47	&	2.73	&			B0/1 Iab &			0.469$\pm$0.012 &	2133$^{+54}_{-39}$	& -6.030$\pm$0.011	&	-0.700$\pm$0.012	&	35	&	4 & 5  & \textit{14} & 101 &	Y? & N & 132\\
DR3\,N72 &	269.01 &-0.76  &  $-$	& 0.336$\pm$0.035 & 2927$^{+605}_{-242}$ & -5.457$\pm$0.040	&	0.090$\pm$0.040	 &	42 & 11 & 12 & \textit{30} & -39 & Y? & N & 112\\
DR3\,N60 &	270.91 & -0.73  & $-$	&  0.419$\pm$0.039 & 2410$^{+245}_{-155}$ & -6.329$\pm$0.046	&	-1.430$\pm$0.042 &			$-$	& 10 & $-$ &  15 & -31 & Y & N & 119\\
HD\,298353 &	273.12 & -1.96  & O7V	& 0.377$\pm$0.012 & 2680$^{+89}_{-82}$  &	-6.163$\pm$0.015	&	-1.110$\pm$0.015    &		$-$		& 9 & $-$ & 21	 & -92 & Y	& N & 93\\
\hline
\end{tabular}
}
\tablefoot{
(1) name of star, (2), (3) Galactic coordinates of star, (4) spectral type, (5) \textit{Gaia}\,DR3 corrected parallax and parallax uncertainty, (6) distance estimated by \citet{bailer21} with uncertainty, (7) (8) \textit{Gaia}\,DR3 galactic proper motion and proper motion uncertainty, (9) radial velocity, (10) and (11) transverse and peculiar velocity relative to LSR, respectively, (12) transverse or peculiar (\textit{italic})) velocity relative to Vel\,OB1, (13) distance from the Galactic plane, (14) observed bow shock, (15) alignment status of observed bow shock, and (16) mis-alignment angle. Two objects are indicated as DR3\,N72 and N60 based on their last numbers of \texttt{source\_id} from \textit{Gaia} DR3. Objects in common with \citet{Carretero23} are in bold. (P) - objects included in \citet{Patten25}.}
\end{table*}

\begin{table*}
\centering
\caption[]{Estimate of the kinematical age of the OB runaways in Vel~OB1.}
\resizebox{0.85\textwidth}{!}{
\label{tab:parrunori}
\begin{tabular}{l *{11}{c}}
\hline\hline\noalign{\smallskip}
\centering
Name	&	l & b &	 D & V$_r$ & V$_{pecLSR}$ & Runaway direction & Age$_{\rm kinGP}$ & Parent & D$_{\rm par}$ & Age$_{\rm kinpar}$  &V$_{\rm pecpar}$ \\
	$-$ &	(deg) & (deg)	&	(pc) &  (km\,s$^{-1}$)		&	(km\,s$^{-1}$)		&	$-$ & (Myr) & $-$ &  (pc) & (Myr) &  (km\,s$^{-1}$)		\\
\hline\noalign{\smallskip}
(1)     &       (2)     &       (3)     &       (4) 	&	 (5) 	& (6) & (7) & (8) & (9) & (10)  & (11) & (12)\\
\hline \noalign{\smallskip}
\textbf{HD\,75222} &     258.29 & 4.18 & 	2017$^{+84}_{-76}$  & 64 & 72/\textbf{75} &  A  & 4.9; 3.7; 2.5 &  Bochum\,7 & 2245$^*$ & 5.5 & \textbf{67} \\
CD-41\,4637	&	263.02 & 2.30	& 2280$^{+76}_{-62}$  & 	$-$ &  33 & A? & 11; 6.5; 1.5 & 	RCW\,34 & 1967$^{+18.4}_{-18.0}$ & 2 & 35 \\
GSC\,08152-02059 &	265.88 & -0.39 & 1507$^{+43}_{-40}$  & $-$	 & 14 &	A & 1.0; $-$; $-$ & RCW\,38 &  1785$^{+4.2}_{-4.1}$	& 3-4 &		4	\\
\textbf{HD\,76968} &        270.22  & -3.37 & 	2199$^{+97}_{-83}$ & -22 & 53/\textbf{74} & A? & 9.7; 6.8; 4.0 &  Bochum\,7 & 2245$^*$ &4 & \textbf{75} \\
HD\,298310	&	272.58 & -1.72 & 1900$^{+37}_{-35}$ & -27 &	59/\textbf{80} & A & 5.7; 2.4; $-$ & RCW\,38 &  1785$^{+4.2}_{-4.1}$ & 3.3 &	52 	\\

\hline
HD\,71649 &     255.76 & 1.00      	&	2256$^{+80}_{-73}$ & $-$ &  40 & P & $-$ & $-$ & $-$ & $-$ & $-$ \\
\textbf{HD\,77581 (Vela\,X-1)}	&	263.06 & 3.93 &	 1960$^{+57}_{-53}$  &	-3.2	&	51/\textbf{65}	 &	A & 5.6; 4.2; 2.8 &		Vel\,OB1/RCW\,38? &  1785$^{+4.2}_{-4.1}$	 & 2.3/4-5 &	\textbf{60}/53 \\
\textbf{HD\,74194} &        264.04 & -1.95 &  	2220$^{+125}_{-73}$ & -2.7 & 38/\textbf{54} & A & 3.2; 1.6; $-$ & [FSR2007]1452 & 2549$^*$ & 4 & \textbf{35}\\
HD\,75860	& 264.14 & 0.27& 2276$^{+227}_{-201}$ &	16.8	&	25/\textbf{31} & T & $-$	&	Cl\,1594? & 2084$^{+43.0}_{-41.3}$ & 1.2 & 24 	\\
HD 76031 &	264.50	& 0.30 &	1752$^{+64}_{-63}$	& $-$	&	11 & T & $-$ & $-$ & $-$ & $-$ & $-$ \\
HD 73658 &	264.68	&	-3.13	& 1783$^{+86}_{-66}$	& 52	&	10/\textbf{23} & A? & 7.9; 5.4; 2.9 & $-$ & $-$ & $-$ & $-$ \\	
CD-44 4865	&	264.69	& -0.37	&1937$^{+82}_{-79}$	&39	& 14/\textbf{15} & P &  $-$	& $-$ & $-$ & $-$ & $-$ \\	
CD-45\,4394 & 264.87 & -2.58 & 	1826$^{+37}_{-34}$  & 78 & 6/\textbf{46} & A? & 8.5; 5.2; 1.9 & $-$ & $-$ & $-$ & $-$ \\
HD 74180 & 265.28 & -2.95 & 2532$^{+600}_{-397}$ & 31 & 23/\textbf{22} & A & 5.0; 3.3; 1.6 & $-$ & $-$ & $-$ & $-$ \\
\textbf{HD\,74920} &        265.29 & -1.95 &	1753$^{+63}_{-68}$ &$-$ & 50 & P& $-$	 & Cluster? & 2020$^{+13}_{-12}$	 & 0.2-0.3 & 38  \\
CD-45 4606	&	265.44	&	-0.94	&	2149$^{+57}_{-60}$	& 6	& 9/\textbf{30} & P? & $-$	 & $-$ & $-$ & $-$ & $-$ \\
CD-45 4635	&	265.67	&	-0.90 &	2235$^{+65}_{-67}$	& 18 & 9/\textbf{19} & P? & $-$	 & $-$ & $-$ & $-$ & $-$ \\	
CD-44\,4967 &	266.01 & 0.00   & 1663$^{+30}_{-25}$ & $-$	& 20 &	T & $-$	 &	 RCW\,38? &  1785$^{+4.2}_{-4.1}$	&3-4 & 16\\
HD 78927 &	266.02	&	3.10 &	2169$^{+76}_{-87}$	& 27	&	16/\textbf{17}  & P & $-$	 & $-$ & $-$ & $-$ & $-$ \\	
HD 75991 &	267.17	&	-2.07 &	1618$^{+39}_{-39}$	& 9	& 5/\textbf{20} & A? & 7.4; 3.8; 0.3 & $-$ & $-$ & $-$ & $-$ \\
CD-47 4551 & 267.98 & -1.36 & 1697$^{+95}_{-75}$ & 136 & 2/\textbf{107} & A & 4.4; 1.2; $-$ & $-$ & $-$ & $-$ & $-$ \\
CD-47\,4564	&	268.46	&	-1.65	&	1718$^{+44}_{-52}$	 & -2	&	8/\textbf{31} &	A & 3.4; 1.3; $-$ & $-$ & $-$ & $-$ & $-$ \\
HD 78344 &	268.89	&	-0.38 &	2169$^{+82}_{-83}$	&	-5.7 & 	17/\textbf{42} & A & 0.6; $-$; $-$ & $-$ & $-$ & $-$ & $-$ \\
HD\,77207  & 268.96 & -1.90  & 1784$^{+40}_{-50}$ & $-$ & 8 & A & 4.2; 2.0; $-$ &  $-$ &  $-$ & $-$ &  $-$ \\
CPD-47\,3051 &	 269.21 & -0.91  & 2069$^{+70}_{-77}$ & $-$ & 4 & P & $-$	 & $-$ &  $-$ & $-$ & $-$	\\
HD 85356	& 274.37 & 4.60 & 1754$^{+103}_{-96}$ & $-$ &  27 &  P & $-$	 & $-$ & $-$ & $-$ & $-$ \\
\textbf{HD 298429}	&	274.47	&	-0.25 &	1831$^{+46}_{-44}$	& 25	& 18/\textbf{20} & P & $-$	 & $-$ & $-$ & $-$ & $-$ \\

\hline \noalign{\smallskip}
IRAS\,08351-3951 &	259.62 & 0.55  & 3975$^{+916}_{-599}$  & 35 & 71/\textbf{74} & T & $-$	 & IRAS\,08337-4028? & 1603$^{+31.7}_{-30.5}$	& 1 &	109 \\
CD-42\,4694 &	263.61 & 0.55 &	2466$^{+148}_{-137}$  &	$-$	&	13 & T & $-$	 &	PN\,K\,2-15? & 1787$^{+23.0}_{-22.4}$ & 2-3 & 25	\\
\textbf{HD\,75211} &  263.96       & -0.47 &	1550$^{+31}_{-36}$& 20	& 27/\textbf{30} & A & 0.9; $-$; $-$ & Muzzio\,1 & 1797$^*$ &4& 122 \\
CD-45 4447 &	265.18	&	-2.26 &	888$^{+337}_{-221}$	& $-$	& 11 & P & $-$	  & $-$ & $-$ & $-$ & $-$  \\
HD 74711	&	265.73	&	-2.61 &	1272$^{+37}_{-35}$	&  $-$	& 23 & A & 3.8; 2.4; 0.9 & $-$ & $-$ & $-$ & $-$ \\	
HD 78958 &	266.47	&	2.73 &	2133$^{+54}_{-39}$	& 35	&	4/\textbf{5}  & P & $-$	 & $-$ & $-$ & $-$ & $-$ \\
DR3\,N72 &	269.01 & -0.76  & 2927$^{+605}_{-242}$ & 42 & 11/\textbf{12} & P & $-$	 & $-$ &  $-$ & $-$ & $-$ \\
DR3\,N60 &	270.91 & -0.73  & 2410$^{+245}_{-155}$ & $-$	& 10 & A & 1.8; $-$; $-$ & $-$ &  $-$ & $-$ &  $-$\\
HD\,298353 &	273.12 & -1.96  & 2680$^{+89}_{-82}$  & $-$ & 9 & A? & 6.4; 3.1; $-$ &   $-$ &  $-$  &$-$ &  $-$ \\
\hline
\end{tabular}
}
\tablefoot{
(1) Name of the OB runaway; (2) and (3) Galactic coordinates; (4) Distance with uncertainty (\citet{bailer21}); (5) Radial velocity; (6) Transverse/ peculiar velocity (thus including radial velocity in \textit{bold})) relative to the LSR; (7) Direction of the runaway: A $-$ away from the Galactic plane, P $-$ parallel to the Galactic plane, T $-$ toward the Galactic plane; (8) Kinematic age assuming that the OB runaway originates at b=0$^{\circ}$, b=1$^{\circ}$, or b=2$^{\circ}$, respectively; (9) Parent cluster; (10) Distance of the parent cluster, (11) Kinematic age if ejected from the proposed parent cluster; and (12) Peculiar transverse velocity of the star relative to its origin, and when the radial velocity is included (\textit{bold}). The [KPS2012]\,MWSC\,1594 open cluster is indicated as Cl\,1594. The ID of DR3\,N72 and N60 is based on their last numbers of \texttt{source\_id} from \textit{Gaia} DR3. Objects in common with \citet{Carretero23} are in bold. (*) The distances of these clusters are taken from \citet{Cantat20,Poggio21,Tarricq21}.}
\end{table*}

\section{Membership, distance, and age of young stellar clusters in Vel~OB1 based on {\it Gaia} DR3}
\label{appB}

\begin{figure*}
\centering
\includegraphics[width=0.77\linewidth]{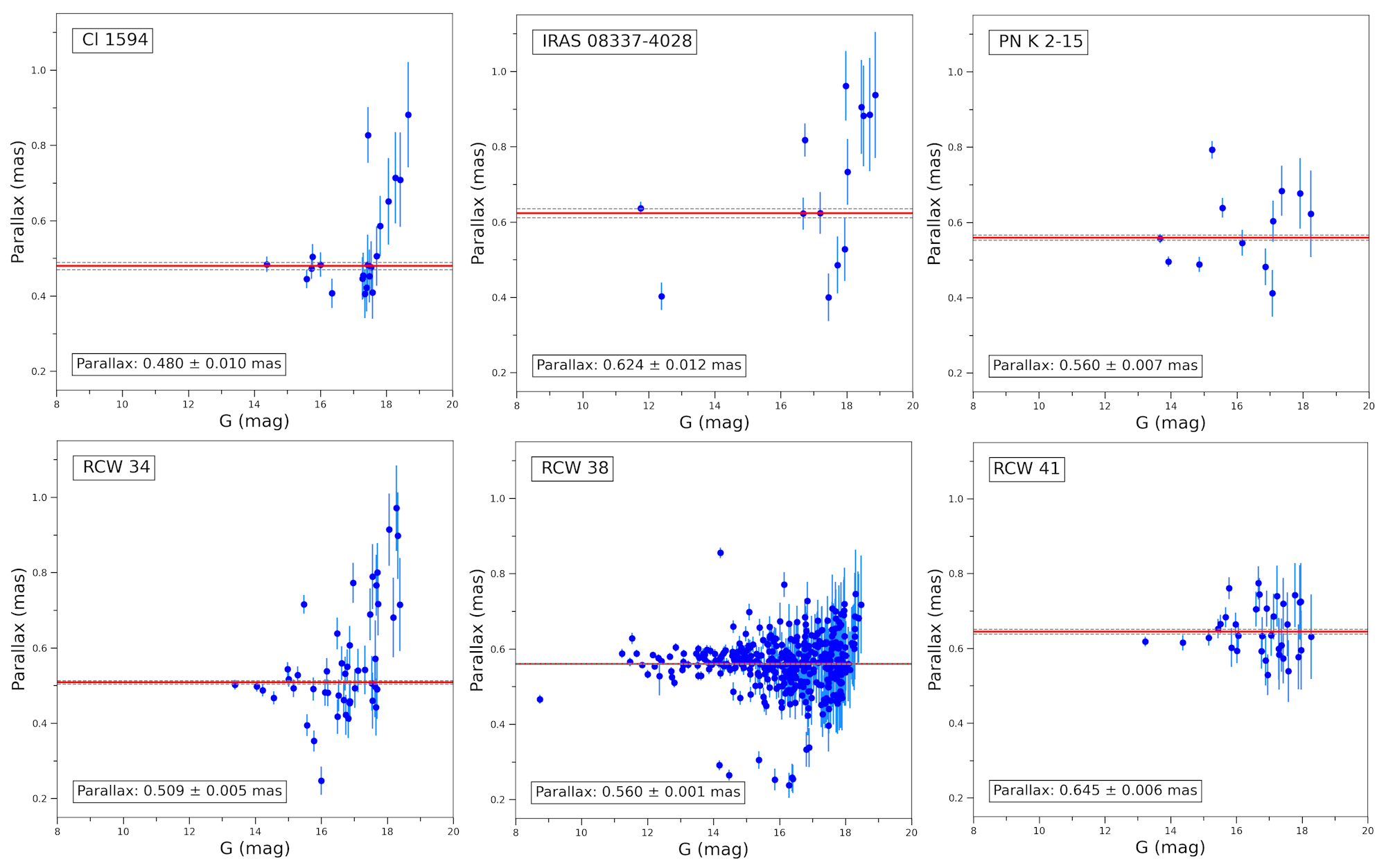}
\caption{Parallax distribution as a function of the G-magnitude for each stellar cluster. The best-fit parallax and 1$\sigma$ uncertainty for the cluster members are shown with the red and grey lines, respectively. The [KPS2012]\,MWSC\,1594 open cluster is indicated as Cl\,1594.}
\label{fig:plxG_sep}
\end{figure*}

In this section we provide the figures used for the analysis of the membership, distance and age of the six young stellar clusters studied in Vel~OB1. In Fig.~\ref{fig:plxG_sep}, we show the parallax distribution of the members of each cluster against their G-magnitude. The best-fit distance corresponds to the maximum of the log-likelihood function. The $16^{\rm th}$ and $84^{\rm th}$ percentiles of the distance distribution derived from the likelihood function provide the 1$\sigma$-uncertainty range. The best-fit parallax and the 1$\sigma$ uncertainties on the parallax range of each cluster are shown with the red solid and grey dashed lines in each panel of Fig.~\ref{fig:plxG_sep}, respectively.

\begin{figure*}
\centering
\includegraphics[width=0.73\linewidth]{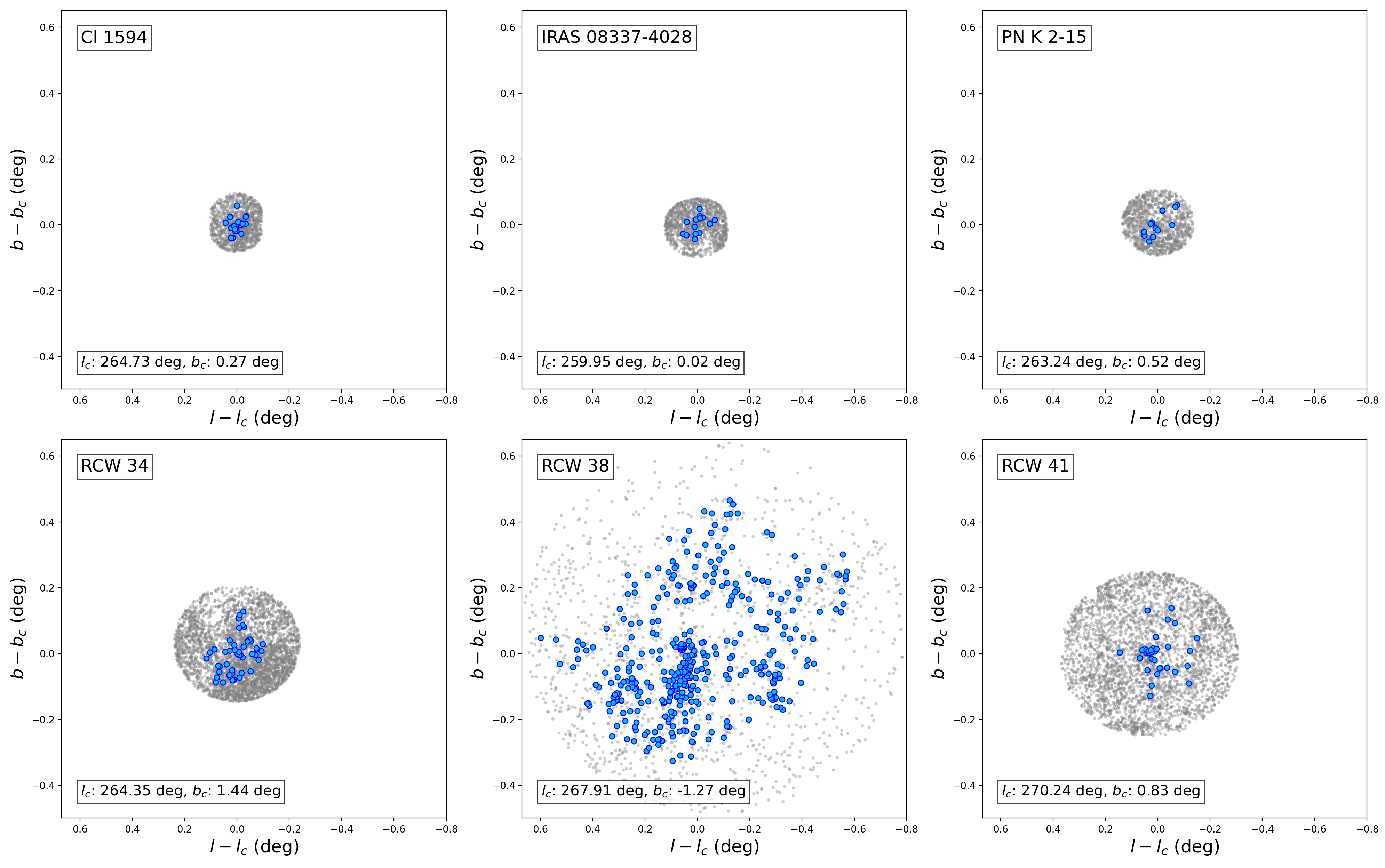}
\caption{Distribution of members for each stellar cluster with field stars. Members and field stars are in blue and grey, respectively. The median value of the cluster coordinates is taken as the centre of each panel. Visual inspection of the image of RCW\,38 shows that the extinction is low where the members show high density which we called "condensation". The [KPS2012]\,MWSC\,1594 open cluster is indicated as Cl\,1594.}
\label{fig:lb_sep}
\end{figure*}

The distribution on the sky of the members of each young cluster is shown in Fig.~\ref{fig:lb_sep} in Galactic coordinates. Members and field stars are in blue and grey, respectively. The median value of the coordinates of the clusters are taken as the centre of each panel.

\begin{figure*}
\centering
\includegraphics[width=0.3\linewidth]{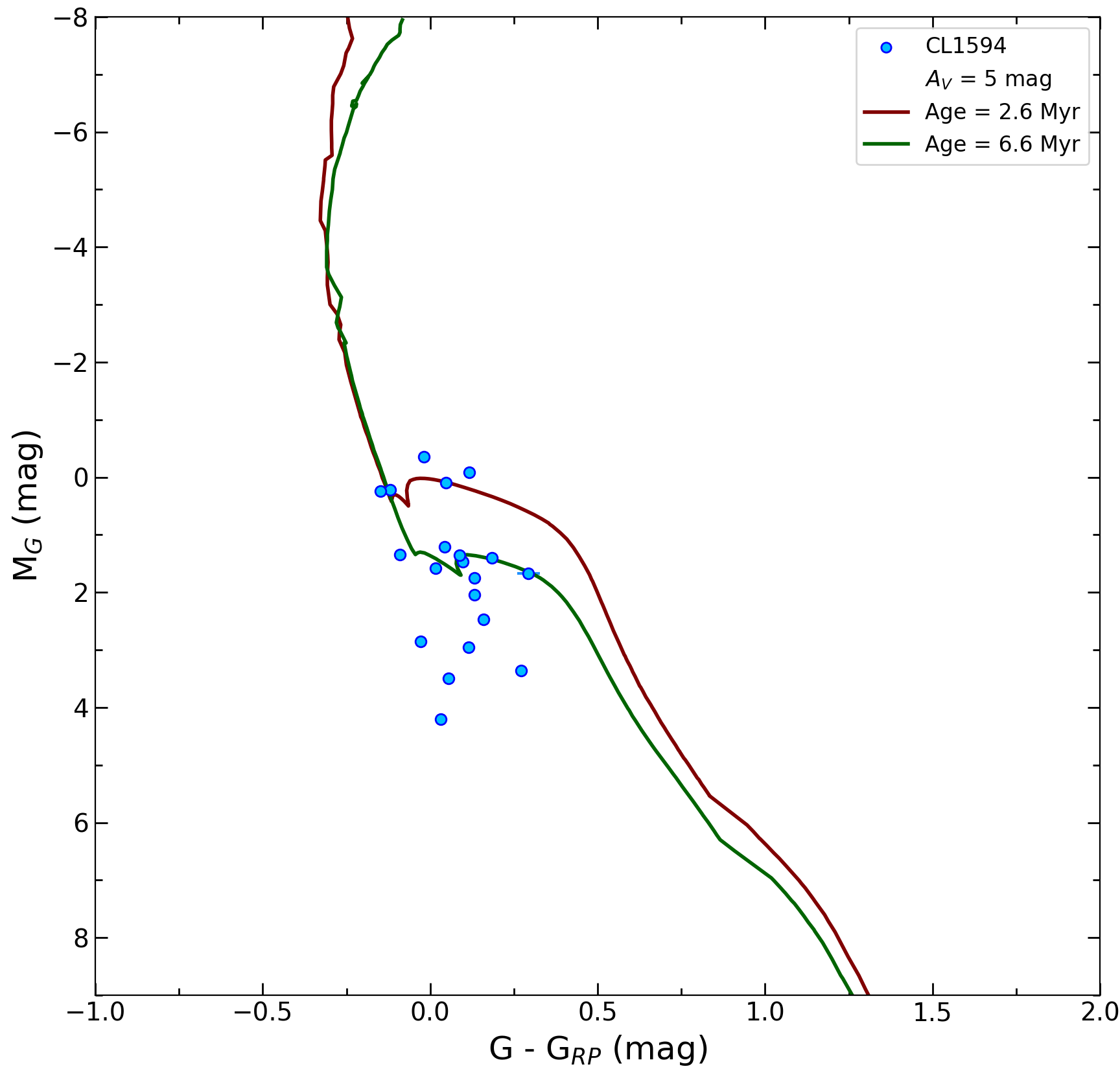}
\includegraphics[width=0.3\linewidth]{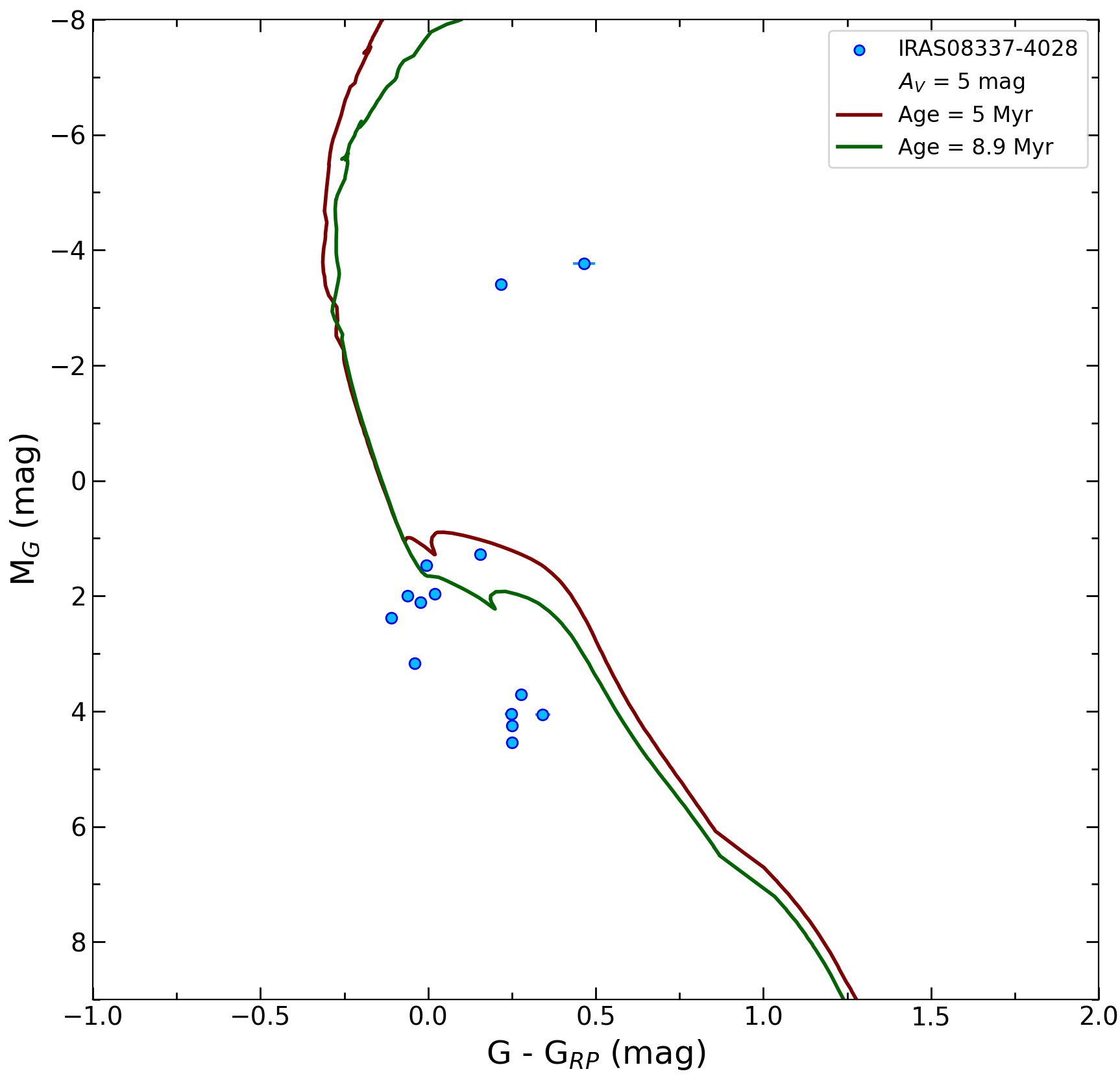}
\includegraphics[width=0.3\linewidth]{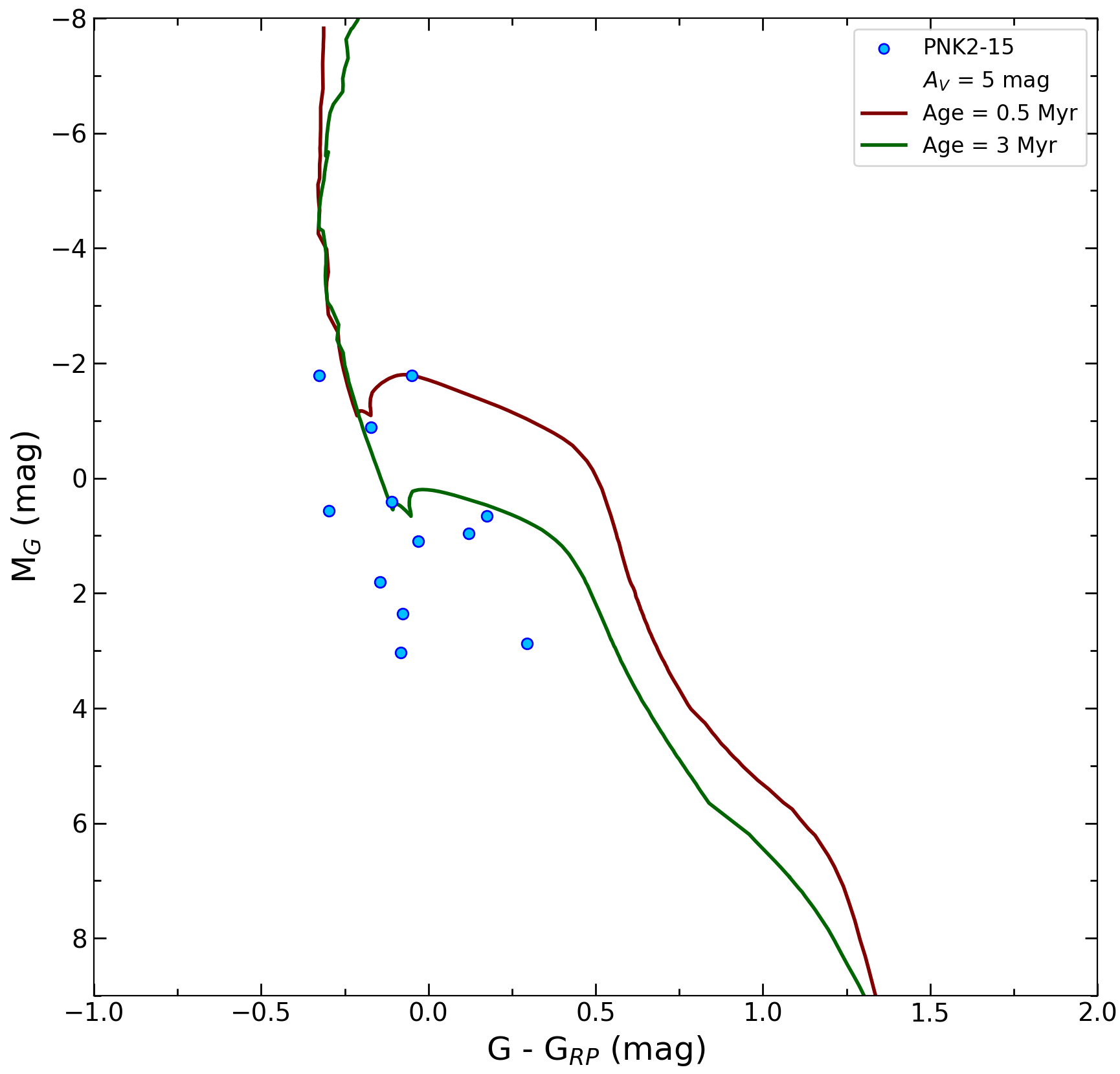}
\includegraphics[width=0.3\linewidth]{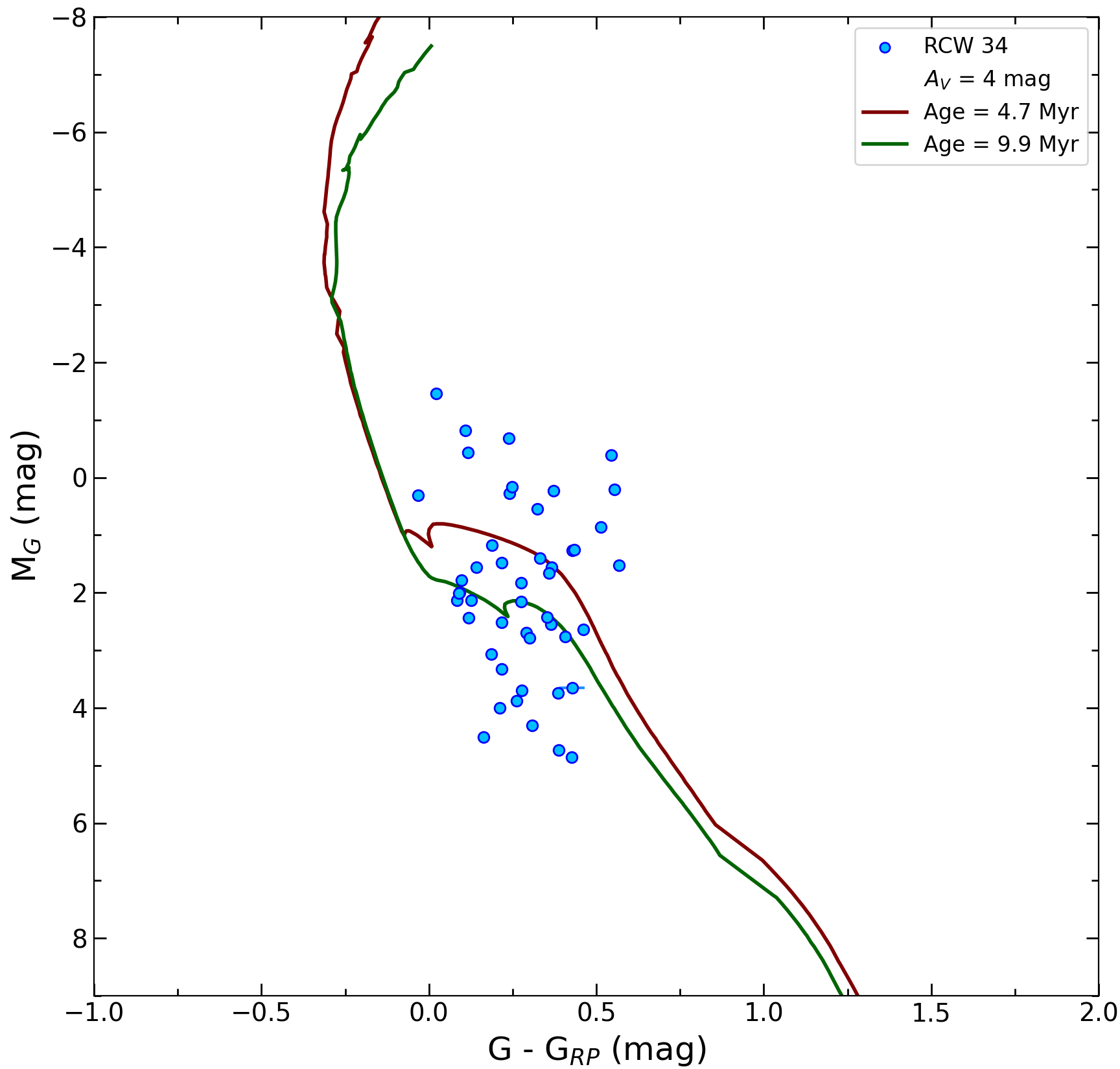}
\includegraphics[width=0.3\linewidth]{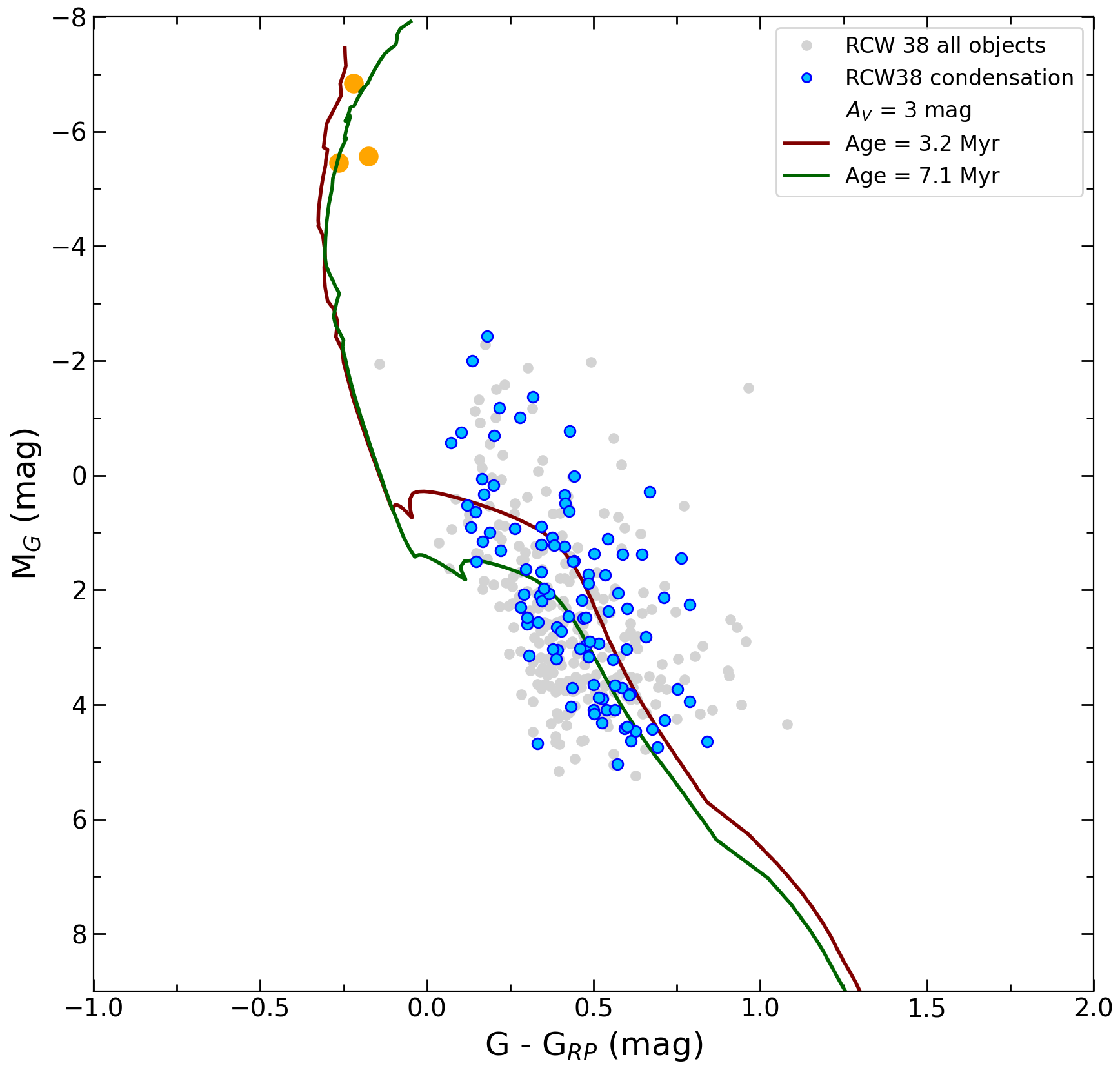}
\includegraphics[width=0.3\linewidth]{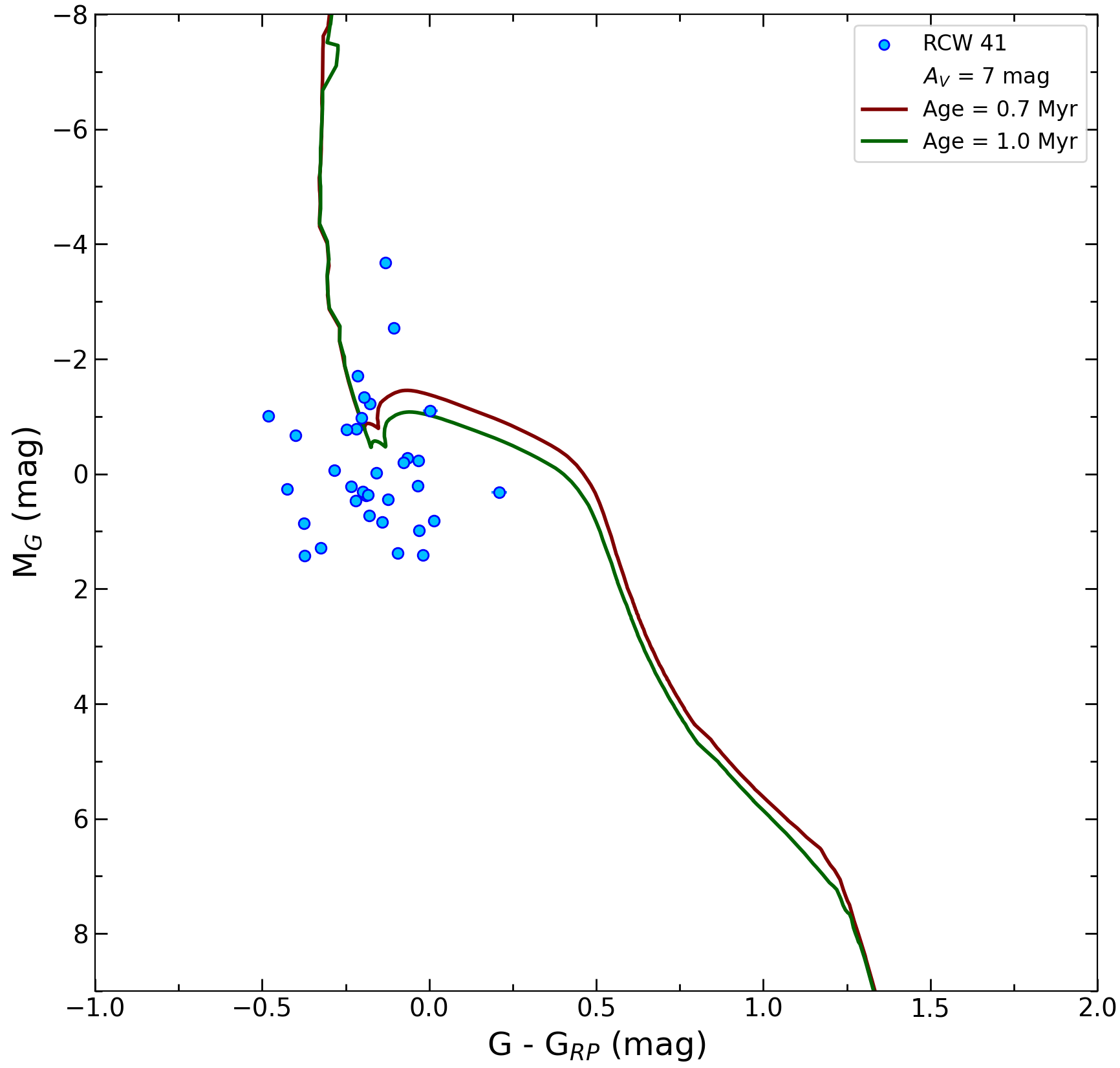}
\caption{Colour-absolute magnitude diagrams for members of each parent cluster. The members
are blue dots. In RCW\,38 the fitting is carried out for objects with similar A$_v$ (blue dots) named "condensation" (see Fig. \ref{fig:lb_sep}) not with all members (grey dots). The upper and lower age isochrones are based on 100 iterations and are represented by the red and green curves, respectively. Orange circles are OB-star members of RCW\,38 (see Tab.~\ref{tab:par1}). Absolute magnitudes are determined from individual parallaxes. The [KPS2012]\,MWSC\,1594 open cluster is indicated as Cl\,1594.}
\label{fig:isoch}
\end{figure*}

Figure \ref{fig:isoch} presents colour-absolute magnitude diagrams for the individual young clusters contained in Vel~OB1 based on the \textit{Gaia} membership analysis; cluster members are represented by the blue dots (including error bars). We estimated the mean reddening of the cluster based on a near-infrared colour-colour diagram of the 2MASS counterparts of the {\it Gaia} DR3 members and took this value to obtain the best-fit isochrone. The mean reddening $A_{V}$ and best age are indicated in the legend of each figure. The extinction in the Vel~OB1 region varies strongly; in Tab.~\ref{tab:parori} we provide the range measured in the near-infrared colour-colour diagram for each cluster. For comparison, in each panel we plot isochrones corresponding to a minimum (red curve) and maximum (green curve) age based on the 100 iterations for each cluster.

\section{AllWISE images OB runaways with no evidence of wind bow shock}

\begin{figure*}
\centering
\includegraphics[width=0.95\linewidth]{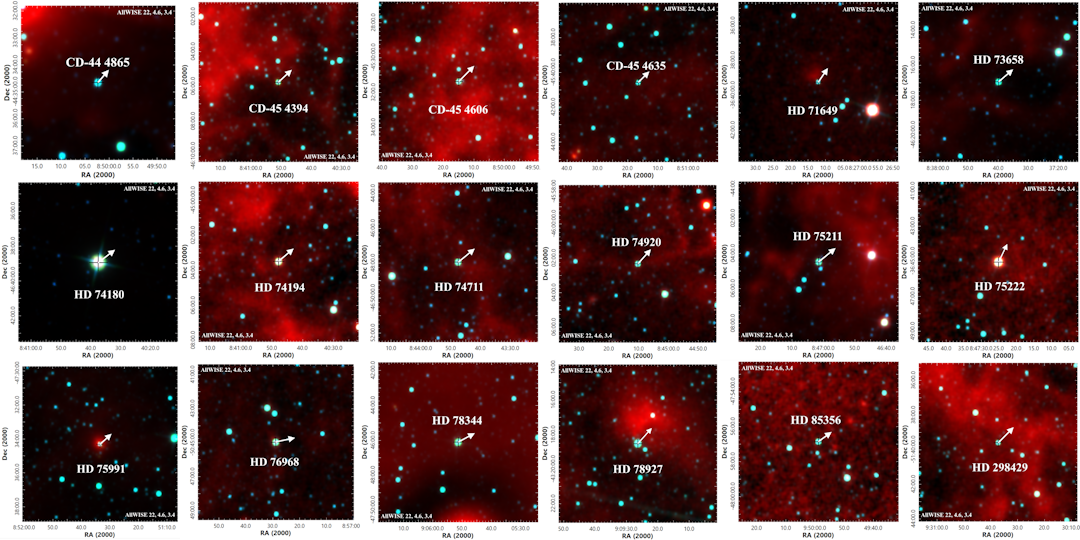}
\caption{AllWISE colour-composite images of runaway stars with high peculiar velocity in the Vel\,OB1 association. The runaway stars are marked with a cross. The white arrow indicates the direction of current proper motion of each star.
}
\label{fig:noBow}
\end{figure*}

We explored the AllWISE survey in the vicinity of all OB runaway stars associated with Vel~OB1 to search for arc-like structures. Figure~\ref{fig:noBow} presents AllWISE colour-composite images of 18 OB runaway stars in Vel~OB1. Although in some cases (e.g. HD~75991, HD~71649, HD~78927) emission features are present close to the stars and could be related to their stellar winds, the available data are insufficient to draw firm conclusions.

\section{Notes on individual objects}
\label{appD}

HD\,75222 is a B0~Ia supergiant compiled in the list of \citet{reed00} (Tab.~\ref{tab:par1}) and identified as an OB runaway star based on its 2D peculiar velocity of 72~km~s$^{-1}$, in agreement with the findings of \citet{Carretero23}. It is moving away from us with a radial velocity of 64~km~s$^{-1}$ at its current distance of 2017~pc and height above the plane of 147~pc. When reconstructing its projected path on the sky, one obtains a match with the young open cluster Bochum~7 \citep[$d \simeq 2000$~pc, age $\sim 27$~Myr, ][]{Song22,gaia23} 5.5~Myr ago. 

CD-41\,4637 is a massive binary with an unresolved companion \citep{aldoretta15} and is not included in the list of \citet{reed00} (Tab.~\ref{tab:par1}). The spectral type of the brightest component is O6\,Ib(f)(n) \citep{vijapurkar93} and its current distance is 2280~pc \citep{bailer21} at a height of 92~pc above the Galactic plane. The AllWISE colour-composite image (Fig. \ref{fig:runaways}) presents a bow shock aligned with the runaway velocity of CD-41\,4637 confirming its supersonic motion. \citet{aldoretta15} suggest that CD-41\,4637 is not associated with any cluster or association, that it is not a runaway star, and that its distance is 3.7\,kpc. The peculiar transverse velocity of CD-41\,4637 star (V$_{pec}$=35\,km\,s$^{-1}$), however, demonstrates its runaway nature. \citet{denoyelle77} included CD-41\,4637 in their list of young stars with A$_v$\,=\,3.25\,mag and obtained a distance of 2.09\,kpc. \citet{kilkenny93} and \citet{klare77} estimated a distance of 2.3\,kpc and 2.66\,kpc, respectively, in good agreement with the distance based on the \textit{Gaia} parallax.

The path of CD-41\,4637 suggests that it escaped from RCW\,34 about 2\,Myr ago. The estimated age range of the RCW\,34 H\,II region and cluster is 4.7 -- 9.9~Myr (Tab. \ref{tab:parori}), consistent with the kinematic age of CD-41\,4637 suggesting a dynamical ejection origin of the young binary \citep{denoyelle77}. \citet{bik10} mention that the central star \citep[vdBH 25a, ][]{Bergh75} of the H\,II region might not have created an associated bubble since it is near the northern edge. Possibly the runaway O supergiant CD-41\,4637 has contributed to the formation of this bubble since it moved away from the cluster (but see \ref{Impact}).

HD\,75860 is a B1.5~Iabp blue supergiant with a mass of 15.5\,M$_{\sun}$ \citep{hohle10} and a peculiar velocity of 31\,km\,s$^{-1}$. \citet{denoyelle77} included the star in their list of young stars (A$_v$\,=\,2.67\,mag) and obtained a distance of 1.7\,kpc. The \textit{Gaia} distance is 2.27$^{+0.20}_{−0.23}$\,kpc \citep{bailer21}. The AllWISE colour-composite image (Fig. \ref{fig:runaways}) shows the aligned bow shock. HD\,75860 is currently located in the Gum\,18 H\,II region. Therefore, the determined Str\"omgren radius (Tab. \ref{tab:estpar1}) may be spurious and the extraction of the extended H\,II region created by UV emission of only HD\,75860 is challenging. The observed bow shock is one of three arc-like molecular structures in Gum\,18 H\,II region \citep{elia07}. The path of HD\,75860 matches with an origin in the [KPS2012]\,MWSC\,1594 (indicated as Cl~1594 in this work) open cluster 1.2\,Myr ago. The parameters of the cluster: distance 1.9\,kpc, age $\sim$\,130\,Myr, and $\mu_{\alpha}^*$ of -4.94\,mas\,yr$^{-1}$ and $\mu_{\delta}$ of 6.18\,mas\,yr$^{-1}$ \citep{kharchenko13} differ from our results (see Tab.~\ref{tab:parori}).  \citet{froebrich07} identified 579 stars in the cluster; we found only 21 members and arrive at a distance of 2.084$^{+0.043}_{-0.041}$\,kpc. Thus, the Cl~1594 cluster and HD\,75860 are located at the same distance. From the age range of Cl~1594 (2.6$-$6.6\,Myr), we propose that HD\,75860 underwent a dynamical ejection. Exploring the region surrounding the Cl~1594 open cluster, an unknown star-forming region may be aligned with the open cluster. There are several reflection nebulae in this region \citep{Magakian03}. Taking into account the proper motion and age discrepancy between our results (Tab. \ref{tab:parori}) and the data known from the literature, probably the parent cluster of HD\,75860 is the star-forming region (with our estimated parameters) rather than the Cl~1594 open cluster.

HD\,76968 is an O9.2~Ib supergiant that has also been identified as an OB runaway by \cite{Carretero23} who report a peculiar tangential velocity of 56~km~s$^{-1}$, and we arrive at a similar value of 53~km~s$^{-1}$ (Tab.~\ref{tab:parrunori}). We do not detect a wind bow shock and obtain a 3D space velocity of 80~km~s$^{-1}$ with respect to the LSR and a Z distance of 126~pc below the Galactic plane. Also for this object Bochum~7 could be a potential parent cluster: the kinematical age would then be 4~Myr.

HD\,298310 is a B0\,V type star with an estimated mass of $\sim$\,20\,M$_{\sun}$ \citep{kobulnicky19} and a peculiar tangential velocity of about 60~km~s$^{-1}$. \citet{denoyelle77} included HD\,298310 in their list of young stars with A$_v$\,=\,3.71\,mag and obtained a distance of 1.91\,kpc in agreement with the \textit{Gaia} distance of 1.9$^{+0.04}_{−0.04}$\,kpc \citep{bailer21}. The AllWISE colour-composite image of HD\,298310 (Fig. \ref{fig:runaways}) shows a bow shock associated with it, although it does not seem to be well aligned with its current motion. The cluster RCW~38 is the likely origin of HD~298310; it moves with a velocity of 52~km~s$^{-1}$ with respect to RCW~38 and has escaped from the cluster 3.3~Myr ago.

IRAS\,08351-3951 is an M8 type star with a space velocity of 74\,km\,s$^{-1}$. The AllWISE colour-composite image (Fig. \ref{fig:runaways}) shows an unaligned, not earlier reported arc-like feature close to it. Only in this case, the peak of the arc-like feature emission changes depending on the wavelength range, i.e. the direction from the star to the peak of the 12\,$\mu$m emission (dashed green arrow) is different from that measured in the 22\,$\mu$m emission (green arrow). The misaligned bow shock could be explained by large-scale ISM motions \citep{Povich08}. The path of IRAS\,08351-3951 star matches with the IRAS\,08337-4028 H\,II region about 1\,Myr ago. IRAS\,08337-4028 region contains 39\,M-type young stellar objects \citep{helou88}. The age of IRAS\,08337-4028 ranges between 1 and 7\,Myr \citep{prisinzano18}. We identify only 14 members and estimate a cluster age of 0.7$-$1.0\,Myr (Tab. \ref{tab:parori}). However, the difference in distance of IRAS\,08351-3951 (3.97$^{+0.9}_{−0.6}$\,kpc) and the IRAS\,08337-4028 region (1.6$^{+0.032}_{-0.031}$\,kpc) raises suspicions about their connection in the past. 

CD-42\,4694 is an O9\,V type star \citep{vijapurkar93} and has a peculiar tangential velocity of 25\,km\,s$^{-1}$. \citet{denoyelle77} included the star in their list of young stars (A$_v$\,=\,2.9\,mag) and obtained a distance of 3.5\,kpc. The AllWISE colour-composite image (Fig. \ref{fig:runaways}) shows a misaligned bow shock which could be explained by large-scale ISM motions. The constructed path of the star matches with an origin in the PN\,K\,2-15 cluster about 2-3\,Myr ago. \citet{massi10} identified more than 200 members of the PN\,K\,2-15 cluster and estimated an age of 1-2\,Myr. We identify only 12 members and estimate an age of 0.5$\pm$3.0\,Myr. The age range is in agreement with the kinematical age of CD-42\,4694. However, the different distance of the star \citep[2.47$^{+0.15}_{−0.14}$\,kpc, ][]{bailer21} compared to that of the PN\,K\,2-15 cluster (1.79$^{+0.023}_{-0.022}$\,kpc, Tab. \ref{tab:parori}) raises suspicions about their connection in the past. 

GSC\,08152-02059 is an OB type star \citep{muzzio77}. The AllWISE colour-composite image (Fig. \ref{fig:runaways}) shows a not earlier reported small arc-like feature very close to it and aligned with its motion (14~km~s$^{-1}$). We have insufficient information to show that this feature is due to a wind bow shock (Sect.~\ref{DisBS}), and therefore its runaway nature remains to be confirmed. Its path crossed the boundary of RCW~38 about 3--4~Myr ago; its current velocity with respect to RCW~38 is only 4~km~s$^{-1}$ which may just be sufficient to have escaped from the cluster (but see \ref{Impact}).

DR3\,N72 with \texttt{source\_id} of 5326969910562195072 from \textit{Gaia}\,DR3 is indicated as N\,72 in this work. The AllWISE colour-composite image (Fig. \ref{fig:runaways}) shows an unaligned, not earlier reported arc-like feature close to it. The highest emission of the arc-like feature is not correlated with its symmetry. The arc-like feature remains unaligned even if we take the centre of symmetry. The existence of the unaligned bow shock could be explained by large-scale ISM motions \citep{Povich08}. 

DR3\,N60 with \texttt{source\_id} of 5325934754726652160 from \textit{Gaia}\,DR3 is indicated as N\,60 in this work. Close to it, the AllWISE colour-composite image (Fig. \ref{fig:runaways}) shows an unaligned bow shock-like structure not reported previously, which could be explained by large-scale ISM motions.

\end{document}